\documentclass[12pt]{article}
\pdfoutput=1
\usepackage{tikz}
\usetikzlibrary{matrix,arrows,decorations.pathmorphing,decorations.pathreplacing,decorations.markings,patterns,shapes}
\usepackage{pgfplots}
\usepackage{jheppub}
\usepackage{amsmath,amssymb,euscript,array,mathrsfs,appendix}
\usepackage{arydshln}
\usepackage{todonotes}
\usepackage{graphicx}
\newcommand{\IM}{\operatorname{Im}}
\newcommand{\Tr}{\operatorname{Tr}}
\newcommand{\RE}{\operatorname{Re}}

\def\a{\alpha}

\def\d{\delta}

\def\l{\lambda}
\def\m{\mu}
\def\n{\nu}

\def\p{\pi}
\def\r{\rho}
\def\s{\sigma}
\def\t{\tau}

\def\w{\omega}

\def\D{\Delta}

\def\RR{{\mathfrak R}}

\def\Dbarslash{\,\,{\raise.15ex\hbox{/}\mkern-12mu {\bar D}}}
\def\Dslash{\,\,{\raise.15ex\hbox{/}\mkern-12mu D}}
\def\delslash{\,\,{\raise.15ex\hbox{/}\mkern-9mu \partial}}
\def\delbarslash{\,\,{\raise.15ex\hbox{/}\mkern-9mu {\bar\partial}}}

\def\RR{{\mathfrak R}}

\def\rta{\rightarrow}
\def\dprime{\prime\prime}

\newcommand{\EQ}[1]{\begin{equation}\begin{split} #1 \end{split}\end{equation}}

\newcommand{\SP}[1]{\begin{equation}\begin{split} #1
\end{split}\end{equation}}

\title{Causality Violation, Gravitational Shockwaves and UV Completion}

\author{Timothy J. Hollowood and Graham M. Shore}
\affiliation{Department of Physics,\\ Swansea University,\\
Swansea,\\ SA2 8PP, UK.}
\emailAdd{t.hollowood@swansea.ac.uk, g.m.shore@swansea.ac.uk}

\abstract{The effective actions describing the low-energy dynamics of QFTs involving gravity generically exhibit
causality violations. These may take the form of superluminal propagation or Shapiro time advances and allow the
construction of ``time machines'', i.e.~spacetimes admitting closed non-spacelike curves. 
Here, we discuss critically whether such causality violations may be used as a criterion to identify unphysical 
effective actions or whether, and how, causality problems may be resolved by embedding the action in a fundamental,
UV complete QFT. We study in detail the case of photon scattering in an Aichelburg-Sexl gravitational shockwave 
background and calculate the phase shifts in QED for all energies, demonstrating their smooth interpolation
from the causality-violating effective action values at low-energy to their manifestly causal high-energy limits.
At low energies, these phase shifts may be interpreted as backwards-in-time coordinate jumps as the photon
encounters the shock wavefront, and we illustrate how the resulting causality problems emerge and are
resolved in a two-shockwave time machine scenario.
The implications of our results for ultra-high (Planck) energy scattering, in which graviton exchange is modelled
by the shockwave background, are highlighted.}

\pgfdeclareimage[interpolate=true,width=8cm]{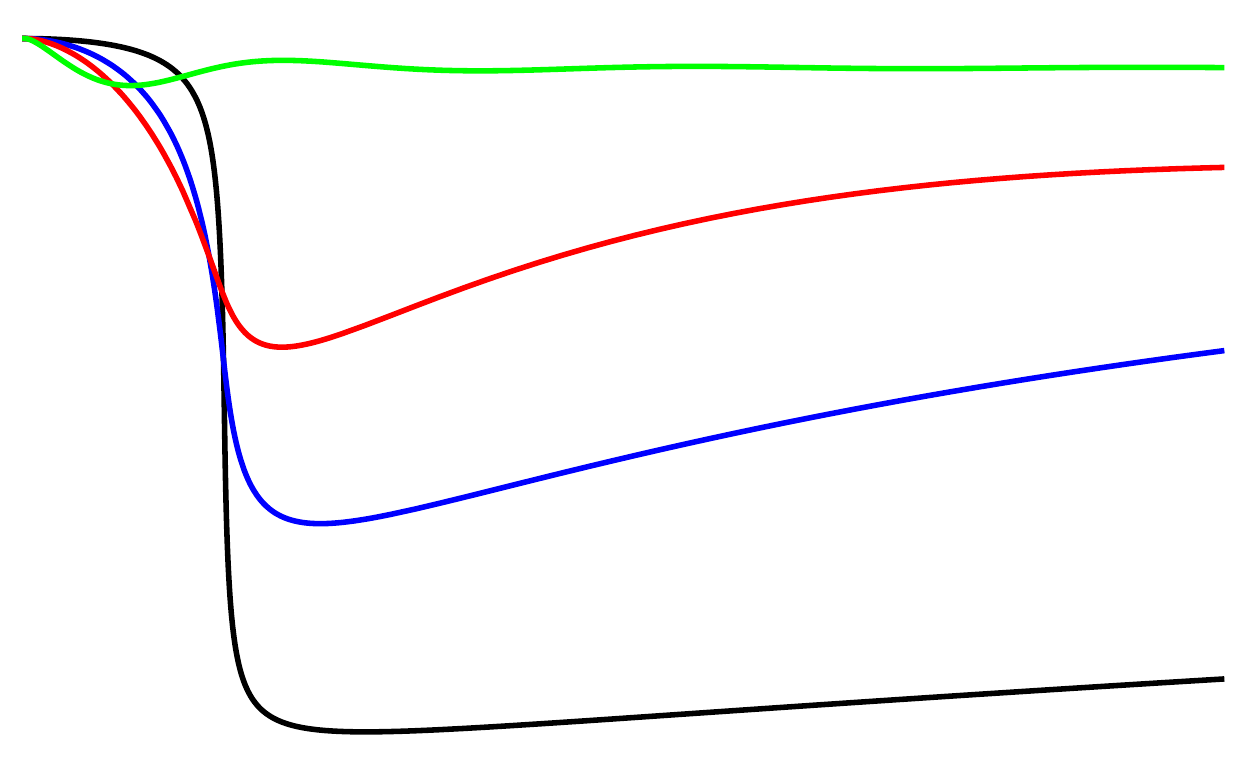}{npic1}
\pgfdeclareimage[interpolate=true,width=8cm]{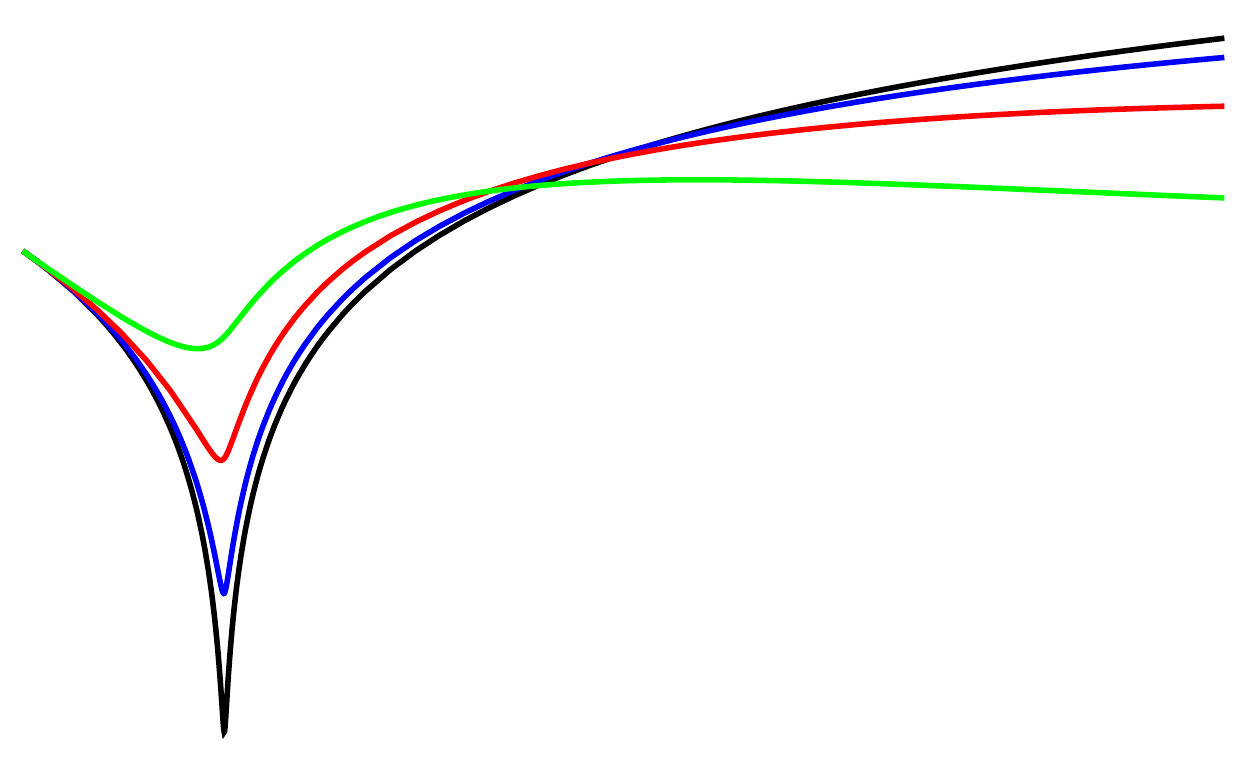}{npic2}
\pgfdeclareimage[interpolate=true,width=8cm]{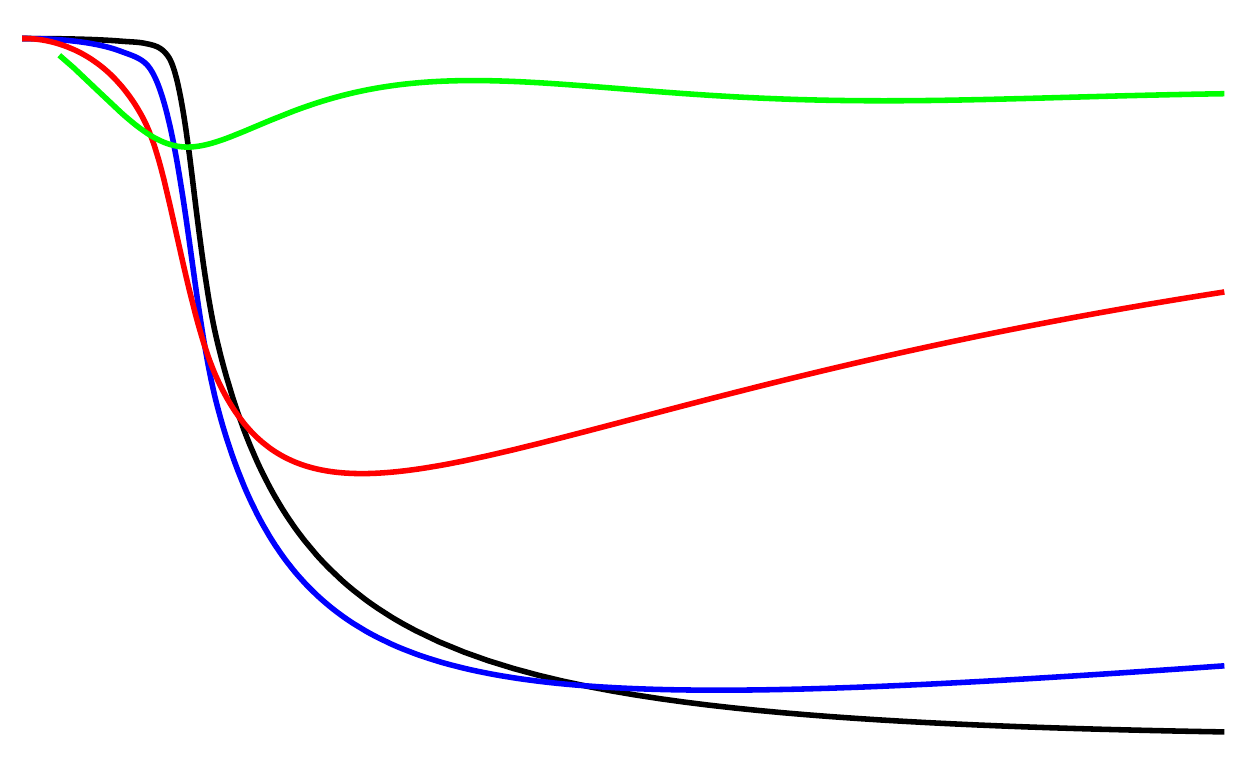}{npic3}
\pgfdeclareimage[interpolate=true,width=8cm]{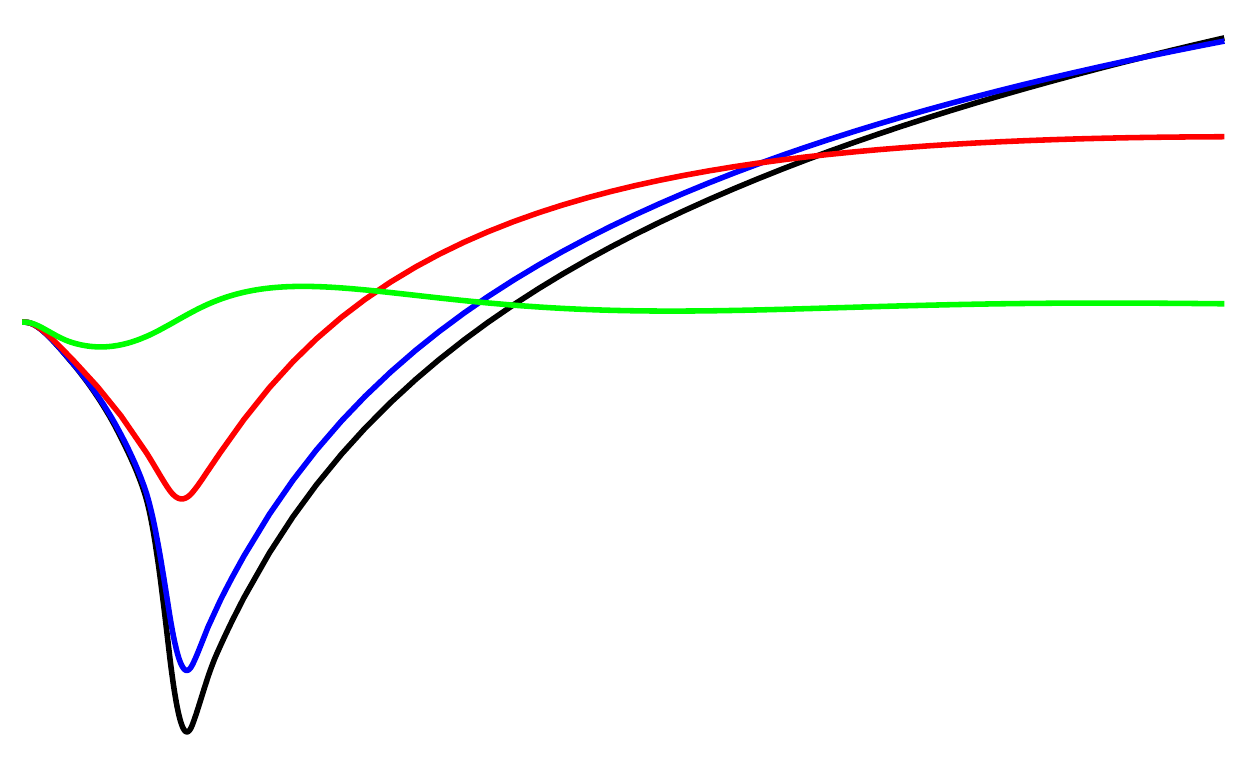}{npic4}
\pgfdeclareimage[interpolate=true,width=8cm]{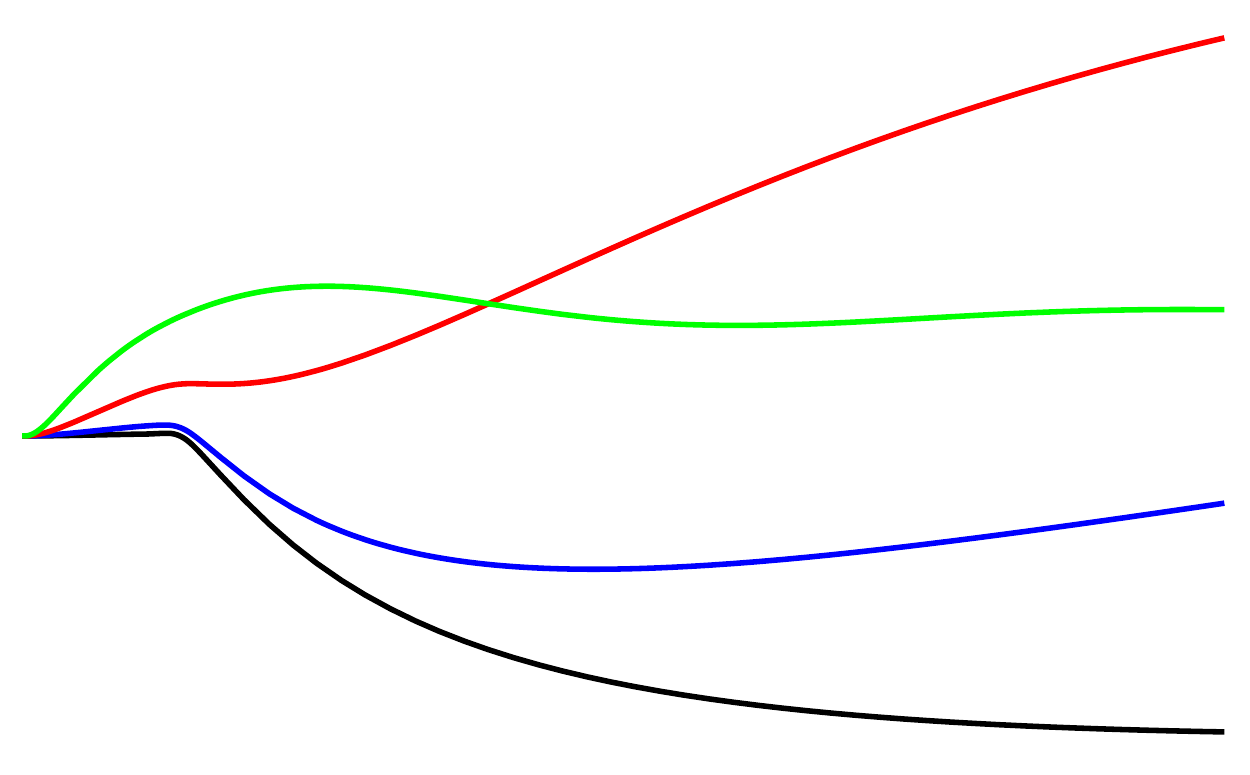}{npic5}
\pgfdeclareimage[interpolate=true,width=8cm]{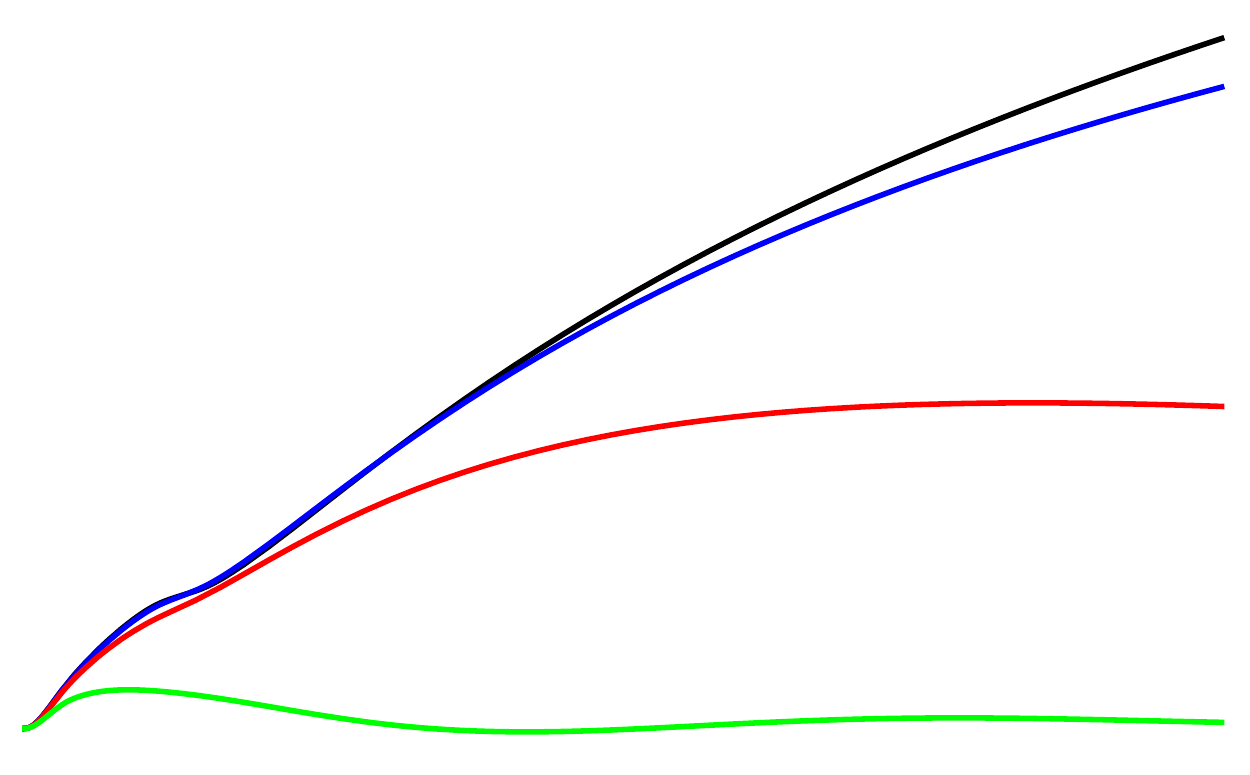}{npic6}
\pgfdeclareimage[interpolate=true,width=8cm]{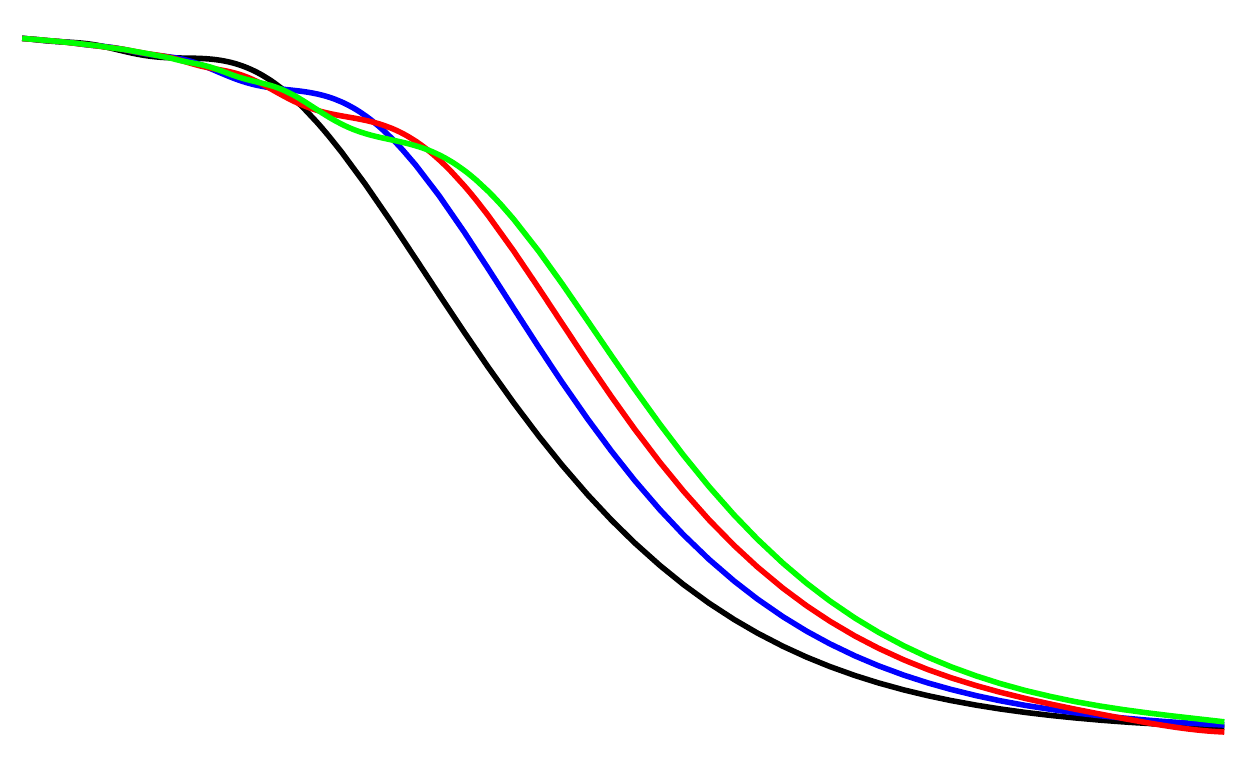}{npic7}
\pgfdeclareimage[interpolate=true,width=8cm]{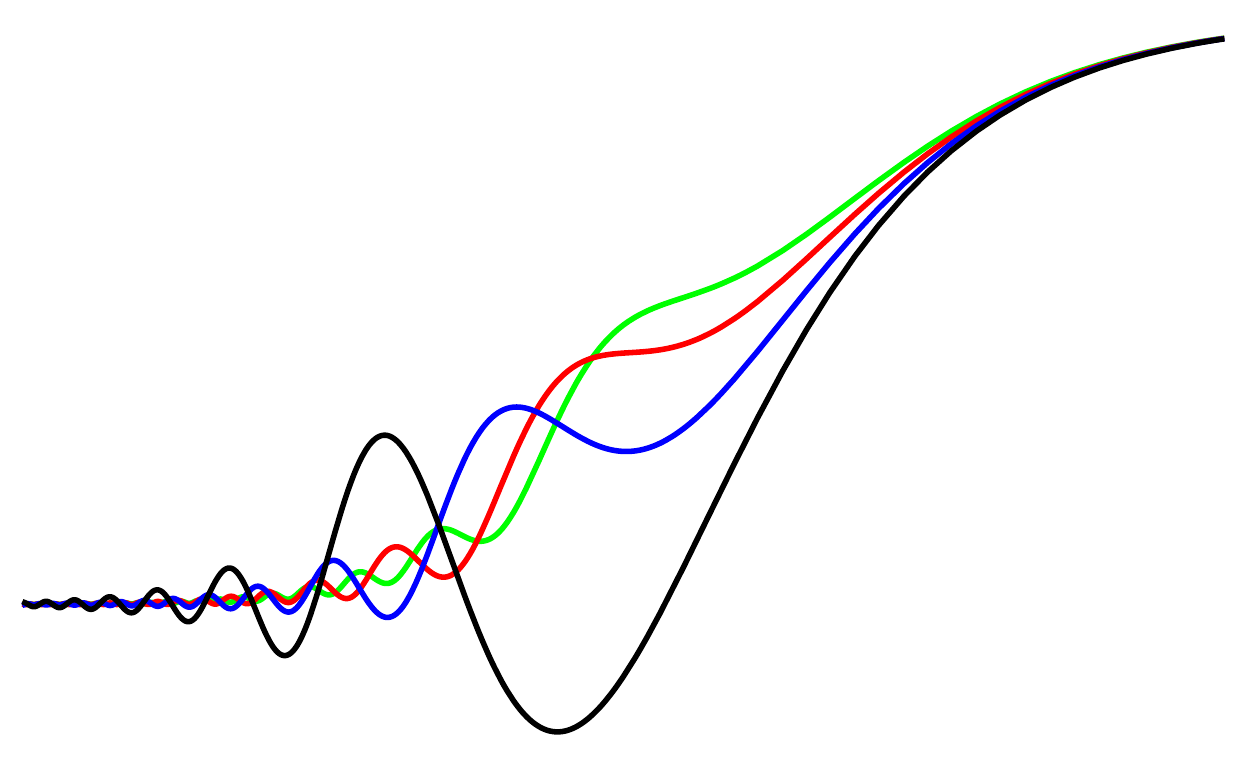}{npic8}
\pgfdeclareimage[interpolate=true,width=8cm]{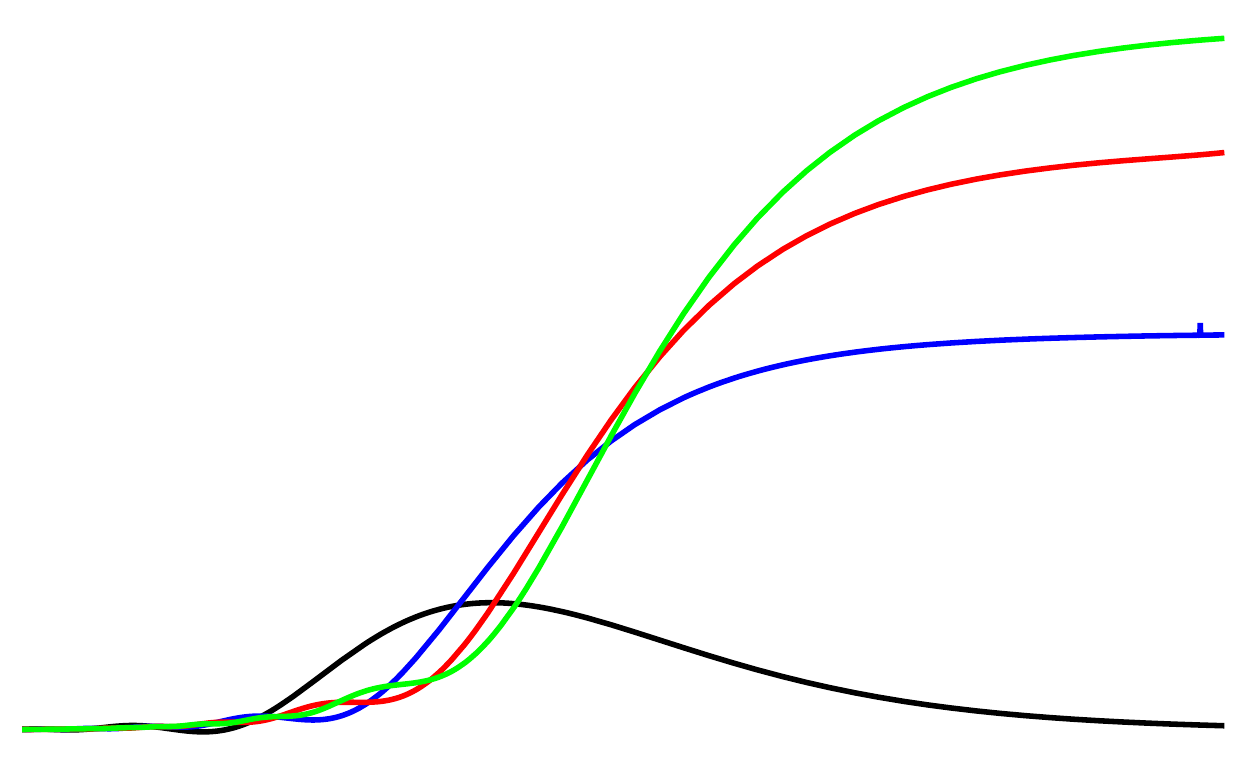}{npic9}
\pgfdeclareimage[interpolate=true,width=8cm]{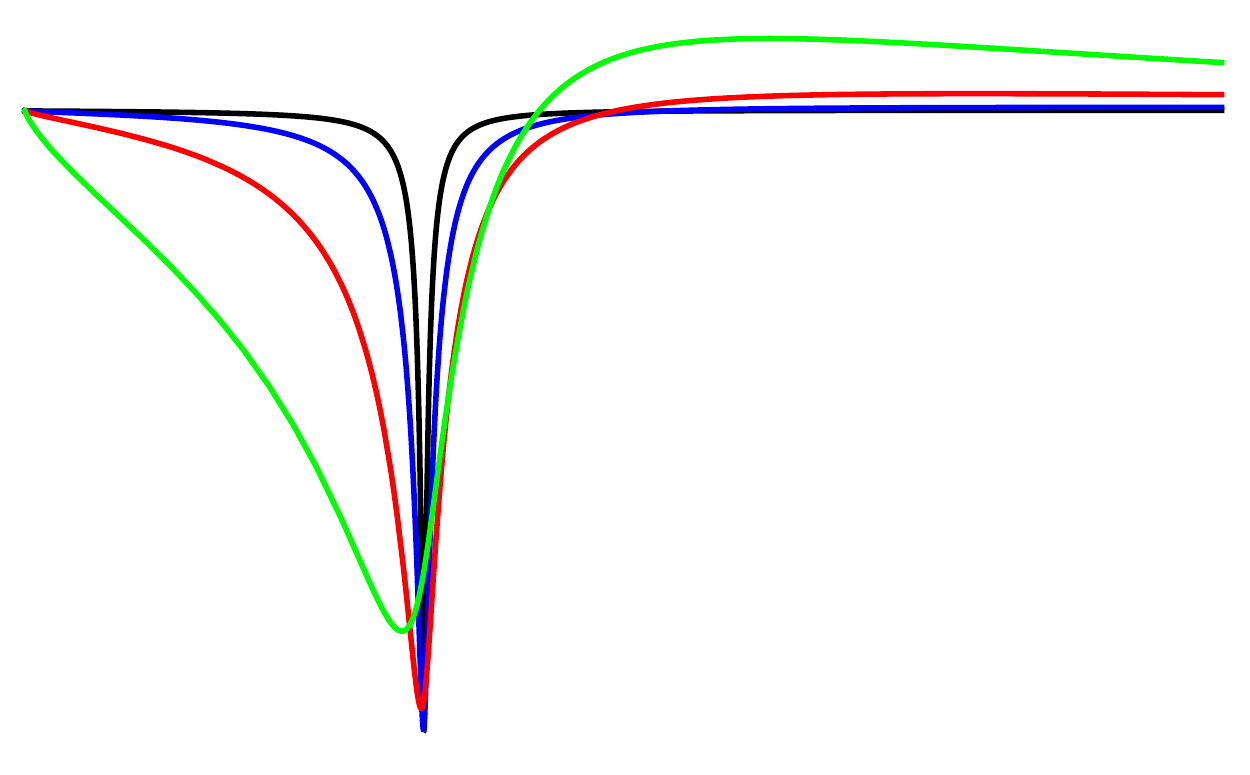}{npic10}
\pgfdeclareimage[interpolate=true,width=8cm]{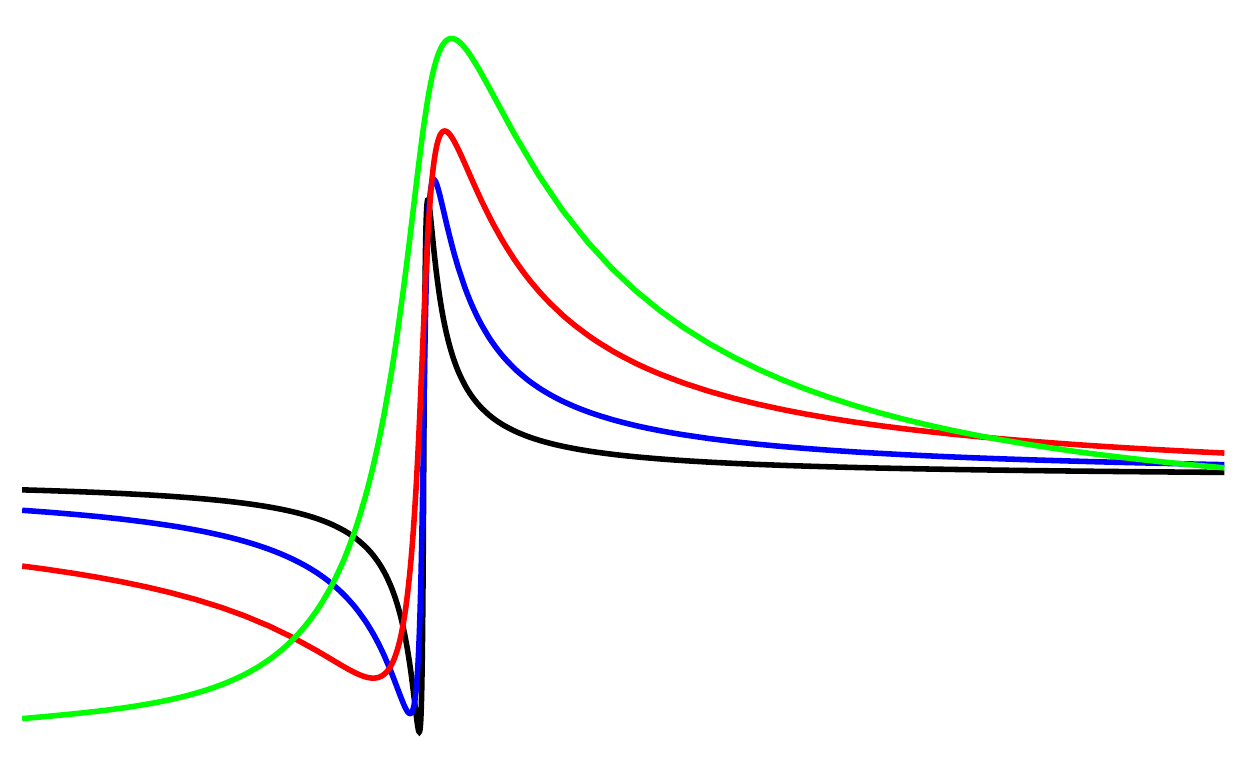}{npic11}
\pgfdeclareimage[interpolate=true,width=8cm]{npic12}{npic12}
\pgfdeclareimage[interpolate=true,width=8cm]{npic13}{npic13}
\pgfdeclareimage[interpolate=true,width=8cm]{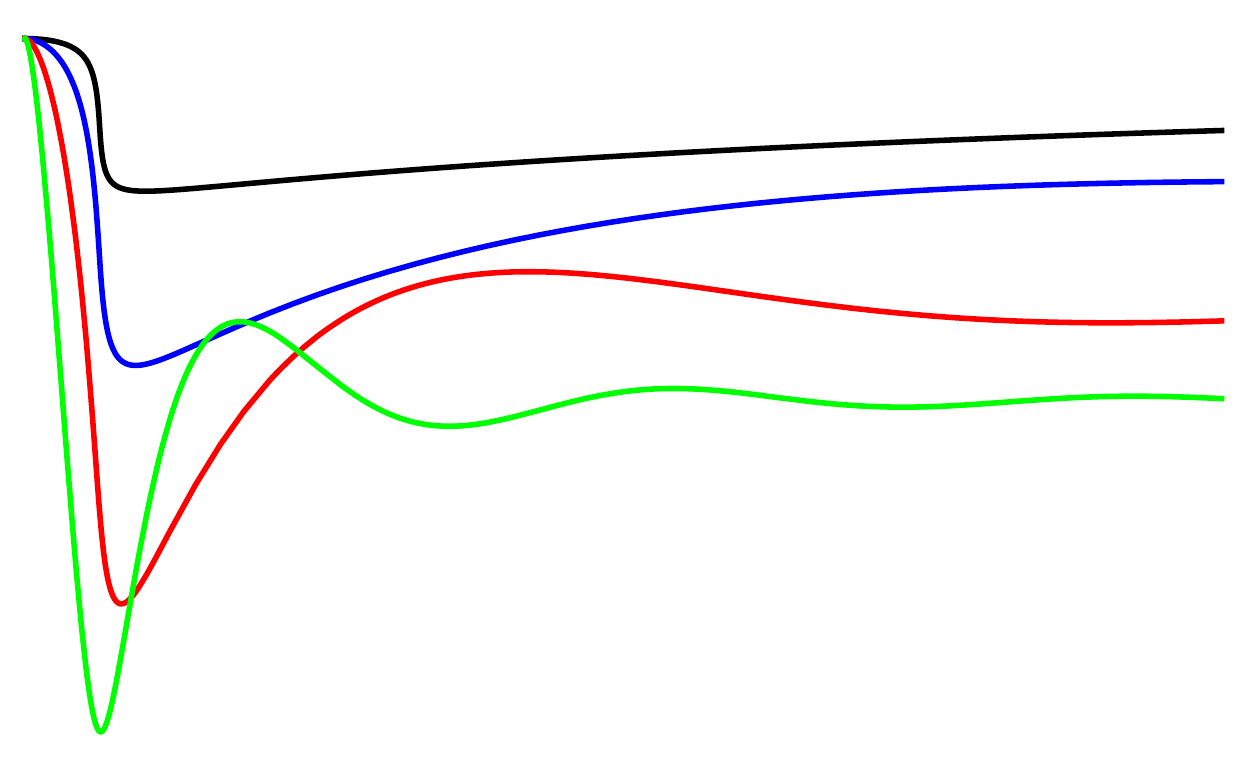}{npic14}
\pgfdeclareimage[interpolate=true,width=8cm]{npic15}{npic15}

\pgfdeclareimage[interpolate=true,width=8cm]{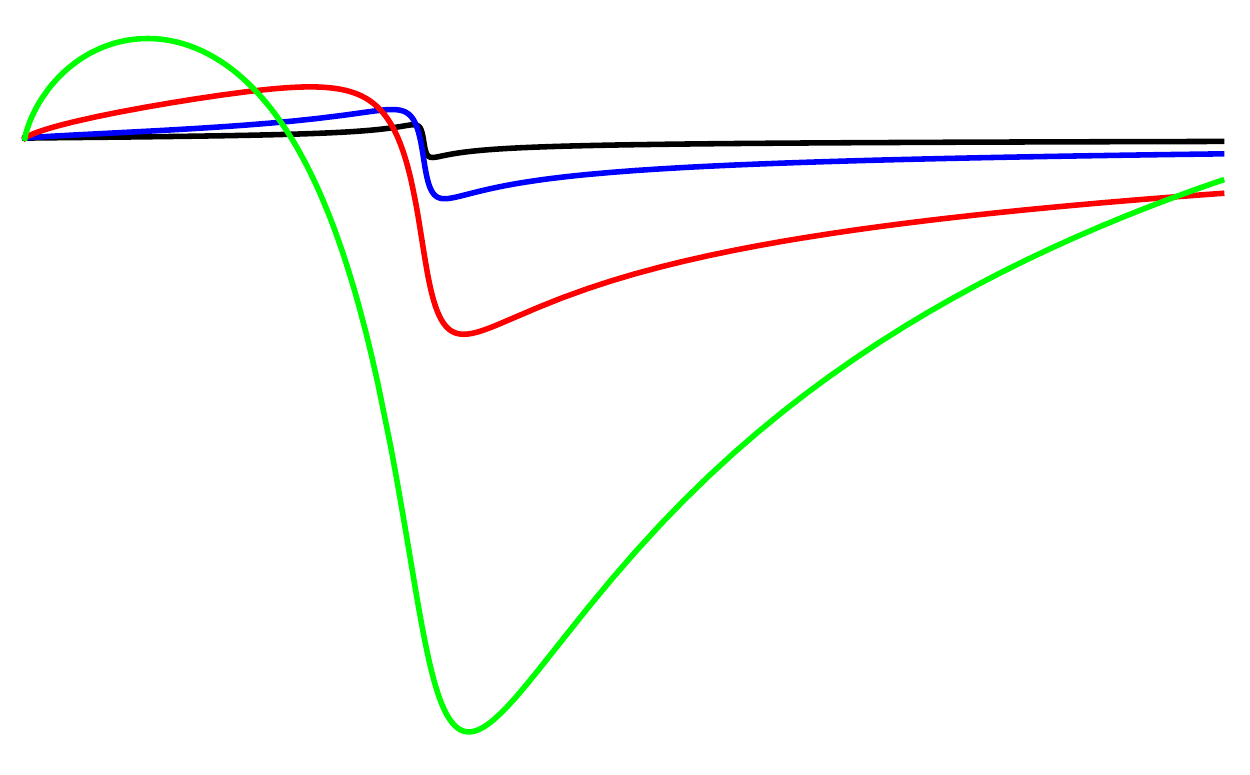}{spic1}
\pgfdeclareimage[interpolate=true,width=8cm]{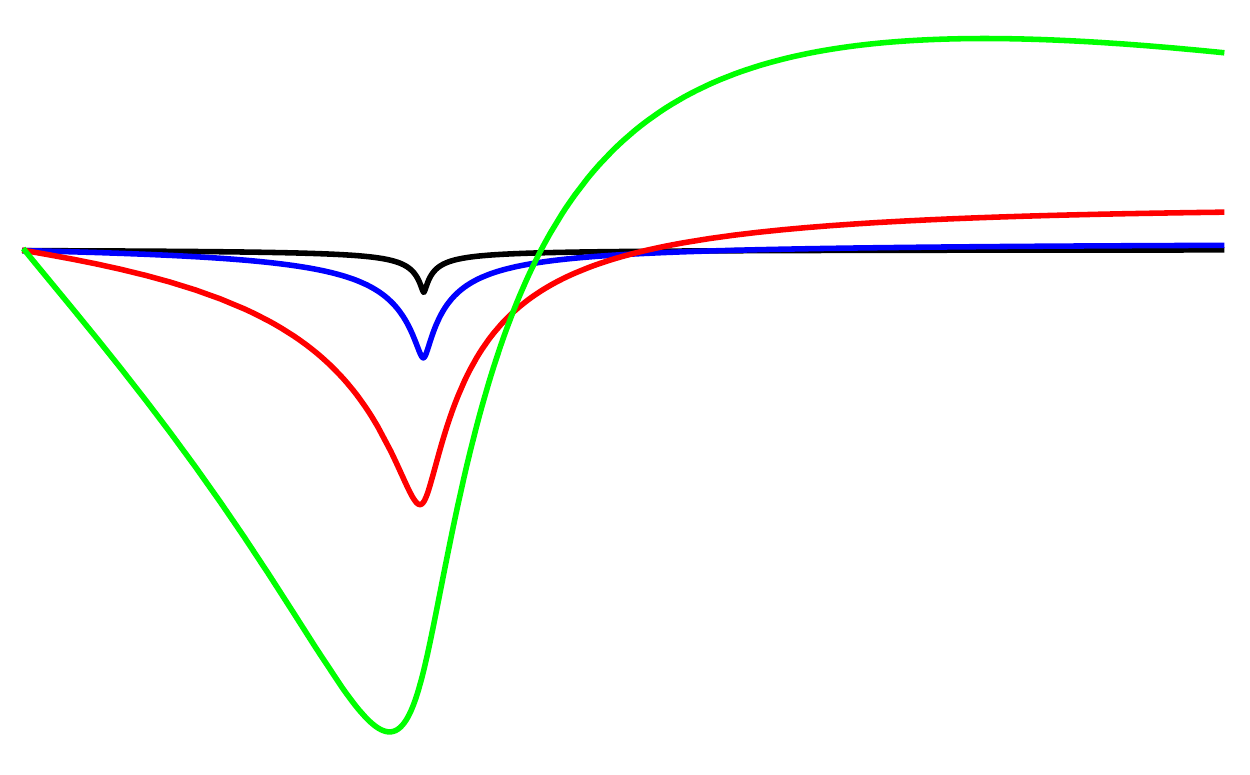}{spic2}
\pgfdeclareimage[interpolate=true,width=8cm]{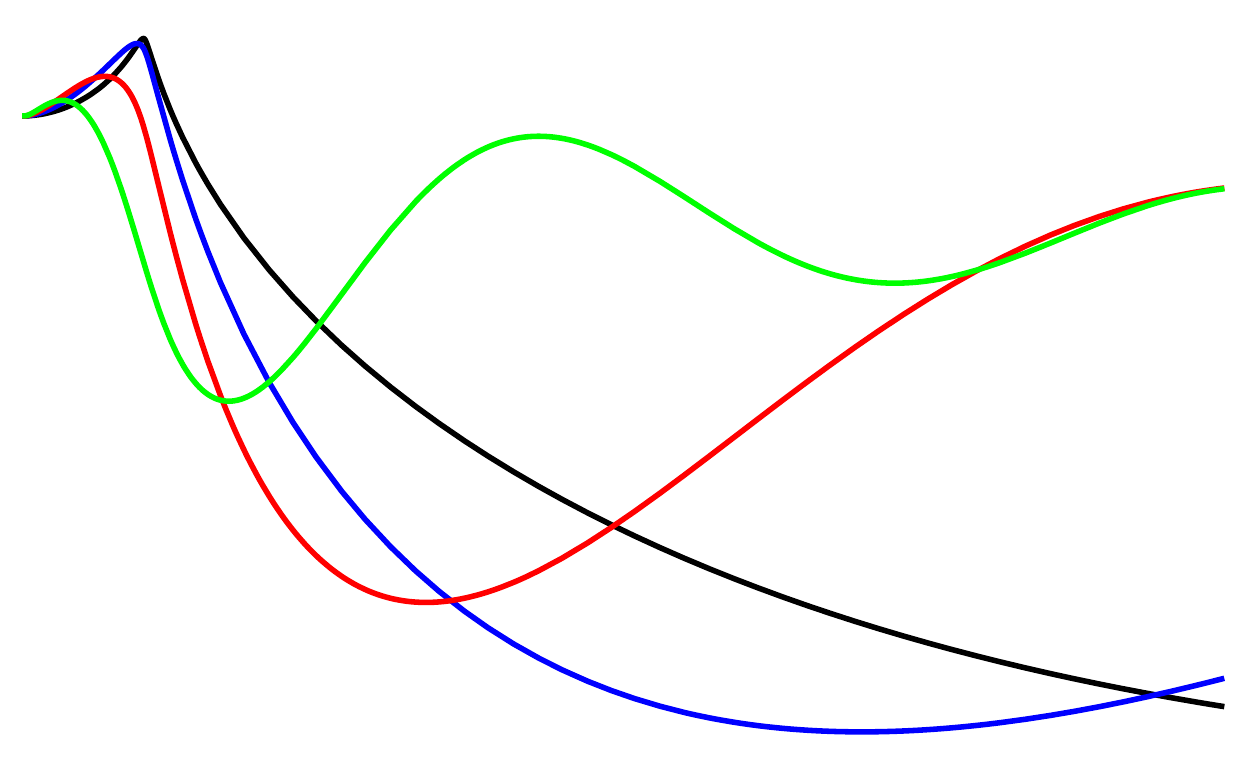}{spic3}
\pgfdeclareimage[interpolate=true,width=8cm]{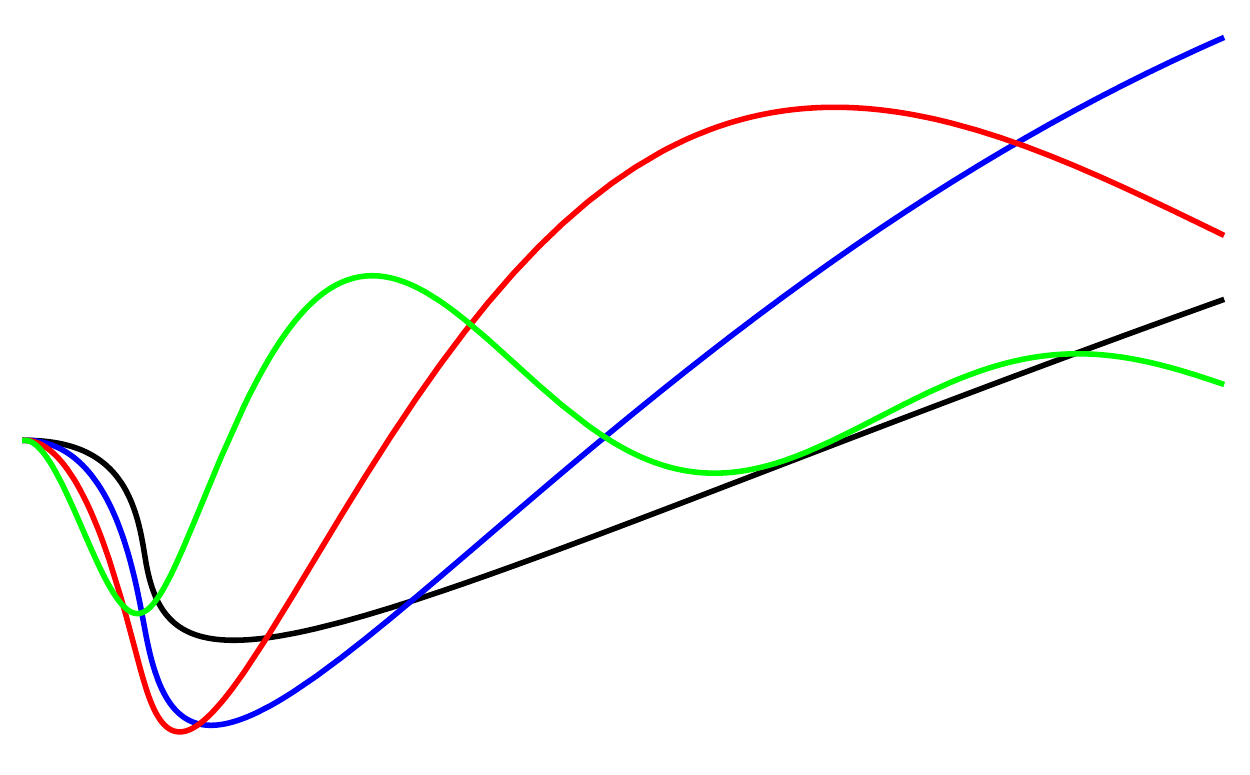}{spic4}
\pgfdeclareimage[interpolate=true,width=8cm]{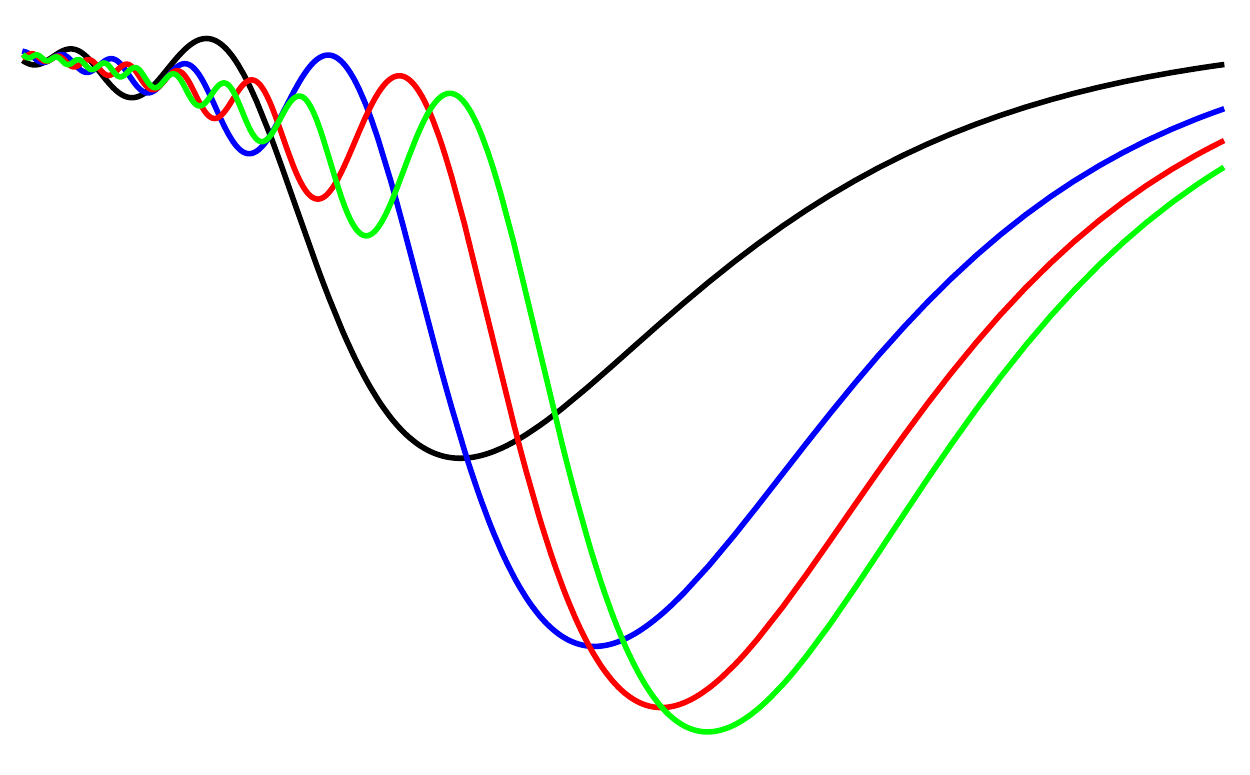}{spic5}
\pgfdeclareimage[interpolate=true,width=8cm]{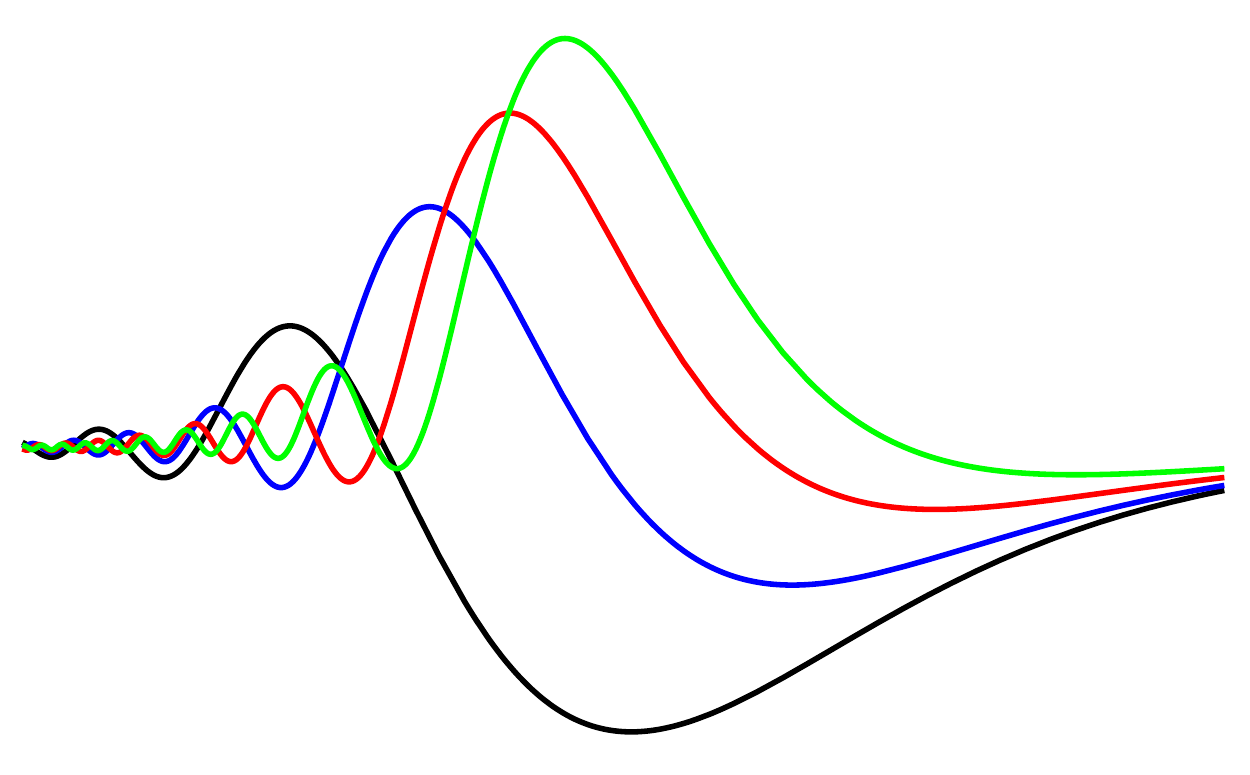}{spic6}
\pgfdeclareimage[interpolate=true,width=8cm]{spic7}{spic7}

\notoc
\begin{document}
\maketitle

\newpage

\section{Introduction}\label{s1}

% This is the first of a series of papers in which we investigate the relation of IR effective theories to their UV
% completion in quantum field theories in gravitational backgrounds. The focus will be on causality, in particular 
% how the apparent causality violations which generically arise in low-energy effective theories in curved spacetime
% can be resolved in a fundamental, UV complete theory.

The relation of IR effective theories to their UV completion in quantum field theories involving gravity is a rich and
far-reaching topic which raises many fundamental issues. In this paper, our focus will be on causality,  in particular 
how the apparent causality violations which generically arise in low-energy effective theories in curved spacetime
can be resolved in a fundamental, UV complete theory.

This work was inspired in part by the idea that causality may restrict the class of physical low-energy effective 
theories by placing constraints on the allowed values of the couplings \cite{Adams:2006sv}.
More recently, it has been proposed that circumventing the causality problems present in an effective theory in the
IR may be used as a guide to constructing a consistent, causal UV completion, especially for gravity 
itself \cite{Camanho:2014apa}.

The potential causality problems in effective theories may take the form of superluminal propagation, or the closely 
related {\it Shapiro time advances\/}, in certain gravitational backgrounds. Shapiro time advances can at first sight
be used to construct ``time machines", that is closed null or timelike trajectories for particles propagating in 
specifically engineered gravitational backgrounds.

The immediate question is whether these apparent causality violations do indeed imply that the effective theory
is unphysical, or whether, and how, causality is realised when the effective theory is embedded in a consistent,
causal UV completion. To test this, we consider a theory that has a known UV completion with sound causal 
properties,\footnote{Fundamentally, causality is guaranteed by the vanishing of the retarded Green functions outside
the backward light-cone. This is known to be true in QED, even in curved spacetime \cite{Hollowood:2008kq}.}
namely QED in curved spacetime, but which does display superluminal propagation (the Drummond-Hathrell effect 
\cite{Drummond:1979pp}) in its low-energy effective action, {\it i.e.}~at scales below the electron mass.
The spacetime is chosen to be the Aichelburg-Sexl gravitational shockwave \cite{Aichelburg:1970dh}, which even at the 
classical level admits null geodesics with discontinuous Shapiro time advances. The propagation of photons, dressed 
by vacuum polarization, in a gravitational shockwave spacetime therefore provides an excellent template for how causality 
problems that are manifest in an IR effective theory may be resolved if a consistent UV completion exists.

As demonstrated in our previous investigations of the realisation of causality in curved spacetime
\cite{Hollowood:2007kt, Hollowood:2007ku,Hollowood:2008kq,Hollowood:2009qz,Hollowood:2010bd,Hollowood:2011yh}, 
in order to verify that causality is respected we need to demonstrate that the phase velocity, which may be superluminal for low
frequencies, is equal to $1$ in the high-frequency limit \cite{Leontovich, Shore:2003jx,Shore:2007um,Hollowood:2008kq}. 
This implies constraints on the phase shift of the photon modes as they scatter from the shockwave.
Here, we complement this approach by using the Shapiro time advances in the effective theory to engineer 
potential time machines in a spacetime describing the collision of two shockwaves \cite{Shore:2002in, Adams:2006sv,
Camanho:2014apa}. We will show explicitly how causality problems emerge and are resolved in these scenarios.

The propagation of a massless particle in a gravitational shockwave background is of considerable importance in 
its own right as a model of Planck energy scattering. The scattering of particles at ultra-high energies is dominated
by graviton exchange and is therefore an important theoretical laboratory to test fundamental ideas in quantum field 
theory, string theory and quantum gravity (see refs.~\cite{Amati:1987wq,Gross:1987kza,Gross:1987ar,
 'tHooft:1987rb,Muzinich:1987in, Amati:1987uf, Amati:1988tn,Amati:1990xe,Amati:1992zb,Amati:1993tb,Veneziano:2004er,
Amati:2007ak,Giddings:2007bw} for a selection of papers). The results derived here for the energy-dependence of
the phase shifts for a photon propagating in the shockwave background can therefore be directly translated to
the amplitudes for Planck energy scattering. The interpretation of our results in terms of Planck energy scattering
in QFT and associated issues involving causality and unitarity are the subject of a companion paper \cite{HS2}.

The relation of IR and UV theories may also be studied directly using dispersion relations, especially the 
Kramers-Kr\"onig identity which relates the phase velocity, or refractive index, of photons at high and low frequency.
Indeed, the conventional flat-space Kramers-Kr\"onig relation, with the usual analytic properties of the relevant
Green functions, would imply that the UV theory necessarily inherits the causal problems of the low-energy theory.
However, in our previous work \cite{Hollowood:2008kq, Hollowood:2011yh}, 
we have shown how the novel analytic structure induced by geometric properties of the curved spacetime background 
imply a re-interpretation of the usual flat-space dispersion relations, with important consequences for causality and the 
optical theorem. In another paper in this series \cite{HS3}, we return to these issues and present a
new analysis of dispersion relations for QFT in curved spacetime. In that work, we will show how the dispersion relation is violated by non-analyticity in the upper-half plane. In flat space, that would imply a breakdown of {\it micro-causality\/}, the non-vanishing of the retarded Green function outside the backward lightcone. But in curved space the shape of the lightcone is non-trivial and this allows for upper-half-plane non-analyticity whilst preserving micro-causality.

A central role in our work is therefore played by the Aichelburg-Sexl metric \cite{Aichelburg:1970dh},
\begin{equation}
ds^2 = -2 du\, dv + f(r) \d(u) du^2 + dx_1^2 + dx_2^2 \ ,
\label{aa}
\end{equation}
which describes a shockwave localised on the lightcone $u=0$ and satisfies the Einstein equations
\begin{equation}
R_{uu} = 8\pi G T_{uu} = -\tfrac{1}{2} \D f(r) \ .
\end{equation}
For an ultra-high energy particle as the source, $T_{uu} = \r(r) \d(u)$ with 
$\r(r) = \mu \d^2(\underline{x})$, which gives the profile function\footnote{Here, 
$r_0$ is some UV cut off scale. One way to understand this is to smear the particle energy density in the 
transverse directions over a scale $r_0$. This gives rise to the ``beam" shockwave \cite{Ferrari:1988cc}, 
which is described in detail  in section \ref{s2}. 
Then $f(r)$ as above describes the geometry outside the beam $r>r_0$.}
\begin{equation}
f(r) = -4G\mu \log(r/r_0)^2 \ .
\end{equation}

The null geodesics for a massless particle propagating in the opposite direction to the shockwave, initially with
$v =0$ and impact parameter $r=b$, are well known. Explicitly,
\begin{eqnarray}
v &=& \frac{1}{2} f(b) \vartheta(u) + \frac{1}{8} f'(b)^2 u \vartheta(u)  \ ,  \nonumber \\
r &=& b + \frac{1}{2} f'(b) u \vartheta(u) \ .
\label{ab}
\end{eqnarray}
In Aichelburg-Sexl coordinates, therefore, the photon experiences a discontinuous jump in the null coordinate $v$, 
\EQ{
\D v_{\small\text{AS}}=\frac12f(b)= -4G\mu\log\frac {b}{r_0}\ ,
}
which is negative, since $b>r_0$, and so backwards in time. 
The fact that this jump in the null coordinate $\Delta v_{\small\text{AS}}$ is negative, that is a 
Shapiro {\it time advance},  is the first indication that issues regarding causality are subtle in shockwave 
spacetimes. This is one reason why the shockwave provides a perfect stage on which to confront issues with causality
in QFT with gravity.  
\begin{figure}
\begin{center}
\begin{tikzpicture}[scale=0.5,fill=red,decoration={markings,mark=at position 0.5 with {\arrow{>}}}]
%\pgfsetfillopacity{0.85}
\draw[line width=2mm,color=black!20] (0,0) -- (-7,7);
\draw[very thick] (-7,0) -- (-4.9,2.1) -- (-3.4,0.6) -- (-1.7,6.4); 
%\draw[very thick] (-3.4,0.6) -- (0,4); 
\draw[very thick, densely dashed] (-4.5,4.5) -- (-4.9,2.1);
%\draw[very thick] (-4.9,2.1) -- (-3.4,0.6);
\draw[-,dashed] (-3.1,0.9) -- (0.7,4.8);
\draw[-,dashed] (-4.6,2.4) -- (-0.8,6.2);
%\draw[<-] (0.7,4.7) -- (0.4,5);
\node at (3.2,5.7) {$\D v_{\small\text{AS}}=-4G\mu\log\dfrac b{r_0}$}; 
\node at (-5.2,3.2) {$b$};
\draw[<->] (4,2) -- (5,1) -- (6,2);
\draw[->] (5,1) -- (5.1,2);
\node at (6.5,2.5) {$u$};
\node at (3.5,2.5) {$v$};
\node at (5.1,2.5) {$x^i$};
\end{tikzpicture}
\caption {\footnotesize The geodesic of the massless particle involves an instantaneous shift in the 
null coordinate $\D v_{\small\text{AS}}$ as it passes the shockwave at $u=0$ as well as a deflection in the transverse space.}
\label{f2}
\end{center}
\end{figure}
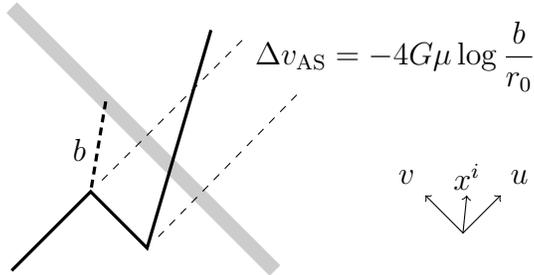

However, as it stands, the fact that the null coordinate is shifted backwards does not constitute 
a {\it prima facie} violation of causality. The geodesics (\ref{ab}) describe straight, null trajectories in both
half-planes $u<0$ and $u>0$ with a discontinuous coordinate shift $\D v_{\text{AS}} = \tfrac{1}{2} f(b)$ 
and a deflection angle $\phi$, with $\tan\phi/2 =  -\tfrac{1}{2}f'(b)$, at $u=0$.
The full shockwave spacetime can therefore be viewed as
two Minkowski half-planes patched together along the surface $u=0$ with a displacement $\D v_{\text{AS}}$. 
The classical Shapiro time advance (for $f(b) < 0$) depends on this patching, which at this geometric level 
is arbitrary. Indeed, the null geodesics are continuous through the shockwave expressed in terms of
the adapted (or Rosen) coordinate $V$ defined in section \ref{s2}. Assigning a physical meaning to the Shapiro 
time advance depends on a physically motivated identification of the past and future Minkowski half-planes,
{\it i.e.}~the asymptotic definitions of time. This will be important when we come to discuss the interpretation 
of our results for Planck scattering amplitudes (see section \ref{s8}).

\begin{figure}
\begin{center}
\begin{tikzpicture}[scale=0.6,fill=red,decoration={markings,mark=at position 0.5 with {\arrow{>}}}]
\draw[line width=2mm,color=black!20] (0,0)  -- (7,7);
\draw[line width=2mm,color=black!20] (7,0)  -- (3.9,3.1);
\draw[line width=2mm,color=black!20] (3.25,3.75)  -- (0,7);
\draw[very thick,postaction={decorate}] (0.7,0.2) -- (0.2,6.4);
\draw[very thick,postaction={decorate}] (0.2,6.4) -- (3,3.6);
\draw[very thick,postaction={decorate}] (3.7,2.9) -- (6.4,0.2);
\draw[very thick,postaction={decorate}] (6.4,0.2) -- (6.9,6.4);
\draw[very thick,postaction={decorate}] (6.9,6.4) -- (0.7,0.2);
\node at (3,7) (a1) {$\D v_{\small\text{AS}}$};
\draw[->] (a1) -- (2.8,2.8);
\draw[->] (a1) -- (1.8,5);
\node[opacity=0.4] at (-1,6) {shock 1};
\node[opacity=0.4] at (8.3,6) {shock 2};
\end{tikzpicture}
\hspace{1cm}
\begin{tikzpicture}[scale=0.6,fill=red,decoration={markings,mark=at position 0.5 with {\arrow{<}}}]
\draw[fill,color=black!20] (0,0) -- (0.4,0) -- (3,7) -- (1.6,7) -- (0,0);
\draw[fill=black!20,color=black!20] (5,0) -- (6.4,0) -- (4.2,7) -- (3.8,7) -- (5,0);
\draw[very thick,postaction={decorate}] (0.3,0.2) -- (2.5,6.4);
\draw[very thick,postaction={decorate}] (2.5,6.4) -- (5.8,0.4);
\draw[very thick,postaction={decorate}] (5.8,0.4) -- (4.2,6.4);
\draw[very thick,postaction={decorate}] (3.15,4.85) -- (0.3,0.2);
\draw[very thick,postaction={decorate}] (4.2,6.4) -- (3.38,5.15);
\draw[fill=black!20] (0.2,0) ellipse  (0.2cm and 0.1cm);
\draw[fill=black!20] (2.3,7) ellipse  (0.7cm and 0.2cm);
\draw[fill=black!20] (5.7,0) ellipse  (0.7cm and 0.2cm);
\draw[fill=black!20] (4,7) ellipse  (0.2cm and 0.1cm);
\draw[densely dashed] (1.5,3.5) -- (4.8,3.5);
\node at (3,3.1) {$L$};
\node[opacity=0.4] at (0,6) {shock 2};
\node[opacity=0.4] at (6.3,6) {shock 1};
\end{tikzpicture}
\caption{\footnotesize A closed trajectory for a massless particle made from two shockwaves moving in 
opposite directions with some impact parameter of the same order as the shifts $\D v_{\small\text{AS}}$ at each shockwave. 
Mirrors are placed at at the points just before and just after the photon gets close the shockwaves to direct the photon 
in the right direction. The right-hand picture is a side view showing the non-vanishing impact parameter.}
\label{f3}
\end{center}
\end{figure}
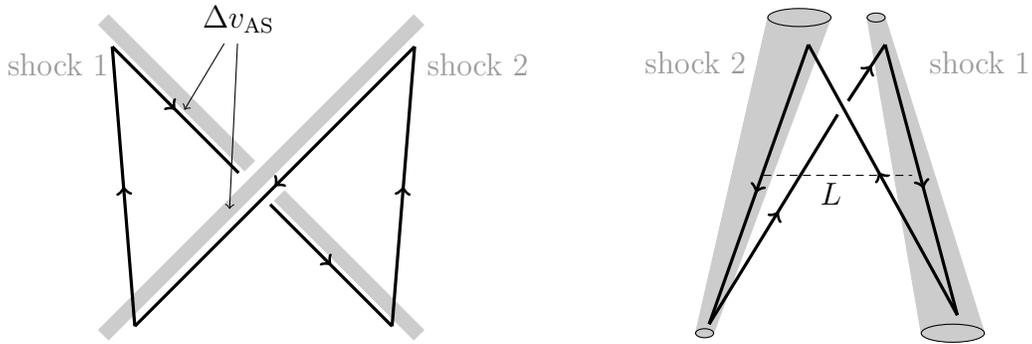

A particularly striking way to highlight these causality issues is to use the time advance to engineer a ``time machine". 
In the present context, a natural idea is to consider the propagation of the photon in the background of two 
shockwaves which are moving in opposite directions with some non-vanishing impact parameter $L$, 
illustrated in fig.~\ref{f3}. 
The two shockwave time machine was first discussed in \cite{Shore:2002in} and then 
in \cite{Adams:2006sv,Camanho:2014apa}. In order to ensure that the the gravitational interaction between 
the shockwaves is small, we need $G\mu/L\ll1$. As long as this separation $L$ is of the order of the 
shift $\D v_{\small\text{AS}}$, it seems a closed non-spacelike trajectory can be constructed, 
as illustrated in fig.~\ref{f3}. 
In fact, such a time machine does not actually work because the shockwave 1, say, also  induces a 
shift $\Delta v_{\small\text{AS}}$ on the wavefronts of the second shockwave and this effectively cancels out the effect of the shift on the photon. 
This is just the equivalence principle in action; equivalently, the time shifts may be seen as an artefact of working in Aichelburg-Sexl coordinates.
So there are no closed non-space-like curves in the two-shockwave geometry, as we describe in section \ref{s3}, following \cite{Shore:2002in}. 

According to  Camanho {\it el al.\/}~\cite{Camanho:2014apa}, the plot thickens if one now considers the effect on the 
scattered particle of additional terms in its (effective) action over and above the usual minimal coupling to gravity. 
This reference considered two cases:
\begin{enumerate}\item The particle is a graviton and the additional terms in the action correspond to the Gauss-Bonnet term in $D>4$. 
\item The particle is a photon (a $U(1)$ gauge boson) and the additional interactions involve coupling to the 
curvature:\footnote{In fact \cite{Camanho:2014apa} only considered the Riemann tensor term.}
 \begin{equation}
S= \int d^4x \sqrt{g} \left[ -\frac{1}{4} F_{\m\n}F^{\m\n} + a R_{\m\n}F^{\m\l}F^\n{}_\l 
+ \tilde{a} R_{\m\n\l\r}F^{\m\n}F^{\l\r} \right]\ .
\label{ac}
\end{equation} 
\end{enumerate}
We will concentrate on the second possibility in this work. 

In QED, the photon trajectories are realised in the eikonal, or geometric optics, approximation, where solutions 
to the field equations in the shockwave geometry are written in terms of a rapidly-varying phase $\Theta(x)$,
with the tangent vector field $\partial^\mu\Theta(x)$ defining a collection of rays, {\it i.e.}~a congruence of 
null geodesics. The new curvature-dependent terms in the effective action now lead to additional shifts 
in the null coordinate as the photon passes the shockwave.\footnote{There is a hidden assumption here, 
that the geometric optics limit applies so we can describe the scattering by a particle trajectory. 
This requires that the frequency of the photon $\omega\gg G\mu/b^2$, the transverse curvature scale.} 
We find that, for the two physical polarizations, the additional curvature-coupling induced shifts are 
\EQ{
\Delta v=\pm\frac{32G\mu}{b^2}\tilde a\ .
\label{vv1}
}
The fact that this result is independent of the Ricci tensor term is because the particle shockwave is 
Ricci flat even in the transverse directions along the wavefront (and of course the curvature vanishes except on $u=0$).

The implication is that one of the polarization states has $\D v<0$. The additional shift in the null coordinate 
is now a genuine Shapiro time advance that is not just an artefact of the choice of coordinates. 
As we show in section \ref{s3}, it is now in principle possible to engineer a two-shockwave time machine 
and causality is apparently violated.

The question that Camanho {\it et al.\/}~\cite{Camanho:2014apa} posed was how could this apparent causality violation 
be remedied by embedding the effective action in a more fundamental, UV complete theory, possibly involving new physics. 
One proposal, for the graviton scattering example, is to add new massive particles to the theory. It turns out that this 
can restore causality if the new states form an infinite tower of higher spin massive particles as in string theory.
This has the effect of Reggeizing the amplitude and this solves the causality problem associated with the 
original action \cite{Camanho:2014apa, D'Appollonio:2015gpa}.
What is very striking here is that a potential causality violating effect in an effective action can be fixed by introducing 
a tower of particles of the form we have in string theory. This introduces a new scale into the problem in the form of 
$\lambda_s = \sqrt{\alpha'}$. 
The moral is that even actions which on their own exhibit causality violations may be the low-energy effective actions
for some causal, UV complete theory.

There are other issue that are relevant in the Gauss-Bonnet gravity example; namely, whether the two shockwave spacetime can actually be engineered. Papallo and Reall \cite{Papallo:2015rna} have argued that in Gauss-Bonnet gravity, small black holes cannot be boosted close to the speed of light in order to approximate the shockwaves and make the time machine. This constraint is not strictly relevant to our set up, since we are not considering the Gauss-Bonnet gravity theory.

In the present paper, we investigate these issues in the case of QFTs which are known to have consistent UV completions.
In particular, we consider in detail the case of a photon scattering with a gravitational shockwave. 
In that case, it has been shown by Drummond and Hathrell (DH), that QED\footnote{We consider  scalar QED,
where the electron is a complex scalar rather than a Dirac spinor. This is technically simpler than its spinor counterpart, 
although the necessary formalism for the latter is established in \cite{Hollowood:2009qz}.  In section \ref{s7}, 
we also discuss a super-renormalizable scalar theory, which exhibits interesting differences from QED in its UV behaviour.}
produces a term precisely of the form \eqref{ac} in the effective action of the photon when the electron is integrated out.
In this case,
\EQ{
a = \frac{\a}{720\pi m^2}\ ,\qquad\tilde{a} = \frac{\a}{1440\pi m^2}\ ,
\label{uu2}
}
where $m$ is the electron mass. The corresponding Compton wavelength $\lambda_c = 1/m$ provides the fundamental length
scale of the QFT.
For these values of the couplings, we will denote the corresponding shift in the shockwave as $\D v_{\small\text{DH}}$ which, 
for the particle shockwave, equals
\EQ{
\D v_{\small\text{DH}}=\pm\frac\alpha{45\pi}\cdot\frac{G\mu}{b^2m^2}\ .
\label{dhs}
}

QED in a curved spacetime is, of course, a perfectly causal theory, so the question naturally arises: if there are terms 
like \eqref{ac} in the photon's effective action when the electron is integrated out, and these lead to causality violations 
involving the shockwave time machine, how is causality cured? It is the purpose of this paper to answer this question. 
We will show that a resolution of the apparent problems with causality attributed to the effective Lagrangian is obtained
{\it entirely within\/} the framework of the UV completion provided by QED,\footnote{Note that by UV completion we mean 
at the perturbative level. The non-perturbative issue involving the Landau pole will not be relevant in the present discussion.}
even though gravity is involved in an essential way. In particular, this will demonstrate how causality is respected in Planck energy
scattering mediated by graviton exchange in renormalisable QFTs \cite{HS2}. 

Before we explain the mechanism, let us consider the various parameters that we have in the photon-shockwave scattering process. 
The shockwave is described by $\mu$, which gives the energy of the original particle,\footnote{In this paper, we relate the
lightcone coordinates in (\ref{aa}) to the time coordinate by $u = \tfrac{1}{2}(t+z)$, $v=t-z$. With these identifications, $\omega$ 
is the photon energy while $\mu$ is twice the energy of the source particle generating the shockwave. Hence $s = 2 \mu \omega$.} 
and the photon by its frequency $\omega$. The usual Mandelstam parameter is $s=2\mu\omega$.
It will also be useful to define 
\EQ{
\s = \frac{4G\mu}{b^2}\ ,
} 
where $b$ is the impact parameter, which is the curvature scale experienced by the photon (expressed as a mass scale) 
and the dimensionless frequency scale
\EQ{
\hat s=\frac{Gs}{b^2m^2}= \frac{\omega\s}{2m^2}\ .
}

Also note at this point that having a shift $\D v_{\small\text{DH}}$ is not by itself sufficient to engineer a time machine and 
demonstrate a violation of causality. 
The point is that the violation should be observable within the resolution scale provided by the photon mode. 
This means that frequency of the photon needs to be
\EQ{
\omega>\D v_{\small\text{DH}}^{-1}\qquad\implies\qquad\hat s >\frac1\alpha\gg1\ .
\label{kk1}
}
So in order to assess the efficacy of the time machine, we need to work with photons with suitably large enough frequency 
so that $\hat s\gg1$.\footnote{Note that \eqref{kk1} implies that we need $\hat s$ to be non-perturbatively large for observability. 
However, we shall find that observability is violated well before $\hat s$ reaches that scale.}
This is to be expected when discussing causality: it is the high frequency limit that is 
relevant \cite{Leontovich,Shore:2003jx, Shore:2007um,Hollowood:2008kq}.
The DH calculation is only valid at low frequency and so by itself is not adequate to make judgements regarding causal issues.

\begin{figure}[h]
\begin{center}
\begin{tikzpicture}[scale=0.5,fill=red,decoration={markings,mark=at position 0.5 with {\arrow{>}}}]
\draw[line width=2mm,color=black!20] (0,0) -- (-7,7);
\draw[decorate,decoration={snake,amplitude=0.1cm},very thick] (-6.5,0.5) -- (-5.5,1.5); 
\draw[decorate,decoration={snake,amplitude=0.1cm},very thick]  (-2.5,4) -- (-1.5,6.4); 
\draw[very thick,->] (-5.5,1.5) to[out=30,in=-170] (-3.7,3.3);
\draw[very thick,->] (-5.5,1.5) to[out=-60,in=-150] (-3.7,1.2);
\draw[very thick] (-3.7,3.3) to[out=10,in=140] (-2.5,4);
\draw[very thick,] (-3.7,1.2) to[out=30,in=-120] (-2.5,4);
\node at (-4.8,3.6) (i1) {$e^+$};
\node at (-3.3,0.7) (i2) {$e^-$};
\node at (-6.9,0.1) (i3) {$\gamma$};
\node at (-1.1,6.8) (i3) {$\gamma$};
\filldraw[black] (-5.5,1.5) circle (4pt);
\filldraw[black] (-2.5,4) circle (4pt);
\end{tikzpicture}
\caption{\small The one-loop Feynman diagram contributing to the
  vacuum polarization in QED in the curved background of the shockwave.
The figure illustrates the gravitational tidal forces acting on the virtual electron-positron cloud
screening the dressed photon. }
\label{f4}
\end{center}
\end{figure}
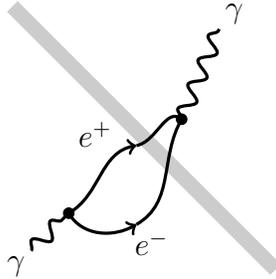

In the context of QED, the calculation of the DH effective action and its extension to the high-frequency regime, 
means that we must take into account the effects of vacuum polarization, namely the fact that the photon is an 
extended object consisting of a bare photon surrounded by a cloud of virtual electron-positron pairs. 
The task before us is therefore to calculate the tidal effect of the background geometry on the dressed photon for all energies.
This is encoded in the self-energy of the photon at one loop with curved space propagators for the 
electron and positron; see fig.~\ref{f4}.
In general, such a calculation would not be tractable. However if we impose the following two conditions there does 
exist a tractable window on high frequencies \cite{Hollowood:2008kq}:
\begin{enumerate}
\item $\omega\gg\sigma$. This is the geometric optics, or eikonal, limit which allows us to analyse the propagation 
of photons in terms of trajectories in spacetime. 
\item $m\gg\sigma$. This is the requirement that the size of the virtual cloud set by the Compton wavelength 
of the electron $\lambda_c=1/m$ is small compared with the scale over which the curvature varies.\footnote{This is rather 
subtle in a shockwave spacetime which has a delta function curvature. However, it is the curvature in the transverse 
directions that is actually relevant and this is determined by the mass scale $\sigma$.}
\end{enumerate}
The key point is that the two limits leave a window on the high frequency regime via the dimensionless 
ratio $\hat{s}=\omega\sigma/2m^2$. We will go beyond the DH result by calculating the full dependence of the
shift on $\hat s$.  We also show that $\Delta v$ is a function of the null distance from the shockwave,
\EQ{
\D v(u,\omega)=\D v_{\small\text{DH}} F(\sigma u,\hat s)\ ,
}
where we can think of $v =\D v(u,\omega)$ as describing the trajectory of the dressed photon in the $(u,v)$ subspace.
In the low frequency limit, $F(\sigma u,\hat s\to0)=\vartheta(\sigma u)$, the Heaviside function. 
In what follows, we determine $F(\sigma u,\hat s)$ for all $\hat s$, including the crucial high-energy limit.

We derive our results in terms of the instantaneous phase $\Theta(u,\omega)$ which characterises the photon modes
as they are scattered by the shockwave. This depends on $(u,\omega)$ via the two dimensionless quantities 
$\sigma u$ and $\hat s$. In turn, $\Theta(u,\omega)$ is derived from a local refractive index\footnote{Strictly speaking, 
this interpretation is only valid if $\Delta n$ remains perturbatively small.}  along the photon's trajectory:
\EQ{
n(u,\omega)=1+\D n(u,\omega)\ ,\qquad \D n(u,\omega)=\frac{1}{\omega}\frac{\partial}{\partial u}\Theta(u,\omega)\ .
}
The corresponding cordinate shift is then identified as\footnote{An alternative definition appropriate to a wave packet 
rather than a single Fourier mode would be 
\begin{equation*}
\D v(u,\omega)=\frac{\partial}{\partial\omega}\Theta(u,\omega)\ .
\end{equation*}
This has essentially the same high-frequency dependence as the
definition (\ref{DV}), as described in section \ref{s5} (see in particular fig.~\ref{g2}).}
\begin{equation}
\D v(u,\omega)=\frac{1}{\omega}\Theta(u,\omega)\ .
\label{DV}
\end{equation}
Note that all the quantities $n(u,\omega)$, $\Theta(u,\omega)$ and $\Delta v(u,\omega)$ actually have both real and 
imaginary parts.
The scattering phase, which determines the amplitude for photon-shockwave scattering, is then obtained in the limit
\EQ{
\Theta_\text{scat.}(s,b)=\Theta(u\to\infty,\omega)\ .
\label{bb1}
}

The paper is organised as follows. In section \ref{s3} we discuss how two colliding shockwaves can potentially be used 
to engineer time machines, that is a spacetime where a particle can follow a closed non-spacelike trajectory. 
Then, in section \ref{s2}, we review the essential features of the geometry of the 
gravitational shockwave and its Penrose limit and evaluate the Van Vleck-Morette matrix, which plays a key role in our analysis. 
Section \ref{s4} describes the basic formalism we apply to analyse photon-shockwave and
contains the formulae for the refractive index and phase shift derived in our earlier work. 
In sections \ref{s5} and \ref{s6}, we analyse the scattering of a photon with a beam and particle shockwave, respectively,
complementing our exact analytical results with a detailed numerical analysis of the phase and coordinate shifts. 
Having obtained their high-frequency limits, we then return to the shockwave 
time machine in section \ref{TM2} and discuss how causality is restored in the UV complete theory.
Section \ref{s7} is devoted to an analysis of a simpler, super-renormalizable scalar theory with vacuum polarization 
to provide a comparison with QED in a theory in which the UV behaviour is rather different. 
Finally, in section \ref{s8}, we draw some conclusions, including a brief discussion of the relation of our results
to scattering amplitudes at ultra-high energies.

\section{Shockwave Time Machines}\label{s3}

The fact that classical null geodesics and quantum loop corrections exhibit lightcone
coordinate shifts $\D v < 0$ characteristic of a Shapiro time {\it advance} naturally raises the question 
of whether we can build a time machine. In this context, by ``time machine'' we mean
a piecewise smooth closed non-spacelike trajectory in spacetime. 

The possibility of using a two-beam shockwave spacetime to construct a time machine was studied in detail
in \cite{Shore:2002in}. Here, we present a closely related analysis by studying
in detail the case of two colliding particle shockwaves at non-vanishing impact parameter. 
This allows us to control the curvature.
First, we present a na\"\i ve argument for the existence of a time machine and then go on to show that a 
proper treatment invalidates one of the implicit assumptions. The conclusion is that 
time machines based on the general relativity shift $\Delta v$ cannot exist. 
In fact this is ensured by the equivalence principle.
However, if additional contributions to the shift coming from quantum corrections are present, 
then the (strong) equivalence principle is broken and a time machine can be constructed.

\subsection{A na\"\i ve analysis}

\begin{figure}[ht]
\begin{center}
\begin{tikzpicture}[scale=0.6,fill=black!20,decoration={markings,mark=at position 0.6 with {\arrow{>}}}]
\draw[line width=2mm,color=black!20] (0,0) -- (7,7);
\draw[line width=2mm,color=black!20] (7,0) --  (0,7);
%\node at (-0.3,3.6) {$\gamma$};
\draw[very thick,postaction={decorate}] (0,4) -- (1.5,5.5);
\draw[very thick,postaction={decorate}] (1.5,5.5) -- (5.5,1.5);
\draw[very thick,postaction={decorate}] (5.5,1.5) -- (6,6);
\draw[very thick,postaction={decorate}] (6,6) -- (1,1);
\draw[very thick,postaction={decorate}] (1,1) -- (1.4,3.2);
\node at (0.6,5.7) {$S$};
\node at (6.2,1.6) {$P$};
\node at (6.5,5.6) {$Q$};
\node at (0.3,1.2) {$R$};
%\node at (3.5,2.8) {$O$};
\node[rotate=45] at (6,6.8) {$v=0$};
\node[rotate=-45] at (1,6.8) {$u=0$};
\node[opacity=0.4] at (-1,6) {shock 1};
\node[opacity=0.4] at (8.3,6) {shock 2};
\node at (-0.7,3.5) {I};
\node at (7.5,3.5) {II};
\node at (3.5,7) {IV};
\node at (3.5,0) {III};
\end{tikzpicture}
\caption{\footnotesize The proposed time machine consisting of two shockwaves moving in opposite directions 
that collide with some impact parameter $L$. The photon collides with the first at $S$, experiences a shift back to $P$ 
which then allows it to catch up with shockwave 2 with a jump back to $R$ in the past lightcone of $S$.}
\label{f6}
\end{center}
\end{figure}
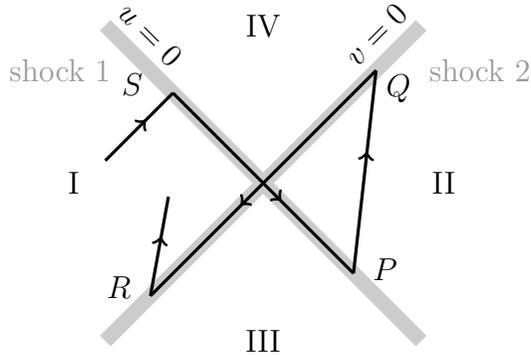

The putative time machine is sketched in fig.~\ref{f6}.
Consider two shockwaves travelling in the opposite direction along the $z$-axis with $u=0$ and $v=0$,
and profile functions $f_1(r)$ and $f_2(r)$, respectively. They collide with some impact parameter $L$, so shockwaves 
1 and 2 have $x_i=(0,0)$ and $x_i=(L,0)$, respectively.

A photon coming in following shock 2 hits the wavefront of shock 1 at point $S$ with $u=0$, $v=v_S >0$ and 
impact parameter $x_1=b$. It then jumps back by an amount $\Delta v_{\small\text{AS}}=\frac12 f_1(b)<0$, 
which we can arrange to be greater than $v_S$, to point $P$. It is clear that we can then connect $P$ 
to a point $Q$ lying on the wavefront of shock 2 at impact parameter $b'$ by a timelike or null trajectory.
A photon at this point can then be made to jump back by an amount $\Delta u_{\small\text{AS}}=\frac12 f_2(b')$ 
in the null coordinate $u$ (for shock 2 the coordinates $u$ and $v$ are reversed) to a point $R$ which is in the 
past lightcone of point $S$. So a time machine has been engineered.
In fact, in \cite{Shore:2002in} a completely closed geodesic trajectory of a single photon was constructed in the case 
of zero impact parameter $L=0$.

\subsection{A consistent analysis}

However, before accepting this construction as a true time machine, we need to critically analyse the assumption that the shockwaves 
are non-interacting \cite{Shore:2002in}. The shockwave geometry can be analysed in terms of the four regions I, II, III and IV 
shown in fig.~\ref{f6}. The geometry in regions I, II and III is actually flat whereas in region IV the collision curves the 
spacetime in a way which is difficult to analyse \cite{D'Eath:1992hb}. However, if we take the shockwaves to be particle shockwaves
(having the same energy $\mu$ for simplicity) and the impact parameter such that $G\mu/L\ll1$, then we expect that the 
curvature in region IV will be small. 

So working in this regime, one would suspect that the shockwaves have a negligible effect on each other. However, 
each shockwave carries with it a wavefront located at $u=0$, for shockwave 1, and $v=0$, for shockwave 2. 
These wavefronts are extended in the transverse directions $x_i$. So even though shockwave 2 has a large impact parameter 
$L\gg G\mu$ relative to shockwave 1, its wavefront ${\cal W}_2(u)$ extends out infinitely in the transverse directions. 

\begin{figure}[ht]
\begin{center}
\begin{tikzpicture}[scale=1.5,fill=black!40,decoration={markings,mark=at position 0.5 with {\arrow{>}}}]
\draw[->] (-2,0) --(3,0);
\draw[->] (0,-0.2) -- (0,2.8);
\draw[line width=2mm,color=black!20] (0,0) -- (0,2.6);
\draw[line width=2mm,color=black!20] (1.6,2.8) to[out=-100,in=30] (0,0);
\draw[densely dashed] (0,2.2) -- (1.45,2.2);
\draw[densely dashed] (0,1) -- (1,1);
\draw[postaction={decorate}] (-2,2.2) -- (-1,2.2);
\draw[postaction={decorate}] (-2,1) -- (-1,1);
\draw[postaction={decorate}] (-1,2.2) -- (0,2.2);
\draw[postaction={decorate}] (-1,1) -- (0,1);
\draw[postaction={decorate}] (1.45,2.2) -- (2.2,2.1);
\draw[postaction={decorate}] (1,1) -- (1.9,0.6);
\node at (3.3,0) {$z$};
\node at (0,3.1) {$x_1$};
\filldraw[black] (-0.8,1.6) circle (0.8pt);
\filldraw[black] (0.8,1.6) circle (0.8pt);
\filldraw[black] (1.8,1.6) circle (0.8pt);
\draw[postaction={decorate},very thick] (-2,1.6) -- (-0.8,1.6);
\node at (-2.2,1.6) {$\gamma$};
\draw[densely dashed,very thick] (-0.8,1.6) -- (0.8,1.6);
\node at (1.6,0.3) (a1) {${\cal W}_2(0^+)$};
\node at (-0.95,0.3) (a2) {${\cal W}_2(0^-)$};
\draw[->] (a1) -- (0.8,0.5);
\draw[->] (a2) -- (-0.1,0.5);
\node at (-0.7,1.85) {$S$};
\node at (0.9,1.85) {$P$};
\node at (1.9,1.85) {$P'$};
\end{tikzpicture}
\caption{\footnotesize The plot shows the photon and wavefront of shockwave 2 in the $(z,x_1)$ plane. 
The photon is behind the wavefront. When the wavefront collides with shockwave 1, it gets shifted 
by an amount $\Delta v_{\small\text{AS}}=\frac12f(x_1)<0$. This corresponds to jump forward in $z$ and 
backwards in time. At $u=0^+$, the wavefront becomes curved as shown. Since the photon collides with 
shockwave 1 at some time later and for $z<0$ at $S$ it gets shifted froward to P which lies behind the 
wavefront of shockwave 2. If the photon receives an additional $\Delta v_{\small\text{DH}}(b)<0$ then it 
can then jump to point $P'$ in front of shockwave 2 and a time machine can then be constructed.}
\label{f7}
\end{center}
\end{figure}
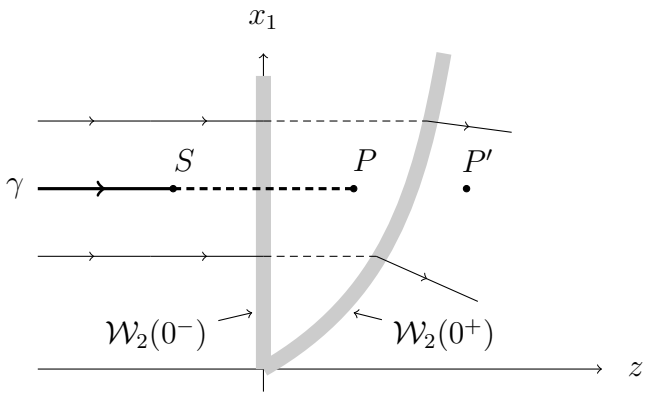

The wavefronts are generated by null geodesics, so as it moves in the geometry of shockwave 1, each point in shockwave 2's 
wavefront moves according to \eqref{bc}. 
So we can describe the evolution of the wavefront in terms of the coordinates $(u,v,r,\phi)$, with $x_1=r\cos\phi$ and $x_2=r\sin\phi$, as
\EQ{
{\cal W}_2(u):\quad &v(u)=\frac12f(r_1)\vartheta(u)+\frac18f'(r_1)^28u\vartheta(u)\ ,\\
&r(u)=r_1+\frac12f'(r_1)u\vartheta(u)\ ,\qquad\phi=\phi_1 \ .
}
At $u=0$, the wavefront of shockwave 2 passes the wavefront of shockwave 1 and so in shockwave 1's Aichelburg-Sexl 
coordinates the point $(r_1,\phi_1)$ experiences a shift $\Delta v_{\small\text{AS}}=\frac12f(r_1)$.
The jump in the wavefront is shown in fig.~\ref{f7}, which shows the $(z,x_1)$ plane. The photon, also shown, hits the 
wavefront of the first shockwave at $S$, which is at $u=0$, $v=v_S$ and $x_1=b$. It jumps to point $P$ which lies 
{\it behind} the wavefront ${\cal W}_2(0^+)$ \cite{Shore:2002in}.

The fact the wavefront ${\cal W}_2(u)$ experiences a Shapiro time advance all along its length is, of course, 
just the equivalence principle in action: if photon experience a Shapiro time advance then so should the 
shockwaves themselves. 

We now prove that the point $P$ is spacelike separated from the future evolution of the wavefront 
${\cal W}_2(u)$, $u>0$, implying that the photon can never catch up with the second shockwave. In order to 
show this, we will assume that the curvature in region IV, where $P$ is located, is small and can be neglected. 
Using the flat metric, and coordinates $(u,v,r,\phi)$,
the spacetime separation between an arbitrary point on the wavefront 
$K=(u,\frac12f(r_1)+\frac18f'(r_1)^2u, r_1+\frac12f'(r_1)u, \phi_1)$ and the photon at 
$P=(0,v_S+\frac12f(b),b,0)$
is
\EQ{
\Delta s^2_{KP}=2uv_S+8G\mu u \Big( f(b) -f(r_1) - (b-r_1)f'(r_1) \Big) +(b-r_1\cos\phi_1)^2+(r_1\sin\phi_1)^2\ .
}
Since $u>0$ and $v_S>0$, and noting that for the particle shockwave, the function in the bracket is positive semi-definite:
\begin{equation}
f(b) -f(r_1) - (b-r_1)f'(r_1)   = \frac{b}{r_1}-1-\log\frac b{r_1} \ge 0 \ ,
\end{equation}
we have $\Delta s^2_{KP}>0$. 
Therefore, as claimed, $P$ is spacelike separated from any point on the wavefront ${\cal W}_2(u)$.

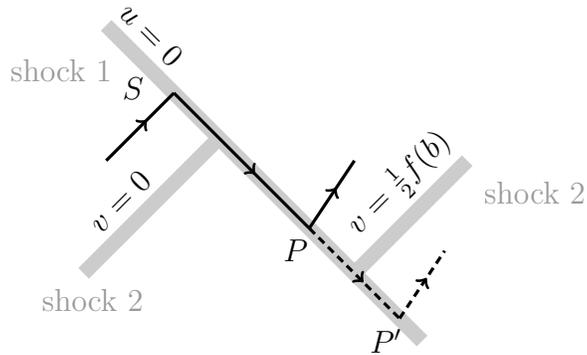
\begin{figure}[ht]
\begin{center}
\begin{tikzpicture}[scale=0.6,fill=black!20,decoration={markings,mark=at position 0.6 with {\arrow{>}}}]
\draw[line width=2mm,color=black!20] (-0.5,1.5) -- (2.5,4.5);
\draw[line width=2mm,color=black!20] (5.5,1.5) -- (8,4);
\draw[line width=2mm,color=black!20] (7,0) -- (0,7);
\draw[very thick,postaction={decorate}] (0,4) -- (1.5,5.5);
\draw[very thick,postaction={decorate}] (1.5,5.5) -- (4.5,2.5);
\draw[very thick,postaction={decorate}] (4.5,2.5) -- (5.5,4);
\draw[very thick,densely dashed,postaction={decorate}] (4.5,2.5) -- (6.5,0.5);
\draw[very thick,densely dashed,postaction={decorate}] (6.5,0.5) -- (7.5,2);
\node at (0.6,5.6) {$S$};
\node at (4.2,2) {$P$};
\node at (6.2,0) {$P'$};
%\node at (7,5.8) {$Q$};
%\node at (-0.1,0.8) {$R$};
%\node at (2.5,3.6) {$O$};
\node[rotate=45] at (0.3,3.1) {$v=0$};
%\node[rotate=-45] at (2,5.8) {$u=0$};
\node[rotate=45] at (6.5,3.5) {$v=\frac12f(b)$};
\node[rotate=-45] at (1,6.8) {$u=0$};
\node[opacity=0.4] at (-1,6) {shock 1};
\node[opacity=0.4] at (9.5,3.3) {shock 2};
\node[opacity=0.4] at (-0.3,0.9) {shock 2};
%\node at (-0.3,3.6) {$\gamma$};
\end{tikzpicture}
\caption{\footnotesize In the true picture, both shockwave 2 and the photon undergo the {\it same} shift 
$\Delta v_{\small\text{AS}}=\frac12f(b)<0$. This is illustrated in the figure, which shows a cross-section at $x_1=b$,
the impact parameter of the photon.
It is clear that the photon can, therefore never catch up with the shockwave 2 to complete the circuit shown in fig.~\ref{f6}. 
However, an additional shift $\Delta v_{\rm DH}$ can take the photon to point $P'$, in which case a time machine can be constructed.}
\label{f8}
\end{center}
\end{figure}

So fig.~\ref{f6} should be replaced by fig.~\ref{f8} which shows a cross section in the transverse direction at $x_1=b$.

\subsection{Drummond-Hathrell shifts}

The next issue is whether the obstruction to the time machine construction can be circumvented 
if we include the additional discontinuous coordinate shift $\Delta v_{\small\text{DH}}(b)<0$
implied by the effective Lagrangian \eqref{ac}. 

With the additional shift, the point $P$ in fig.~\ref{f7} can become $P'$, now in front of the wavefront ${\cal W}_2(0^+)$. 
In that case, the spacetime interval between $P'$ and the point $K$ on the evolution of the wavefront is
\EQ{
\Delta s^2_{KP'}&=2u\big(v_S+\Delta v_{\small\text{DH}}(b)\big)+8G\mu u\Big( f(b) -f(r_1) - (b-r_1)f'(r_1)\Big)\\ &
\qquad+(b-r_1\cos\phi_1)^2+(r_1\sin\phi_1)^2 \ .
}
Since the Drummond-Hathrell coordinate shift $\Delta v_{\small\text{DH}}(b)<0$, we see that $\Delta s^2_{KP'}$ can now be negative. 
For instance, this can be achieved by taking $x_1=b$ and $v_S<|\Delta v_{\small\text{DH}}(b)|$.
The implication is that the photon can reach the wavefront of the second shockwave and then be shifted back in $u$ to make a time machine. 

At the level of the effective action, therefore, a two-shockwave time machine can be constructed and causality is apparently violated.
The next question is whether causality is restored and the time machine fails when the effective action is embedded in the full UV 
complete theory. We return to this issue in section \ref{TM2}, after determining the dependence of the coordinate shifts $\Delta v$ 
on the photon frequency in QED itself.

\section{Geometry of Gravitational Shockwaves and the Penrose Limit}\label{s2}

Our results on photon propagation in the gravitational shockwave background are written entirely in terms of
geometrical quantities characterising the spacetime and its null geodesic congruences.
In this section, we briefly review the essential features of the geometry of the Aichelburg-Sexl shockwave and its 
Penrose limit that we need for our analysis.  In particular, we will focus on geodesic deviation and the construction 
of the Van Vleck-Morette (VVM) determinant, which plays a key role in the discussion.

The Aichelburg-Sexl metric for a gravitational shockwave is given in \eqref{aa},
\begin{equation}
ds^2 = -2 du \,dv + f(r) \d(u) du^2 + dx_1^2 + dx_2^2 \ .
\label{ba}
\end{equation}
We consider four-dimensional spacetime in this work. The profile function $f(r)$ is determined by the 
Ricci curvature $R_{uu} = 8\pi G T_{uu}$
and depends on the nature of the matter source for the shockwave. We consider two sources, an
infinitely boosted particle with $T_{uu} = \rho(\underline{x}) \d(u)$ with 
$\rho(\underline{x}) = \m \d^2(\underline{x})$ and a homogeneous beam with $T_{uu} = \rho \d(u)$,
$\rho = {\rm const.}$ \cite{Ferrari:1988cc}. The corresponding 
profiles follow from the relation $R_{uu} = - \tfrac{1}{2} \D f(r)$,
where $\D$ is the two-dimensional Laplacian, so we find
\EQ{
f(r) =\begin{cases} - 4 G \m \log (r/r_0)^2   &   \text{(particle)} \\
- 4\pi G \rho r^2   &    \text{(beam)}\ ,\end{cases}
\label{bb}
}
where $\m(r) = \pi \r r^2$ gives the energy density of the beam within radius $r$. In the particle case, 
the solution depends on an arbitrary constant $r_0$ which should be thought of an a UV cut off and so $r>r_0$. 
One way to make this concrete is to consider the particle as a beam with a finite size. 
This would correspond to a profile function
\EQ{
f(r)=\begin{cases}-4G\mu (r/R)^2 & r\leq R\\ -4G\mu\log(r/r_0)^2 & r\geq R\ .\end{cases}
}
Matching the solutions at $r=R$ fixes $r_0=e^2R$. So $r_0$ can be identified with the scale of the size of the beam. 
Taking this small then gives the profile function of the particle shockwave.

The null geodesics corresponding to the trajectories of a massless particle, the photon, propagating
in the $u$-direction in this background are well-known and, as we have discussed, display a discontinuous jump in the
Aichelburg-Sexl $v$ coordinate as the photon crosses the shockwave (see fig.~\ref{f2}). 
In polar coordinates for the transverse space,
\EQ{
v &= V + \frac{1}{2} f(R) \vartheta(u) + \frac{1}{8} f'(R)^2 u \vartheta(u)  \ ,\\
r &=  R + \frac{1}{2} f'(R) u \vartheta(u) \ , \\
\phi &=\Phi   \ ,
\label{bc}
}
where $V,R,\Phi$ are constants labelling the individual geodesics in a null congruence.
They are therefore natural ``adapted coordinates'', in terms of which the Aichelburg-Sexl metric
can be rewritten as
\begin{equation}
ds^2 ~=~ - 2 du \,dV + \left[1 + \frac{1}{2}f^{\dprime}(R) u \vartheta(u)\right]^2 dR^2
+\left[1 + \frac{1}{2R} f'(R) u \vartheta(u)\right]^2 R^2 d\Phi^2
\label{bd}
\end{equation}

Now, as discussed extensively in our earlier work, the effect of vacuum polarization on the propagation
of a photon in a curved spacetime background depends on the geometry of geodesic deviation.
This is precisely the feature of the background that is encoded in the Penrose limit \cite{Penrose}.
The Penrose limit is a plane-wave spacetime which is determined from the original spacetime metric 
{\it and} a preferred geodesic. In a general spacetime, in adapted coordinates with preferred geodesic
$V = X^a = 0$ $(a = 1,2)$, the metric may be written as
\begin{equation}
ds^2 = - 2 du\,dV + C(u,V,X^a) dV^2 + 2 C_a(u,V,X^a) dX^a\, dV + C_{ab}(u,V,X^a) dX^a\, dX^b \ .
\label{be}
\end{equation}
The Penrose limit is then
\EQ{
d{\hat s}^2 = {\rm lim}_{\l\rta0} ~~\frac{1}{\l^2}~ ds^2(u,\l^2 V, \l X^a) = - 2 du \,dV + C_{ab}(u,0,0) dX^a \,dX^b  \ .
\label{bf}
}

For the Aichelburg-Sexl shockwave, we choose a preferred geodesic with impact parameter $b$,
i.e.~$V=0, R=b, \Phi=0$, so that $X^1 = R-b, X^2 = b\Phi$. The Penrose limit is then \cite{Hollowood:2009qz}
\begin{equation}
d{\hat s}^2 =~- 2 du \,dV + C_{ij}(u) dX^i\, dX^j  \ ,
\label{bg}
\end{equation}
with
\begin{equation}
C_{11} = \left[1 + \frac{1}{2} f^{\dprime}(b) u \vartheta(u)\right]^2,  ~~~~~~~~~~
C_{22} = \left[1 + \frac{1}{2b} f'(b) u \vartheta(u)\right]^2   \ .
\label{bh}
\end{equation}
This is written in Rosen coordinates, which are well-suited to describing the geodesic congruence.
An alternative presentation is in terms of Brinkmann coordinates, where the metric is instantly
recognisable as a plane wave:
\begin{equation}
d{\hat s}^2 ~=~ - 2 du\, dv - h_{ij}(u) x^i x^j du^2 + \d_{ij}dx^i dx^j \ .
\label{bi}
\end{equation}
The profile function $h_{ij}(u) = R_{iuju}$ in terms of the Aichelburg-Sexl curvature. This makes
clear the connection with geodesic deviation, since the separation vector $z^i$ between geodesics
in a null congruence satisfies the Jacobi equation
\begin{equation}
\frac{d^2 z^i}{du^2} ~=~ - R^i{}_{uju} z^j \ .
\label{bj}
\end{equation}

Rosen and Brinkmann coordinates are related by
\EQ{
x^i = E^i{}_a X^a \ , \qquad
v =V + \frac{1}{2} \Omega_{ab} X^a X^b \ ,
\label{bk}
}
where $E^i{}_a$ is a zweibein defined from the Rosen metric as $C_{ab}(u) = E^i{}_a(u) \d_{ij} E^j{}_b(u)$
and $\Omega_{ab} = E^i{}_a \Omega_{ij} E^j{}_b$ with $\Omega^i{}_j = E_j{}^a \frac{d}{du}E^i{}_a$,
(with $E_j{}^a$ the inverse zweibein). The profile function is given by 
\begin{equation}
h_{ij} = - E_i{}^a \frac{d^2}{du^2} E_{ja} = -\frac{d}{du}\Omega_{ij} - \Omega_{ik} \Omega^k{}_j \ .
\label{bkk}
\end{equation}

For the shockwave metric (\ref{bg}), the zweibeins are
\begin{equation}
E^1{}_1(u) = 1 + \frac{1}{2} f^{\dprime}(b) u\vartheta(u),   ~~~~~~~~~~
E^2{}_2(u) = 1 + \frac{1}{2b} f'(b) u \vartheta(u)   \ ,
\label{bl}
\end{equation} 
and we find
\begin{equation}
h_{11} = -\frac{1}{2} f^{\dprime}(b) \d(u) ,   ~~~~~~~~~~
h_{22} = -\frac{1}{2b} f'(b) \d(u)   \ ,
\label{bm}
\end{equation}
clearly showing the dependence of the Penrose limit metric on the impact parameter of the chosen 
photon geodesic.
Evaluating for the particle and beam shockwaves, we have
\EQ{
h_{ij} &=\frac{4G\mu}{b^2} \begin{pmatrix}-1&0\\0&1\end{pmatrix} \d(u)   
= \s \begin{pmatrix}-1&0\\0&1\end{pmatrix} \d(u)    ~~~~~~~~{\rm (particle)} \ , \\[3pt]
&= \frac{4G\mu(b)}{b^2} \begin{pmatrix}1&0\\0&1\end{pmatrix} \d(u) 
~=~ \s \begin{pmatrix}1&0\\0&1\end{pmatrix} \d(u)  ~~~~~~~~{\rm (beam)}\ ,
\label{bn}
}
where $\m(b)$ is the energy of the beam within the impact parameter radius $b$.
We see that the particle shockwave gives a {\it Ricci flat\/} plane wave ($R_{uu} = \Tr h_{ij} = 0$)
provided $b\neq 0$,
while the beam gives a {\it conformally flat\/} plane wave ($C_{iuju} = 0$). 
This introduces the key parameter $\s = 4G\m/b^2$ which combines the energy
of the shockwave and the photon impact parameter.

The next step is to derive the Van Vleck-Morette matrix which encodes the geometry of geodesic 
deviation. The VVM matrix is defined from the geodesic interval
\EQ{
\s(x,x') =-\frac12 \int_0^1 d\t\, g_{\m\n}(x) \dot x^\m \dot x^\n \ ,
\label{cci}
}
where $x^\m(\t)$ is the null geodesic joining $x = x(0)$ and 
$x' = x(1)$, and is
\EQ{
\D_{\m\n}(x,x') ={\partial^2 \s(x,x')\over \partial x^\m \,
\partial x'^\n} \ .
\label{ccj}
}
In Rosen coordinates, the elements of the VVM matrix for the transverse directions is
\begin{equation}
\D_{ab}(u,u') = (u-u') \left[\int_{u'}^u du''\, C(u'')\right]_{ab}^{-1} \ .
\label{bo}
\end{equation}
Writing the (diagonal) zweibeins as $E^i{}_a(u) = (1 -\s_i u \vartheta(u)) \d^i{}_a$, 
where $-\s_1 =\s_2 =\s$ for the particle shockwave and $\s_i =\s_2 = \s$ for the beam 
shockwave, we can readily calculate $\Delta_{ab}(u,u')$ in the three separate cases
$(u<0, u'<0)$,  $(u>0, u'<0)$ and $(u>0, u'>0)$. 
The result is most simply expressed in Brinkmann form. Defining,
\begin{equation}
\D_{ij}(u,u')=E_i{}^a(u) \D_{ab}(u,u') E_j{}^b(u') \ ,
\label{bp}
\end{equation} 
we find
\EQ{
\D_{ij}(u,u') = \begin{cases}\dfrac{u-u'}{u-u' +\s_i u u'} \d_{ij}  \qquad\qquad&  (u,u' \quad\text{opposite sides})\ ,\\[3pt]
\d_{ij}  &  (u,u'\quad\text{same side})\ .\end{cases}
\label{bq}
}

We can also evaluate the VVM matrix in the transverse directions directly in Brinkmann coordinates as
\begin{equation}
\D_{ij}(u,u') = - (u-u') \left[A(u,u')^{-1}\right]_{ji} \ ,
\label{br}
\end{equation}
where the matrix $A_{ij}(u,u')$ satisfies the Jacobi equation\footnote{This comes from the
fundamental definition of $A_{ij}(u,u')$ from the solution of the geodesic deviation equation
\begin{equation*}
\frac{d^2 z^i}{du^2} = - h^i{}_j z^j
\end{equation*}
as
\begin{equation*}
z^i(u) = B^i{}_j(u,u') z^j(u') + A^i{}_j(u,u')\dot{z}^j(u') \ .
\end{equation*}   }
\begin{equation}
\frac{d^2}{du^2} A_{ij} + h_i{}^k A_{kj} = 0 \ ,
\label{bs}
\end{equation}
with ``geodesic spray" boundary conditions $A_{ij}(u,u) = 0,~~\frac{d}{du}A_{ij}(u,u')\big|_{u=u'}=\delta_{ij}$.
This definition makes the connection of $\D_{ij}(u,u')$ with geodesic deviation completely transparent.
A short calculation using the expressions (\ref{bn}) for $h_{ij}$ then reproduces the expression
(\ref{bq}) for $\D_{ij}$.

\begin{figure}[ht]
\begin{center}
\begin{tikzpicture}[scale=1,decoration={markings,mark=at position 0.5 with {\arrow{>}}}]
\draw[line width=2mm,color=black!20] (0,-2)  -- (0,2);
\draw[->] (-5,0) -- (5,0);
\node at (5.3,0) {$u$};
\node at (0,2.4) {$x^i$};
\draw[very thick,postaction={decorate}] (-4,0) -- (0,1.5);
\draw[very thick,postaction={decorate}] (-4,0) -- (0,-1.5);
\draw[very thick,postaction={decorate}] (0,1.5) -- (3.5,0);
\draw[very thick,postaction={decorate}] (0,-1.5) -- (3.5,0);
\draw[very thick,postaction={decorate}] (-5,1) -- (0,1);
\draw[very thick,postaction={decorate}] (-5,-1) -- (0,-1);
\draw[very thick,postaction={decorate}] (0,1) -- (2,0);
\draw[very thick,postaction={decorate}] (0,-1) -- (2,0);
\node at (-4,-0.3) {$u'$};
\node at (3.5,-0.3) {$u$};
\node at (2,-0.3) {$\sigma_i^{-1}$};
\end{tikzpicture}
\caption{\footnotesize A pair of conjugate points $(u,u')$ and the focal point at $\sigma_i^{-1}$ where parallel rays 
from $-\infty$ are focussed for the beam shockwave.}
\label{f5}
\end{center}
\end{figure}
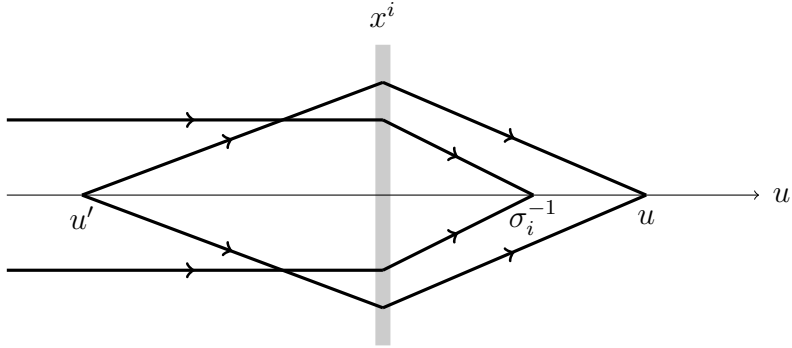

It is clearly important in our analysis that the VVM matrix is only non-trivial when the arguments $u, u'$ 
lie on opposite sides of the shockwave. Another crucial general feature is that $\D_{ij}(u,u')$ becomes singular 
when $u$ and $u'$ correspond to conjugate points on the geodesic congruence. These singularities 
directly affect the analytic properties of the Green functions and the refractive index and phase shift
as functions of the photon energy $\omega$. For the shockwave, with $u>0, u'<0$, there are conjugate
points when
\begin{equation}
\frac{1}{u} + \frac{1}{|u'|} = \s_i\ .
\label{bt}
\end{equation}
associated to the transverse direction $x^i$.
This is just the lens formula and identifies the focal length as $\sigma_i^{-1}$. For $\s_i$ positive,
as is the case for both transverse directions for the beam, but only one for the particle shockwave, the
congruence is converging. The other transverse direction for the particle shockwave is diverging. 
Note that a congruence of parallel geodesics coming in from $-\infty$ will be focussed at the point 
$\s_i^{-1}$ behind the shockwave: see fig.~\ref{f5}.
Recalling that $\s_i$ is independent of $b$ for the beam shockwave, this
implies geodesics of all impact parameters focus at the same point. For the particle shockwave, 
the focal point varies with the impact parameter as $b^2$.

For spacetimes with a smooth curvature, we can expand the VVM matrix for nearby points in terms of the
curvature and its derivatives. We then have
\begin{equation}
\D_{ij}(u,u') ~=~ \d_{ij} + \frac{1}{6} R_{iuju} (u-u')^2  - \frac{1}{12} \dot{R}_{iuju} (u-u')^3 + \cdots
\label{bu}
\end{equation}
where $\dot{R}_{iuju} = \frac{d}{du}R_{iuju}$. This will be used below to relate the general formulae for the
refractive index and phase shift in terms of the vacuum polarization tensor to the effective Lagrangian.
Clearly, however, the expansion (\ref{bu}) is not appropriate for the shockwave since the curvature
$R_{iuju} \sim \d(u)$ is singular in $u$ and since $\D_{ij}(u,u')= \d_{ij}$ unless $u$ and $u'$ are separated
on opposite sides of the shock. This is directly relevant to the interpretation of results inferred from
the low-energy effective Lagrangian.

\section{Photon-Shockwave Scattering}\label{s4}

With these geometrical results in hand, we can now analyse the dynamics of the scattering of a photon
from a gravitational shockwave at the quantum loop level. The main goal is to find an explicit formula for the
instantaneous coordinate shift $\Delta v(u,\omega)$ and local phase shift $\Theta(u, \omega)$. 
This yields the scattering phase shift in the limit \eqref{bb1}.

The phase shift $\Theta(u,\omega)$ actually depends on two dimensionless ratios.
The first is $\hat{s} =  \omega\s/2m^2=Gs/b^2 m^2 = Gs \left(\l_c / b\right)^2$, 
which combines the total energy squared $s = 2\m\omega$ of the collision
and the ratio of the impact parameter $b$ and the Compton wavelength $\l_c = 1/m$ of the `electrons'
in the quantum loop, which characterises the fundamental scale of the quantum field theory.

As anticipated in the introduction, this phase shift also depends on the lightcone distance $u$ the photon 
has travelled beyond the collision; in fact, we find this dependence is entirely on the rescaled variable 
$\hat u = \s u$. Unlike the classical shift, which is discontinuous and localised at $u=0$, the photon still 
experiences the effect of the shockwave even for $u>0$, which we can picture as due to the 
finite size of the vacuum polarization cloud and is made mathematically precise using causal Green
functions in the expressions below. 

This picture, where we view the scattering process as the evolution of the photon field through a fixed 
curved spacetime background, is the quickest and most straightforward way to derive the phase shift.
We build this up in three stages. 

\begin{figure}[ht]
\begin{center}
\begin{tikzpicture}[scale=1,decoration={markings,mark=at position 0.5 with {\arrow{>}}}]
\begin{scope}[xshift=0cm,yshift=0.14cm]
\draw (0,0) circle (0.4cm); 
\end{scope}
\begin{scope}[xshift=3.2cm,yshift=-1cm]
\draw (0,0) circle (0.4cm);  
\end{scope}
%\draw[very thin,postaction={decorate}] (-2,0.9)  to[out=-20,in=170]  (5,-1.4);
\begin{scope}[xshift=0cm,yshift=0.4cm]
\draw[very thin] (-2,0.93)  to[out=-20,in=170]  (5,-1.4);
\end{scope}
\begin{scope}[xshift=0cm,yshift=-0.4cm]
\draw[very thin] (-2,0.9)  to[out=-20,in=170]  (5,-1.4);
\end{scope}
\filldraw[black] (0,0.14) circle (0.07cm);
\filldraw[black] (3.2,-1) circle (0.07cm);
%\node at (3.2,-0.8) {$x$};
%\node at (0.1,0.35) {$x'$};
\draw[decorate,decoration={snake,amplitude=0.1cm},very thick] (-2,0.9) -- (0,0.14); 
\draw[decorate,decoration={snake,amplitude=0.1cm},very thick] (3.2,-1) -- (5,-1.4); 
\draw[very thick,postaction={decorate}] (0,0.14) to[out=5,in=150] (3.2,-1);
\draw[very thick,postaction={decorate}] (0,0.14) to[out=-30,in=180] (3.2,-1);
\node at (1.7,0.3) (i1) {$e^+$};
\node at (1.5,-1.2) (i2) {$e^-$};
\node at (-2.5,1) (i3) {$\gamma$};
\node at (5.5,-1.5) (i3) {$\gamma$};
\draw[->] (3.2,0) -- (3,-0.54);
\draw[->] (2.57,-1.8) -- (2.75,-1.28);
\node at (3.8,0) {$m^{-1}$};
\draw[<->] (4.9,0.6) -- (4,-3);
\node at (4.9,0.8) {$\sigma^{-1}$};
\end{tikzpicture}
\caption{\footnotesize The scale of the $e^\pm$ loop is set by the electron's Compton wavelength $1/m$. 
If this is much smaller than the scale over which the transverse curvature varies, the length scale $1/\sigma$, 
then the full metric may be approximated by a tubular neighbourhood around the photon's null geodesic. 
This is the Penrose limit, which is a plane wave geometry.}
\label{f11}
\end{center}
\end{figure}
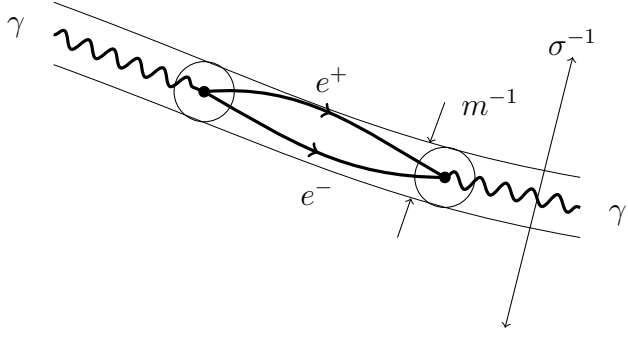

Since we are working in the limit of geometric optics, it is meaningful to analyse the effect of 
vacuum polarization on a particular ray, or null geodesic. 
One of the main insights of our previous work \cite{Hollowood:2007kt, Hollowood:2007ku,
Hollowood:2008kq,Hollowood:2009qz,Hollowood:2010bd,Hollowood:2011yh}, is that, as long 
as $m\gg\sigma$ (the transverse curvature scale), 
we may approximate the geometry in the vicinity of the chosen ray with its associated Penrose limit 
geometry. This is illustrated in fig.~\ref{f11}.
We start with the solution of the classical Maxwell equations $\nabla_\mu F^{\mu\nu}=0$ for a propagating photon 
in the Penrose limit geometry.
The solutions can be found exactly  \cite{Hollowood:2008kq},
\EQ{
A_\mu =\delta_{\mu}{}^aE^i{}_a(u) g(u)^{-1/4}\exp\big[-i(\omega V + p_a X^a + \psi(u))\big] \ ,
\label{cb}
}
where
\begin{equation}
\partial_u\psi(u) = \frac{1}{2\omega} C^{ab}(u) p_a p_b \ .
\label{cc}
\end{equation}
In \eqref{cb}, $i=1,2$ labels the two physical polarization states.

The amplitude is governed by the metric factor $\hat\vartheta = g(u)^{-1/4}$.
This is identified as the expansion, one of the optical scalars occurring in the Raychoudhuri equations,
which describes how the area of the null congruence changes along the photon trajectory.
For our purposes here, we can focus on the solution associated to the null geodesic labelled by $V$ and $X^a=0$, 
and so we will take the transverse momenta $p_a=0$. Then, 
\begin{equation}
A_\mu^{(i)}=\delta_{\mu}{}^aE^i{}_a(u) g(u)^{-1/4}e^{-i \omega V} \ ,
\label{ce}
\end{equation}
where $i=1,2$ labels the polarization.

Next, consider the solution to the field equation arising from the effective Lagrangian \eqref{ac} which includes 
the DH terms linear in the curvature. The solution can be written as 
\begin{equation}
A_\mu^{(i)}=\delta_\mu{}^aE^i{}_a(u) g(u)^{-1/4} e^{-i(\omega V-\Theta_i(u,\omega))} \ ,
\label{ch}
\end{equation}
(no sum over $i$ on the right-hand side) for each polarization state $i=1,2$.\footnote{The result here assumes 
that $h_{ij}$, the profile matrix function of the plane wave is diagonal. In the general case, we 
must replace $\exp i\Theta(u,\omega)$ with $\text{Pexp}\big[i\omega\int^u_{-\infty}du'\,\D n(u',\omega)\big]$ 
for the matrix refractive index.} 
The phase is expressed as the integral of the local matrix quantity $\Delta n_{ij}(u,\omega)$ as follows: 
\EQ{
\Theta_i(u,\omega)=\text{eigenvalues of }\left[\omega\int^u_{-\infty}du'\,\Delta n_{ij}(u,\omega)\right]\ ,
\label{uu1}
}
where the refractive index is\footnote{Strictly speaking $n_{ij}$ is only the refractive index when $\D n_{ij}$ is perturbatively small.} 
\begin{equation}
n_{ij}(u,\omega)= \delta_{ij}+\D n_{ij}(u,\omega)=\delta_{ij} - 2a R_{uu}\delta_{ij} - 8\tilde a R_{iuju}\ .
\label{cj}
\end{equation}
Notice that the DH refractive index is actually independent of $\omega$.

If we apply this formula to the shockwave, we find that the refractive index has a delta function
contribution at $u=0$:
\EQ{
n(u,\omega) = \begin{cases}1\pm8\tilde a\sigma\d(u) & \text{(particle)}\ ,\\[3pt]
1-4\sigma(a+2\tilde a)\d(u) & \text{(beam)}\ .\end{cases}
\label{ckk}
}
The $\pm$ for the particle shockwave case corresponds to the two polarizations, whereas for the beam both 
polarizations propagate in the same way.

Consequently the phase shift takes place discontinuously at the 
collision surface: 
\EQ{
\Theta(u,\omega) = \begin{cases} \pm8\tilde a \sigma\omega \vartheta(u) & \text{(particle)}\ ,\\[3pt]
-4(a+2\tilde a)\sigma\omega\vartheta(u) & \text{(beam)}\ .\end{cases}
\label{cl}
}
The corresponding coordinate shift $\D v_{\small\text{DH}}=\Theta/\omega$ is given in \eqref{vv1} 
for the particle shockwave.

Finally, we come to the complete picture in which the one-loop vacuum polarization
contribution to photon propagation is fully implemented. This has been discussed extensively
in our previous work and we only quote the final results here. In particular, ref.~\cite{Hollowood:2011yh}
gives a careful derivation of the solution in terms of an initial value problem, evaluated using the 
correct causal propagators. The field equation is 
\begin{equation}
\nabla^\nu F_{\nu\mu}(x) =-4 \int d^4 x'\, \sqrt{g'}~ \Pi^\text{ret}_{\mu\nu}(x,x') A^\nu(x')\ ,
\label{cm}
\end{equation}
where $\Pi^\text{ret}_{\mu\nu}(x,x')$ the retarded (Schwinger-Keldysh) vacuum polarization tensor. 

It turns out that since the null coordinate is playing the role of time in the plane wave background, the 
retarded polarization is actually equal to the Feynman polarization when integrated with positive frequency modes
$A_\mu(x)$ as in (\ref{cm}).  At one loop, it is expressed in terms of the Feynman scalar propagators 
of the electron/positron as 
\EQ{
\Pi^\text{ret}_{\mu\nu}(x,x') &= e^2g_{\mu\nu}\delta^{(4)}(x-x')G(x,x)\\ &+2e^2\Big[\partial_\mu G(x,x')
\partial'_\nu G(x,x')-G(x,x')\partial_\mu\partial'_\nu G(x,x')\Big]\ .
\label{cn}
}

The idea is now to solve \eqref{cm} at the one loop level but also within the eikonal approximation. 
The latter should really be termed a re-summation since it involve a perturbative correction to the phase 
rather than $A_\mu(x)$ itself. In this sense, it is in the same spirit as the Wigner-Weisskopf approach to time dependent 
states in quantum mechanics, or the {\it dynamical renormalization group\/} (see, for example, \cite{Boyanovsky:2011xn}). 
The solution takes the form \eqref{ch}, where the phase is expressed in terms of the matrix refractive index \eqref{uu1}, with
\EQ{
\Delta n_{ij}(u,\omega)=\frac{2}{\omega^2}\int_{u'\leq u} d^4x'\,(g'g)^{1/4}\,\Pi^\text{ret}_{ij}(x,x')e^{-i\omega V'}\ ,
\label{uu2}
}
where $x=(u,0,0,0)$.
The fact that the integral over $u'$ is restricted to $u'\leq u$ is just a manifestation of the causal properties of 
$\Pi^\text{ret}_{\mu\nu}(x,x')$ which vanishes when $x'$ lies outside the backward lightcone of $x$. In fact, 
the restriction happens automatically because the integral vanishes when $u'>u$ in a plane wave background 
where the null direction $u$ plays the role of time. 

The integrals in \eqref{uu2} can be evaluated using the explicit expression for the scalar Feynman propagator 
in a plane wave spacetime in the proper time formalism:
\EQ{
G(x,x')=\frac i{(4\pi)^2}\sqrt{\Delta(x,x')}\int_0^{\infty-i0^+} \frac{dT}{T^2}\,\exp\Big[\frac{i\sigma(x,x')}{2T}-im^2T\Big]\ .
}
The integral in \eqref{uu2} over $V'$ yields a delta function and those over $X^{\prime a}$ are Gaussian.
The calculation, described in detail in \cite{Hollowood:2008kq}, yields a very elegant solution in terms of the VVM matrix, 
neatly capturing the insight that the physics of vacuum polarization is determined by the geometry
of geodesic deviation. We find
\EQ{
\D n_{ij}(u,\omega)&= -\frac{i\a}{2\pi\omega}\int_0^1d\xi \,\xi(1-\xi) \\ &\times
\int_{-\infty+i0^+}^u \frac{du'}{(u-u')^2} \,e^{i z (u'-u)} \left[\sqrt{\det  \D(u,u')} \D_{ij}(u,u') 
- \d_{ij} \right] \ ,
\label{ee}
}
with $z = m^2/(2\omega\xi(1-\xi))$. 

Some remarks are in order. The integral over $u'$ in the above can be thought of as the position of one 
of the vertices of the one-loop diagram that lies in the past $u'\leq u$ of the other vertex. 
This expresses the causal nature of the correction. The parameter $\xi$ is a Feynman parameter familiar from a 
one-loop calculation. The $u'$ integral comes with a prescription of how to avoid singularities due to conjugate points 
where the VVM matrix diverges. The prescription requires that these are avoided in the upper-half plane. 

For a general background spacetime with a differentiable curvature, we can use the expansion 
(\ref{bu}) of the VVM matrix in powers of $(u-u')$ to find the low-frequency approximation
to $\D n(u,\omega)$ from this expression. A short calculation, making the convenient change of variable
$t=u-u'$, gives
\EQ{
&\D n_{ij}(u,\omega) =-\frac{i\a}{2\pi\omega}\int_0^1d\xi \,\xi(1-\xi)\\ &\times\int_0^\infty \frac{dt}{t^2}
\,e^{-i z t}\Big\{\frac{1}{12} \big(R_{uu}\delta_{ij}+2R_{iuju}\big)t^2 - \frac{1}{24}\big(\dot R_{uu}\delta_{ij}+
2\dot R_{iuju}\big)t^3 +\cdots \Big\}\\
&=- \frac{\a }{360\pi m^2}\big( R_{uu}\delta_{ij}+2R_{iuju}\big) + \frac{i\alpha\omega}{1680m^4} \big(\dot R_{uu}\delta_{ij}
+2\dot R_{iuju}\big) + \cdots \ ,
\label{cq}
}
recovering the result (\ref{cj}) derived above from the effective Lagrangian, together with the
leading higher derivative correction.\footnote{Notice that, if we assume the scale of derivatives
of the curvature is of the same order as the curvature itself, the expansion parameter here
is $\omega \sqrt{\RR}/m^2$  \cite{Shore:2007um,Hollowood:2009qz} where $\RR$ is a typical 
curvature component. This is the parameter $\omega \s/m^2$ for the shockwave.}

This series is not well defined for the shockwave because of the delta function in the Riemann tensor. 
This is of course really just an idealization, but even for the idealized shockwave we can still use 
the integral expression \eqref{ee}.
Finally, recall that for the gravitational shockwave, the VVM matrix is the identity $\D_{ij}(u,u') = \delta_{ij}$ if
$u$ and $u'$ are on the same side of the shock surface $u=0$. This means that the integral over $u'$ 
in \eqref{ee} actually has an upper limit of $u'=0$ rather than $u'=u$.

\section{The Beam Shockwave}\label{s5}

We begin with the simplest case, the beam shockwave. In this case, where the background is conformally flat,
both polarization states propagate in the same way and so the polarization indices on the refractive index and phase
can be dropped.
 
Inserting the explicit form \eqref{bq} for the VVM matrix into \eqref{ee}, the refractive index is given by
\EQ{
\Delta n(u,\omega) = -\frac{i\a}{2\pi\omega} \int_0^1 d\xi\, \xi(1-\xi) 
\int_{u}^{\infty -i0^+} dt\, e^{-i z t} \left[\big(t +
    \s u(u-t)\big)^{-2} -t^{-2} \right]\ ,
\label{ff1}
}
where $z=m^2/(2\omega\xi(1-\xi))$.
Note that the deformation of the $t$ contour in \eqref{ff1} evades the pole in the VVM determinant 
at $t = (\s u)^2/(\s u - 1)$, which is the location of the conjugate points of the congruence according to \eqref{bt}, 
by veering into the lower-half plane. Note also that $\Delta n(u,\omega)$ vanishes when $u<0$, {\it i.e.}~before 
the shockwave is reached.

The integral over $t$ can be performed analytically, giving the following expression for the 
refractive index in terms of incomplete Gamma functions:
\EQ{
\D n(u,\omega) = -\frac{\a m^2}{4\pi\omega^2}
\int_0^1 d\xi \,\Big\{\Gamma(-1,i uz)-\frac 1{(1-\sigma u)^2}\exp\Big[\frac{i\sigma u^2 z}{1-\sigma u}\Big]
\Gamma\Big(-1,\frac{iuz}{1-\sigma u}\Big)\Big\} \ .
\label{db}
}
This expression makes it clear that 
\EQ{
\D n(u,\omega)=\frac \sigma\omega F(\hat u,\hat s)\ ,\qquad \hat u=\sigma u\ ,
\quad\hat s= \frac{\omega\sigma}{2m^2}\ .
}

\begin{figure}[ht]
\begin{center}
\begin{tikzpicture}[scale=0.84]
\draw[-] (0,0.5) -- (7.66,0.5) -- (7.66,6) -- (0,6) -- (0,0.5);
\draw[-] (2.55,0.5) -- (2.55,0.7);
\node at (2.55,0) {$\sigma^{-1}$};
\node at (-0.3,5.1) {0};
\draw[dashed,black!30] (0,5.1) -- (7.66,5.1);
% \draw[-,thick] (0,5.5) -- (1.7,5.5) -- (1.7,0.3) -- (7.66,0.3);
\pgftext[at=\pgfpoint{-0.15cm}{0.86cm},left,base]{\pgfuseimage{npic10}} 
\node at (3.8,-0.1) {$u$};
\node[rotate=90] at (-0.6,3) {$\RE n$};
\begin{scope}[xshift=5.9cm,yshift=1cm,scale=1]
\draw (-0.95,-0.4) -- (1.6,-0.4) -- (1.6,1.1) -- (-0.95,1.1) -- (-0.95,-0.4);
\node at (0.3,-0.1) {\small $\omega$ increasing};
\draw[->] (-0.3,0.2) -- (-0.3,0.8);
\draw[very thick] (0,0.8) -- (1,0.8);
\draw[very thick,color=blue] (0,0.6) -- (1,0.6);
\draw[very thick,color=red] (0,0.4) -- (1,0.4);
\draw[very thick,color=green] (0,0.2) -- (1,0.2);
\end{scope}
\end{tikzpicture}
\hspace{0.2cm}
\begin{tikzpicture}[scale=0.84]
\draw[-] (0,0.5) -- (7.66,0.5) -- (7.66,6) -- (0,6) -- (0,0.5);
\draw[-] (2.55,0.5) -- (2.55,0.7);
\node at (2.55,0) {$\sigma^{-1}$};
\node at (-0.3,2.75) {0};
\draw[dashed,black!30] (0,2.75) -- (7.66,2.75);
% \draw[-,thick] (0,5.5) -- (1.7,5.5) -- (1.7,0.3) -- (7.66,0.3);
\pgftext[at=\pgfpoint{-0.15cm}{0.86cm},left,base]{\pgfuseimage{npic11}} 
\node at (3.8,-0.1) {$u$};
\node[rotate=90] at (-0.6,3.5) {$\IM n$};
\begin{scope}[xshift=5.9cm,yshift=4.7cm,scale=1]
\draw (-0.95,-0.4) -- (1.6,-0.4) -- (1.6,1.1) -- (-0.95,1.1) -- (-0.95,-0.4);
\node at (0.3,-0.1) {\small $\omega$ increasing};
\draw[->] (-0.3,0.2) -- (-0.3,0.8);
\draw[very thick] (0,0.8) -- (1,0.8);
\draw[very thick,color=blue] (0,0.6) -- (1,0.6);
\draw[very thick,color=red] (0,0.4) -- (1,0.4);
\draw[very thick,color=green] (0,0.2) -- (1,0.2);
\end{scope}
\end{tikzpicture}
\caption{\footnotesize The behaviour of the refractive index as a function of the distance from the shockwave $u$. 
The position of the focal point at $u=\sigma^{-1}$ is very pronounced. As the frequency increases, the real part 
approaches a delta function centred on the focal point, while the imaginary part changes sign (see eqs.\eqref{ff10}, \eqref{ff11}).}
\label{g1}
\end{center}
\end{figure}
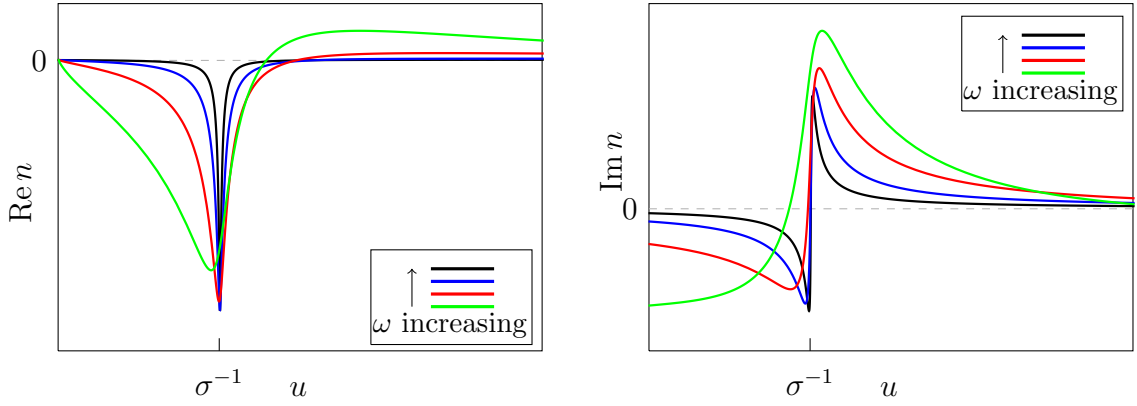

The behaviour of the refractive index as a function of the frequency $\omega$ at fixed $u$ shows a characteristic
oscillatory behaviour, with $\Delta n(u,\omega)$ taking both positive and negative values, before approaching 1 in the 
high-frequency limit as required by causality. Its dependence on $u$ is plotted in fig.~\ref{g1}.  
This shows a striking behaviour near
the focal point of the geodesic congruence at $\s u = 1$, which is explained below. 

\begin{figure}[ht]
\begin{center}
\begin{tikzpicture}[scale=0.84]
\draw[-] (0,0.5) -- (7.66,0.5) -- (7.66,6) -- (0,6) -- (0,0.5);
\draw[-] (1.3,0.5) -- (1.3,0.7);
\node at (1.3,0) {$\sigma^{-1}$};
\node at (-0.3,6) {0};
\draw[dashed,black!30] (0,1.2) -- (7.66,1.2);
\node at (-0.6,1.2) {\footnotesize $-\frac\alpha{12}$};
% \draw[-,thick] (0,5.5) -- (1.7,5.5) -- (1.7,0.3) -- (7.66,0.3);
\pgftext[at=\pgfpoint{-0.15cm}{1.25cm},left,base]{\pgfuseimage{npic1}} 
\node at (3.8,-0.1) {$u$};
\node[rotate=90] at (-0.6,3) {$\RE\Theta$};
\begin{scope}[xshift=5.9cm,yshift=2.35cm,scale=1]
\draw (-0.98,-0.4) -- (1.6,-0.4) -- (1.6,1.1) -- (-0.98,1.1) -- (-0.98,-0.4);
\node at (0.3,-0.1) {\small $\omega$ increasing};
\draw[->] (-0.3,0.2) -- (-0.3,0.8);
\draw[very thick] (0,0.8) -- (1,0.8);
\draw[very thick,color=blue] (0,0.6) -- (1,0.6);
\draw[very thick,color=red] (0,0.4) -- (1,0.4);
\draw[very thick,color=green] (0,0.2) -- (1,0.2);
\end{scope}
\end{tikzpicture}
\hspace{0.1cm}
\begin{tikzpicture}[scale=0.84]
\draw[-] (0,0.5) -- (7.66,0.5) -- (7.66,6) -- (0,6) -- (0,0.5);
\node at (-0.3,6) {0};
\draw[dashed,black!30] (0,1.2) -- (7.66,1.2);
\node at (-0.6,1.2) {\footnotesize $-\frac\alpha{12}$};
\pgftext[at=\pgfpoint{-0.15cm}{1.25cm},left,base]{\pgfuseimage{npic7}} 
\node at (3.8,-0.1) {$\log\omega$};
\node[rotate=90] at (-0.6,3) {$\RE\Theta$};
\begin{scope}[xshift=1.1cm,yshift=2cm,scale=1]
\draw (-0.98,-0.4) -- (1.6,-0.4) -- (1.6,1.1) -- (-0.98,1.1) -- (-0.98,-0.4);
\node at (0.3,-0.1) {\small $u$ increasing};
\draw[<-] (-0.3,0.2) -- (-0.3,0.8);
\draw[very thick] (0,0.8) -- (1,0.8);
\draw[very thick,color=blue] (0,0.6) -- (1,0.6);
\draw[very thick,color=red] (0,0.4) -- (1,0.4);
\draw[very thick,color=green] (0,0.2) -- (1,0.2);
\end{scope}\end{tikzpicture}\\
{}\hspace{0.2cm}\begin{tikzpicture}[scale=0.84]
\draw[-] (0,0.5) -- (7.66,0.5) -- (7.66,6) -- (0,6) -- (0,0.5);
\draw[-] (1.3,0.5) -- (1.3,0.7);
\node at (1.3,0) {$\sigma^{-1}$};
\node at (-0.3,4.2) {0};
\draw[dashed,black!30] (0,4.2) -- (7.66,4.2);
\pgftext[at=\pgfpoint{-0.15cm}{0.86cm},left,base]{\pgfuseimage{npic2}} 
\node at (3.8,-0.1) {$u$};
\node[rotate=90] at (-0.6,3) {$\IM\Theta$};
\begin{scope}[xshift=5.9cm,yshift=1cm,scale=1]
\draw (-0.95,-0.4) -- (1.6,-0.4) -- (1.6,1.1) -- (-0.95,1.1) -- (-0.95,-0.4);
\node at (0.3,-0.1) {\small $\omega$ increasing};
\draw[->] (-0.3,0.2) -- (-0.3,0.8);
\draw[very thick] (0,0.8) -- (1,0.8);
\draw[very thick,color=blue] (0,0.6) -- (1,0.6);
\draw[very thick,color=red] (0,0.4) -- (1,0.4);
\draw[very thick,color=green] (0,0.2) -- (1,0.2);
\end{scope}
\end{tikzpicture}
\hspace{0.25cm}
\begin{tikzpicture}[scale=0.84]
\draw[-] (0,0.5) -- (7.66,0.5) -- (7.66,6) -- (0,6) -- (0,0.5);
\node at (-0.3,1.15) {0};
\draw[dashed,black!30] (0,1.15) -- (7.66,1.15);
\pgftext[at=\pgfpoint{-0.15cm}{0.86cm},left,base]{\pgfuseimage{npic9}} 
\node at (3.8,-0.1) {$\log\omega$};
\node[rotate=90] at (-0.6,3) {$\IM\Theta$};
\begin{scope}[xshift=1.1cm,yshift=4.7cm,scale=1]
\draw (-0.95,-0.4) -- (1.6,-0.4) -- (1.6,1.1) -- (-0.95,1.1) -- (-0.95,-0.4);
\node at (0.3,-0.1) {\small $u$ increasing};
\draw[<-] (-0.3,0.2) -- (-0.3,0.8);
\draw[very thick] (0,0.8) -- (1,0.8);
\draw[very thick,color=blue] (0,0.6) -- (1,0.6);
\draw[very thick,color=red] (0,0.4) -- (1,0.4);
\draw[very thick,color=green] (0,0.2) -- (1,0.2);
\end{scope}
\end{tikzpicture}
\caption{\footnotesize The real and imaginary parts of the phase $\Theta(u,\omega)$
 as a function of $u$ and $\omega$ for the beam shockwave. Notice especially the step function
shift in the high-frequency limit of $\RE\Theta(u,\omega)$ at the focal point $u = \sigma^{-1}$.
Also note that in QED, $\RE\Theta(u,\omega)$ approaches a negative constant for high frequencies.}
\label{g3}
\end{center}
\end{figure}
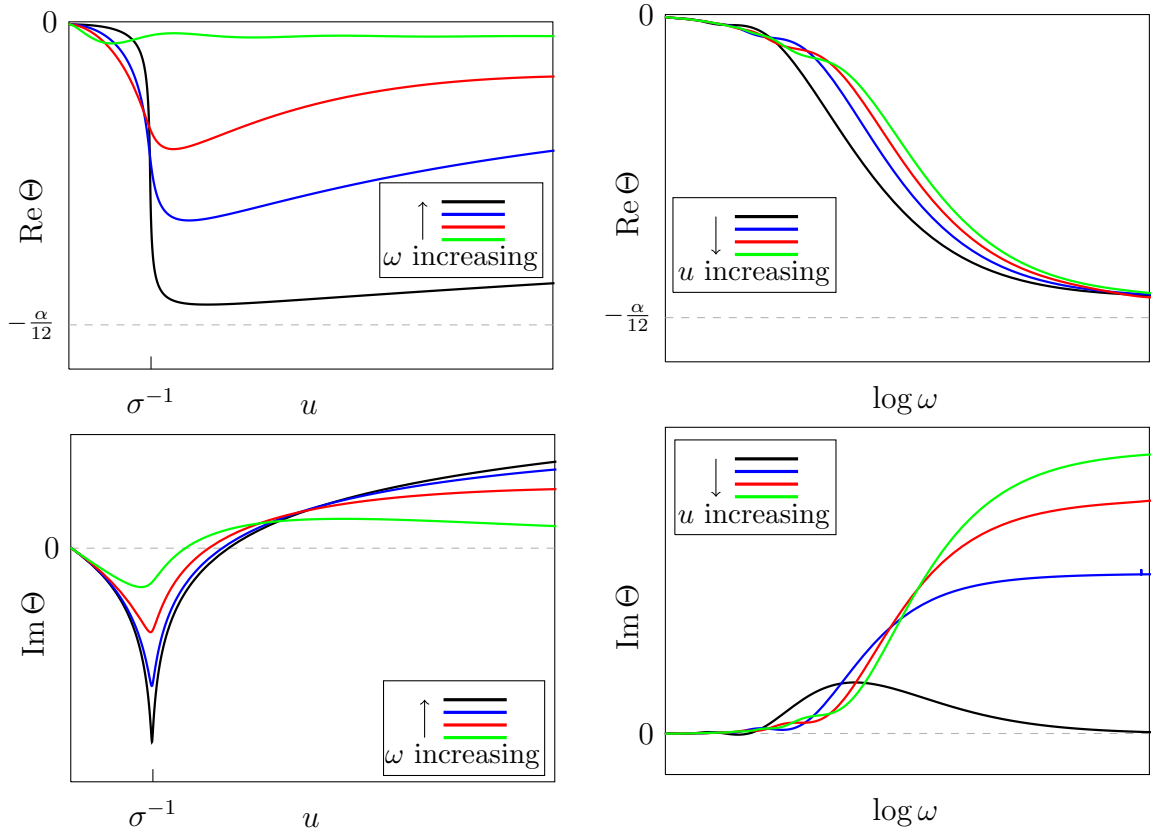

The shift $\Delta v$ and local phase $\Theta(u,\omega)$ can then be obtained by integrating as in \eqref{ee}. 
Because $\Delta n(u,\infty)$ is implicitly only non-vanishing when $u\geq0$, we have
\EQ{
\Delta v(u,\omega)=\int_0^udu'\,\Delta n(u',\omega)\ ,\qquad \Theta(u,\omega)=\omega\Delta v(u,\omega)\ .
\label{ff4}}
The results of a numerical integration for $\Theta(u,\w)$ are shown in fig.~\ref{g3}
as functions of both $u$ and $\w$. 

Before commenting on these figures in detail, it is interesting to study the form of the phase shift 
for small values of $\hat{s} = \omega\s/2m^2$
analytically in order to make contact with the effective Lagrangian and contrast the
corresponding predictions. We can do this by expanding the integrand of \eqref{db} in a 
power series in the curvature $\sigma$. The leading term for the phase is
\EQ{
\Theta(u,\omega)=-\frac{i\alpha\sigma}{2\p}\int_0^1d\xi\,\int_0^udu'\,\int_u^\infty dt\,\frac{2(t-u')u'}{t^3}e^{-izt}+\cdots\ .
}
Performing the integral over $t$ gives
\EQ{
&\Theta(u,\omega)=
-\frac{i\alpha\sigma}{2\p}\int_0^1d\xi\,\xi(1-\xi)\\ &\times\int_0^udu'\,
\Big[(1+iu'z)e^{-iu'z}+iu'z(2+iu'z)\text{Ei}(-iu'z)\Big]+\cdots\ .
}

We can now explicitly perform the integral over $u'$ by using the prescription $u'\to u'-i0^+$ in the
$u\rta \infty$ limit to find the scattering phase shift. We have,
\begin{equation}
\int_0^\infty du'\,
\Big[(1+iu'z)e^{-iu'z}+iu'z(2+iu'z)\text{Ei}(-iu'z)\Big]=\frac1{3iz}  \ ,
\label{df}
\end{equation}
and so finally performing the integral over $\xi$ we find, to linear order in the curvature,
\begin{equation}
\Theta_\text{scat.}(s,b)\equiv\Theta(u\to\infty,\omega) = - \frac{\alpha\sigma\omega}{90\pi m^2}
+\cdots=-\frac{\alpha}{45\pi}\cdot\frac{Gs}{b^2m^2}+\cdots\ .
\label{dg}
\end{equation}
This is precisely the DH phase shift for the beam shockwave, as can be seen by substituting the values 
\eqref{uu2} into \eqref{cl}.
This behaviour is also evident in the plots in fig.~\ref{g3} for $\Theta(u,\w)$ at low frequency, 
where we can see the $u$-independence 
and linear dependence on $\w$ of $\RE\Theta$ given in \eqref {cl}, while $\IM\Theta = O(\w^2)$.

The key point here, however, is that this value of the phase shift is only realised
asymptotically far from the collision surface $u=0$, whereas the DH effective Lagrangian
predicts that it takes place discontinuously at $u=0$. The full quantum field theory smooths out 
the discontinuous effect of the shockwave collision by virtue of its intrinsic scale, in this case the 
size of the vacuum polarisation cloud dressing the photon. Of course, this impacts on the 
question of whether such a phase shift could be used in time machine constructions which
assume a discontinuous Shapiro time advance 
\EQ{
\D v_{\small\text{DH}}=-\frac{\alpha\sigma}{90\pi m^2}<0\ ,
} 
even setting aside the fact that it holds only in the low-energy limit.

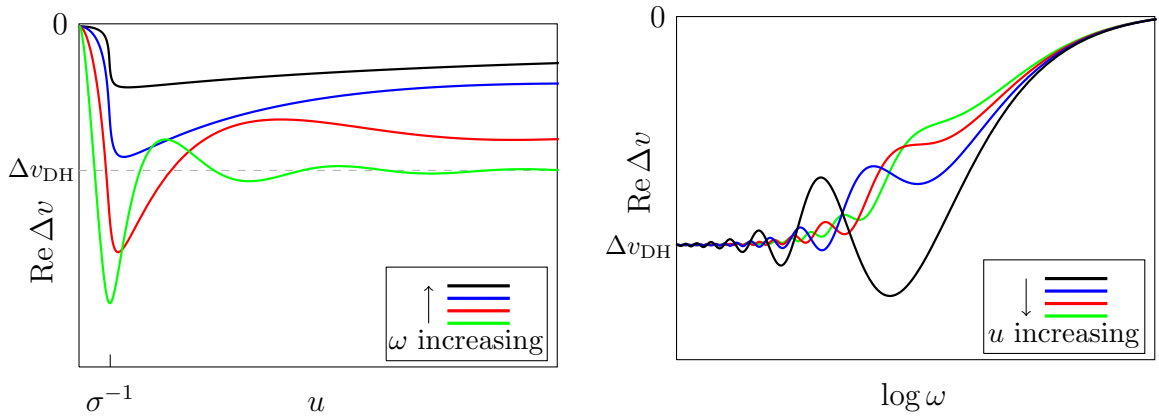
\begin{figure}[ht]
\begin{center}
\begin{tikzpicture}[scale=0.83]
\draw[-] (0,0.5) -- (7.66,0.5) -- (7.66,6) -- (0,6) -- (0,0.5);
\draw[-] (0.5,0.5) -- (0.5,0.7);
\node at (0.5,0) {$\sigma^{-1}$};
\node at (-0.3,6) {0};
\node at (-0.6,3.65) {\footnotesize $\D v_{\small\text{DH}}$};
\draw[dashed,black!30] (0,3.65) -- (7.66,3.65);
% \draw[-,thick] (0,5.5) -- (1.7,5.5) -- (1.7,0.3) -- (7.66,0.3);
\pgftext[at=\pgfpoint{-0.15cm}{1.25cm},left,base]{\pgfuseimage{npic14}} 
\node at (3.8,-0.1) {$u$};
\node[rotate=90] at (-0.6,2.5) {$\RE\D v$};
\begin{scope}[xshift=5.9cm,yshift=1cm,scale=1]
\draw (-0.98,-0.4) -- (1.6,-0.4) -- (1.6,1.1) -- (-0.98,1.1) -- (-0.98,-0.4);
\node at (0.3,-0.1) {\small $\omega$ increasing};
\draw[->] (-0.3,0.2) -- (-0.3,0.8);
\draw[very thick] (0,0.8) -- (1,0.8);
\draw[very thick,color=blue] (0,0.6) -- (1,0.6);
\draw[very thick,color=red] (0,0.4) -- (1,0.4);
\draw[very thick,color=green] (0,0.2) -- (1,0.2);
\end{scope}
\end{tikzpicture}
\hspace{0.1cm}
\begin{tikzpicture}[scale=0.83]
\draw[-] (0,0.5) -- (7.66,0.5) -- (7.66,6) -- (0,6) -- (0,0.5);
\node at (-0.6,2.3) {\footnotesize $\D v_{\small\text{DH}}$};
\node at (-0.3,6) {0};
%\draw[dashed,black!30] (0,2.6) -- (7.66,2.6);
\pgftext[at=\pgfpoint{-0.15cm}{1.25cm},left,base]{\pgfuseimage{npic8}} 
\node at (3.8,-0.1) {$\log\omega$};
\node[rotate=90] at (-0.6,3.5) {$\RE\Delta v$};
\begin{scope}[xshift=5.9cm,yshift=1cm,scale=1]
\draw (-0.98,-0.4) -- (1.6,-0.4) -- (1.6,1.1) -- (-0.98,1.1) -- (-0.98,-0.4);
\node at (0.3,-0.1) {\small $u$ increasing};
\draw[<-] (-0.3,0.2) -- (-0.3,0.8);
\draw[very thick] (0,0.8) -- (1,0.8);
\draw[very thick,color=blue] (0,0.6) -- (1,0.6);
\draw[very thick,color=red] (0,0.4) -- (1,0.4);
\draw[very thick,color=green] (0,0.2) -- (1,0.2);
\end{scope}
\end{tikzpicture}
\caption{\footnotesize The effective coordinate shift $\RE\Delta v(u,\omega)$ as a function of lightcone distance 
$u$ from the beam shockwave for different values of the photon frequency $\omega$ (LH figure) and as a 
function of $\omega$ for different values of $u$ (RH figure).}
\label{g2}
\end{center}
\end{figure}

The behaviour of the real part of $\Delta v$ as a function of $u$ and $\omega$ is shown in fig.~\ref{g2}.
It is clear from the plots that the shift $\D v$ does not occur discontinuously at $u=0$.  
Rather, $\RE\Delta v(u,\omega)$ oscillates before eventually settling to a fixed limit far from the shockwave. 

The frequency dependence of $\RE\Delta v(u,\omega)$ is shown in the right-hand plot in fig.~\ref{g2}.
It shows clearly how the full, UV complete, quantum field theory
reproduces the effective Lagrangian prediction for low collision energy, 
$\Delta v(u,\omega\rta0) = \Delta v_{\small\text{DH}}<0$, but then has an oscillatory dependence
on $\omega$ before vanishing asymptotically for large $\omega$ as $\omega^{-1}$.

\subsection{High frequency limit}

The key regime for a proper discussion of causality is the high frequency limit. In present circumstances this 
corresponds to $\omega\sigma/m^2\gg1$. We can calculate the behaviour in this limit analytically 
by going back to the integral expression 
\eqref{ff1}. The asymptotic high frequency regime is obtained by taking $z=0$ in the integrand. 
The $t$ integral is then trivial and gives
\EQ{
\Delta n(u,\omega\to\infty)&= -\frac{i\a}{2\pi\omega} \int_0^1 d\xi\, \xi(1-\xi) 
\int_{u}^{\infty -i0^+} dt\, \left[\big(t +
    \s u(u-t)\big)^{-2} -t^{-2} \right]\\ &=
\frac{i\alpha}{12\pi\omega}\cdot\frac\sigma{\sigma u-1-i0^+}\ ,
\label{ff10}
}
where the prescription for avoiding the double pole at the focal point follows from the original 
contour deviation in \eqref{ee}, which is determined by causality.
For the real part, we therefore have
\EQ{
\RE \Delta n(u,\omega\to\infty)=-\frac{\alpha}{12\omega}\delta(u-\sigma^{-1})\ ,
\label{ff11}}
which is evident in the left-hand plot of fig.~\ref{g1}. It is interesting, therefore, to compare this with the 
low frequency limit for the real part of the refractive index, that is \eqref{ckk} 
\EQ{
\RE \Delta n(u,\omega\to0)=- \frac{\alpha\sigma}{90\pi m^2}\delta(u)\ .
}
So both involve a delta function contribution, but at low frequency this occurs at the shockwave, while at high frequency 
it occurs at the focal point.

Integrating as in \eqref{ff4} gives the high frequency behaviour of the phase:
\EQ{
\Theta(u,\omega\to\infty)=\frac{\alpha}{12\pi}\Big[-\pi\vartheta(\sigma u-1)+i\log\big|\sigma u-1\big|\Big]\ .
\label{focal}}
The high frequency dependence of the phase is evident in the left-hand plots of fig.~\ref{g3} 
which illustrate the step function shift in $\RE\Theta(u,\w)$ at the focal point, arising from 
integration of the corresponding delta function in $\Delta n(u,\w)$.    For the scattering phase itself,
we find
\EQ{
\Theta_\text{scat.}=\Theta(u\to\infty,\omega\to\infty)=-\frac{\alpha}{12}+\frac{i\alpha}{12\pi}\log(\sigma u)\ .
}
Notice that the requirement of causality that $\Delta n(u,\omega\to\infty)$ goes to zero does not preclude
a non-vanishing value for $\Theta_\text{scat.}$. 
It is particularly noteworthy for the later discussion of causality that $\Theta_\text{scat.}(s,b)$ is a perturbatively small constant.

In this limit, the imaginary part of the phase can be understood as a modulation of the photon amplitude of the form
\EQ{
\big|\sigma u-1\big|^{-\alpha/12\pi}\ .
}
This decreases once the focal point is passed and manifests a real-time wavefunction renormalization of the photon field
which we can interpret as an increased dressing of the photon by the virtual $e^+ e^-$ cloud \cite{Hollowood:2011yh}.

\section{The Particle Shockwave}\label{s6}

In this section, we consider the particle shockwave.  In this case, unlike the beam, the background is not
conformally flat and the photon propagation is polarization dependent, {\it i.e.} displays gravitational birefringence.
However, the conclusions regarding causality and time machines are essentially the same as for the beam shockwave, 
so our discussion will be brief. 

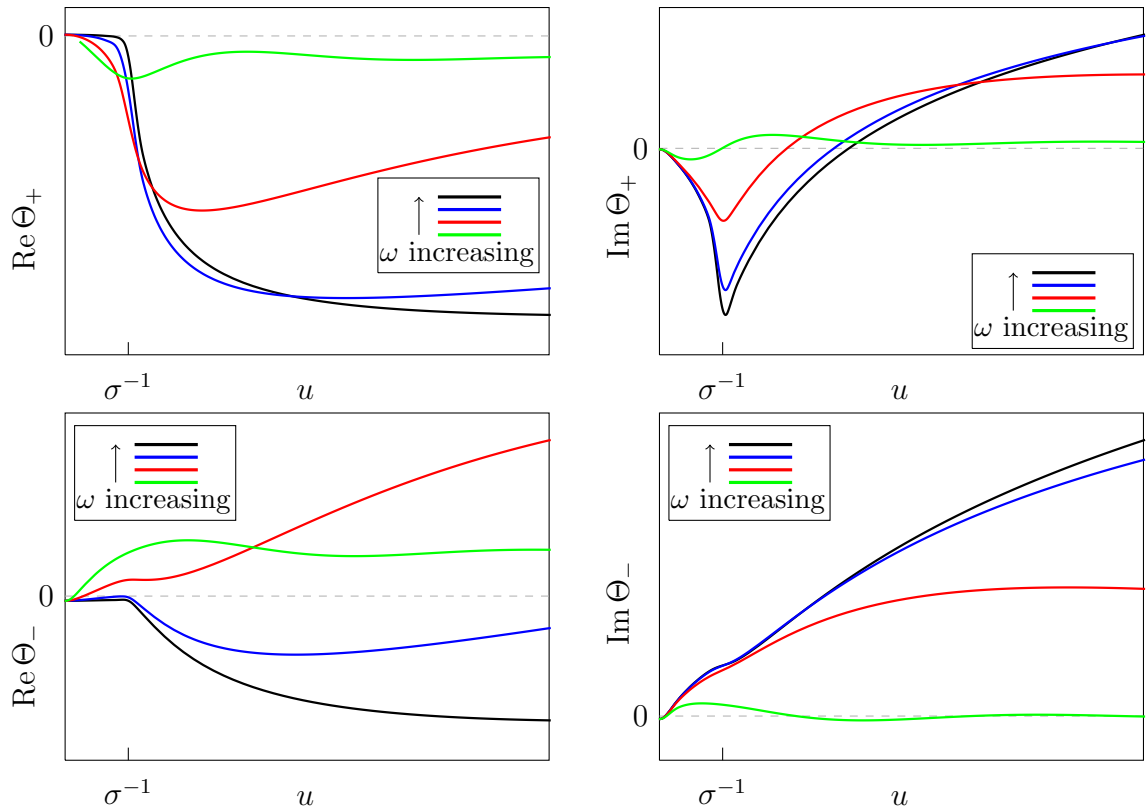
\begin{figure}[ht]
\begin{center}
\begin{tikzpicture}[scale=0.84]
\draw[-] (0,0.5) -- (7.66,0.5) -- (7.66,6) -- (0,6) -- (0,0.5);
\draw[-] (1,0.5) -- (1,0.7);
\node at (1,0) {$\sigma^{-1}$};
\node at (-0.3,5.55) {0};
\draw[dashed,black!30] (0,5.55) -- (7.66,5.55);
\pgftext[at=\pgfpoint{-0.15cm}{0.86cm},left,base]{\pgfuseimage{npic3}} 
\node at (3.8,-0.1) {$u$};
\node[rotate=90] at (-0.6,2.5) {$\RE\Theta_+$};
\begin{scope}[xshift=5.9cm,yshift=2.2cm,scale=1]
\draw (-0.95,-0.4) -- (1.6,-0.4) -- (1.6,1.1) -- (-0.95,1.1) -- (-0.95,-0.4);
\node at (0.3,-0.1) {\small $\omega$ increasing};
\draw[->] (-0.3,0.2) -- (-0.3,0.8);
\draw[very thick] (0,0.8) -- (1,0.8);
\draw[very thick,color=blue] (0,0.6) -- (1,0.6);
\draw[very thick,color=red] (0,0.4) -- (1,0.4);
\draw[very thick,color=green] (0,0.2) -- (1,0.2);
\end{scope}
\end{tikzpicture}
\hspace{0.2cm}
\begin{tikzpicture}[scale=0.84]
\draw[-] (0,0.5) -- (7.66,0.5) -- (7.66,6) -- (0,6) -- (0,0.5);
\draw[-] (1,0.5) -- (1,0.7);
\node at (1,0) {$\sigma^{-1}$};
\node at (-0.3,3.77) {0};
\draw[dashed,black!30] (0,3.77) -- (7.66,3.77);
\pgftext[at=\pgfpoint{-0.15cm}{0.86cm},left,base]{\pgfuseimage{npic4}} 
\node at (3.8,-0.1) {$u$};
\node[rotate=90] at (-0.6,2.7) {$\IM\Theta_+$};
\begin{scope}[xshift=5.9cm,yshift=1cm,scale=1]
\draw (-0.95,-0.4) -- (1.6,-0.4) -- (1.6,1.1) -- (-0.95,1.1) -- (-0.95,-0.4);
\node at (0.3,-0.1) {\small $\omega$ increasing};
\draw[->] (-0.3,0.2) -- (-0.3,0.8);
\draw[very thick] (0,0.8) -- (1,0.8);
\draw[very thick,color=blue] (0,0.6) -- (1,0.6);
\draw[very thick,color=red] (0,0.4) -- (1,0.4);
\draw[very thick,color=green] (0,0.2) -- (1,0.2);
\end{scope}
\end{tikzpicture}\\
\begin{tikzpicture}[scale=0.84]
\draw[-] (0,0.5) -- (7.66,0.5) -- (7.66,6) -- (0,6) -- (0,0.5);
\draw[-] (1,0.5) -- (1,0.7);
\node at (1,0) {$\sigma^{-1}$};
\node at (-0.3,3.1) {0};
\draw[dashed,black!30] (0,3.1) -- (7.66,3.1);
\pgftext[at=\pgfpoint{-0.15cm}{0.86cm},left,base]{\pgfuseimage{npic5}} 
\node at (3.8,-0.1) {$u$};
\node[rotate=90] at (-0.6,2) {$\RE\Theta_-$};
\begin{scope}[xshift=1.1cm,yshift=4.7cm,scale=1]
\draw (-0.95,-0.4) -- (1.6,-0.4) -- (1.6,1.1) -- (-0.95,1.1) -- (-0.95,-0.4);
\node at (0.3,-0.1) {\small $\omega$ increasing};
\draw[->] (-0.3,0.2) -- (-0.3,0.8);
\draw[very thick] (0,0.8) -- (1,0.8);
\draw[very thick,color=blue] (0,0.6) -- (1,0.6);
\draw[very thick,color=red] (0,0.4) -- (1,0.4);
\draw[very thick,color=green] (0,0.2) -- (1,0.2);
\end{scope}
\end{tikzpicture}
\hspace{0.2cm}
\begin{tikzpicture}[scale=0.84]
\draw[-] (0,0.5) -- (7.66,0.5) -- (7.66,6) -- (0,6) -- (0,0.5);
\draw[-] (1,0.5) -- (1,0.7);
\node at (1,0) {$\sigma^{-1}$};
\node at (-0.3,1.2) {0};
\draw[dashed,black!30] (0,1.2) -- (7.66,1.2);
\pgftext[at=\pgfpoint{-0.15cm}{0.86cm},left,base]{\pgfuseimage{npic6}} 
\node at (3.8,-0.1) {$u$};
\node[rotate=90] at (-0.6,3) {$\IM\Theta_-$};
\begin{scope}[xshift=1.1cm,yshift=4.7cm,scale=1]
\draw (-0.95,-0.4) -- (1.6,-0.4) -- (1.6,1.1) -- (-0.95,1.1) -- (-0.95,-0.4);
\node at (0.3,-0.1) {\small $\omega$ increasing};
\draw[->] (-0.3,0.2) -- (-0.3,0.8);
\draw[very thick] (0,0.8) -- (1,0.8);
\draw[very thick,color=blue] (0,0.6) -- (1,0.6);
\draw[very thick,color=red] (0,0.4) -- (1,0.4);
\draw[very thick,color=green] (0,0.2) -- (1,0.2);
\end{scope}
\end{tikzpicture}
\caption{\footnotesize The real and imaginary parts of the phase shift as a function of $u$ for the two polarization states for 
QED in a particle shockwave.}
\label{g4}
\end{center}
\end{figure}

The refractive index, for the the polarization states labelled as $j=\pm$, takes the form
\EQ{
&\Delta n_j(u,\omega) = -\frac{i\a}{2\pi\omega} \int_0^1 d\xi\, \xi(1-\xi) \\ &\times
\int_{u}^{\infty -i0^+} dt\, e^{-i z t} \left[(t +
    \s u(u-t))^{-\frac j2-1}(t-\s u(u-t))^{\frac j2-1} -t^{-2} \right]\ .
\label{gg1}
}
This can be integrated numerically to find the local phase shifts $\Theta_{\pm}(u,\w)$ for the
two polarizations, and the results are illustrated in fig.~\ref{g4}. 

The low-frequency features from \eqref{cl} are again apparent in the plots. Note particularly the equal and opposite
sign values of $\RE\Theta(u\to\infty,\w)$ at low frequency which reproduce the Drummond-Hathrell values.
This feature was first identified as the ``polarization sum rule'' for Ricci-flat spacetimes in \cite{Shore:1995fz}.
The imaginary parts are again of $O(\w^2)$.

Mirroring our discussion of the beam shockwave, we can determine the high frequency limit analytically. 
Setting $z=0$ in the integrand gives us the asymptotic form
\EQ{
\Delta n_\pm(u,\omega\to\infty)=-\frac{i\alpha}{12\pi\omega\sigma u^2}
\Big[\sqrt{\frac{1\pm\sigma u}{1\mp\sigma u}}-1\mp\sigma u\Big]\ .
\label{gg2}
}
In this case, the singularities at the focal point become branch points rather than the poles occurring for the beam shockwave.   
The prescription for dealing with these in the expression above is to take
$\sigma u\to\sigma u-i0^+$. Performing the integral in \eqref{ff4}, we then find the high frequency limit of the phase shift:
\EQ{
\Theta_\pm(u,\omega\to\infty)=\frac{i\alpha}{12\pi}\Big[\mp\frac{1-\sqrt{1-(\sigma u)^2}}{\sigma u}+\log\frac{1+
\sqrt{1-(\sigma u)^2}}2\Big]\ .
\label{gg3}
}

The conclusion for causality is the same as for the beam shockwave. Since $\Delta n(u,\omega)$ is $O(1/\omega)$, the phase
velocity goes to 1 as $\omega^{-1}$.  The high-frequency limit of the phase is a negative constant (for both polarizations)
and is bounded by a perturbatively small amount, ensuring that the coordinate shift goes to zero like $\omega^{-1}$.

\section{The Fate of Time Machines and Causality}\label{TM2}

Given these exact results for the high-frequency limit of the refractive index and phase shifts,
we can now see why the shockwave time machine fails to work. In fact it fails on several counts.

First of all, the real part of the local phase $\Theta(u,\omega)$ is bounded by its high frequency limit 
far from the shockwave:
\EQ{
\Theta(u,\omega)\leq\Theta(u\to\infty,\omega\to\infty)=-\frac\alpha{12}\ .
\label{kk2}
}
In other words, the scattering phase always remains perturbatively small. This means that the 
observability requirement \eqref{kk1} can never be satisfied.
 
This is sufficient in itself to recover causality, but the {\it coup de gr\^ace} for a time machine is provided 
by the fact that the coordinate shift $\D v(u,\omega)$ goes to zero in the high-frequency limit. 
As we have frequently emphasised, in order to discuss causality we need to consider the high-frequency limit
of photon propagation---in this context, to show that the closed null trajectory is realised
by a wave with phase velocity $v_{\rm ph}(\infty)$. However, we have shown that in this limit
the refractive index goes to 1, {\it i.e.}~the phase velocity $v_{\rm ph}(\omega \rta \infty) = 1$. 
It follows that in this limit, there is {\it no\/} coordinate shift from the quantum loop diagrams.  

Yet another reason for the failure of the time machine is clear from fig.~\ref{g2}, which shows
that the shift $\Delta v(u,\omega)$ does not occur instantaneously at the shockwave itself ($u=0$).
In fact, for the high-frequency photons relevant for causality, the shift occurs at the focal point 
$u=\sigma^{-1}$ in front of the shockwave. 
Since the jump is not discontinuous, the coordinate shift
necessarily takes the photon trajectory into the curved region IV 
(see fig.~\ref{f6}) where the Aichelburg-Sexl geodesic equations no longer apply.
The photon never reaches the post-collision point $P$ in fig.~\ref{f6}, or $P'$ in fig.~\ref{f7},
in the time machine trajectory.

In the end then, we see that the implication of the effective Lagrangian that there is a 
causality-violating Shapiro time advance when a photon scatters from a gravitational shockwave,
and that this permits the construction of a closed null curve or time machine, 
does not survive in the full quantum field theory. The consistent UV completion encoded
in the full theory ensures that causality is preserved.

\section{Scalar Field Theory}\label{s7}

Since the resolution of the causality problem arising in the low-energy effective Lagrangian is intimately
related to its UV completion, it is interesting to consider the same issues in a super-renormalizable theory,
for which the UV behaviour differs from that of QED.
We therefore consider a 4-dim theory with a massless scalar field $A$, playing the role of the ``photon", and a massive scalar 
field $\phi$, playing the role of the ``electron'', with an interaction $eA\phi^2$. 
We find that while the causality problem is resolved in a qualitatively similar way to QED, there are significant
differences of detail arising from the different UV power counting.

The analogue of the Drummond-Hathrell curvature-dependent term in the effective action of the scalar photon is
\begin{equation}
S= \int d^4 x\,\sqrt{g} \left[\frac{1}{2}g^{\m\n}\partial_\m A \partial_\n A +
a   R^{\m\n}\partial_\m A \partial_\n A \right]\ .
\label{cf}
\end{equation}
The curvature term arises by integrating out the heavy field $\phi$, and we have
\EQ{
a = \frac{e^2}{1440 m^2}\ ,\qquad\alpha=\frac{e^2}{4\pi m^2}\ .
}
where $\alpha$ is a dimensionless coupling.
The curvature coupling leads to a local refractive index 
\EQ{
n(u,\omega) = 1 - a R_{uu} \ .
\label{cj}
}
So for the beam shockwave, there is a singular contribution to the refractive index:
\begin{equation}
n(u,\omega) = 1 -  \frac{\a}{720\pi}\frac{\s}{m^2} \d(u) \ .
\label{ck}
\end{equation}
This leads to a negative shift in the null coordinate
\EQ{
\Delta v_{\small\text{DH}}=-  \frac{\a}{720\pi}\frac{\s}{m^2} 
}
occurring discontinuously at the shockwave $u=0$.

\begin{figure}[ht]
\begin{center}
\begin{tikzpicture}[scale=0.84]
\draw[-] (0,0.5) -- (7.66,0.5) -- (7.66,6) -- (0,6) -- (0,0.5);
\draw[-] (2.55,0.5) -- (2.55,0.7);
\node at (2.55,0) {$\sigma^{-1}$};
\node at (-0.3,4.95) {0};
\draw[dashed,black!30] (0,4.95) -- (7.66,4.95);
% \draw[-,thick] (0,5.5) -- (1.7,5.5) -- (1.7,0.3) -- (7.66,0.3);
\pgftext[at=\pgfpoint{-0.15cm}{0.86cm},left,base]{\pgfuseimage{spic1}} 
\node at (3.8,-0.1) {$u$};
\node[rotate=90] at (-0.6,3) {$\RE n$};
\begin{scope}[xshift=5.9cm,yshift=1cm,scale=1]
\draw (-0.95,-0.4) -- (1.6,-0.4) -- (1.6,1.1) -- (-0.95,1.1) -- (-0.95,-0.4);
\node at (0.3,-0.1) {\small $\omega$ increasing};
\draw[->] (-0.3,0.2) -- (-0.3,0.8);
\draw[very thick] (0,0.8) -- (1,0.8);
\draw[very thick,color=blue] (0,0.6) -- (1,0.6);
\draw[very thick,color=red] (0,0.4) -- (1,0.4);
\draw[very thick,color=green] (0,0.2) -- (1,0.2);
\end{scope}
\end{tikzpicture}
\hspace{0.2cm}
\begin{tikzpicture}[scale=0.84]
\draw[-] (0,0.5) -- (7.66,0.5) -- (7.66,6) -- (0,6) -- (0,0.5);
\draw[-] (2.55,0.5) -- (2.55,0.7);
\node at (2.55,0) {$\sigma^{-1}$};
\node at (-0.3,4.22) {0};
\draw[dashed,black!30] (0,4.22) -- (7.66,4.22);
% \draw[-,thick] (0,5.5) -- (1.7,5.5) -- (1.7,0.3) -- (7.66,0.3);
\pgftext[at=\pgfpoint{-0.15cm}{0.86cm},left,base]{\pgfuseimage{spic2}} 
\node at (3.8,-0.1) {$u$};
\node[rotate=90] at (-0.6,3) {$\IM n$};
\begin{scope}[xshift=5.9cm,yshift=1cm,scale=1]
\draw (-0.95,-0.4) -- (1.6,-0.4) -- (1.6,1.1) -- (-0.95,1.1) -- (-0.95,-0.4);
\node at (0.3,-0.1) {\small $\omega$ increasing};
\draw[->] (-0.3,0.2) -- (-0.3,0.8);
\draw[very thick] (0,0.8) -- (1,0.8);
\draw[very thick,color=blue] (0,0.6) -- (1,0.6);
\draw[very thick,color=red] (0,0.4) -- (1,0.4);
\draw[very thick,color=green] (0,0.2) -- (1,0.2);
\end{scope}
\end{tikzpicture}
\caption{The real and imaginary parts of the refractive index $\Delta n(u,\omega)$ 
in the scalar $eA\phi^2$ theory as a function of $u$
at fixed values of the frequency $\omega$.}
\label{h1}
\end{center}
\end{figure}
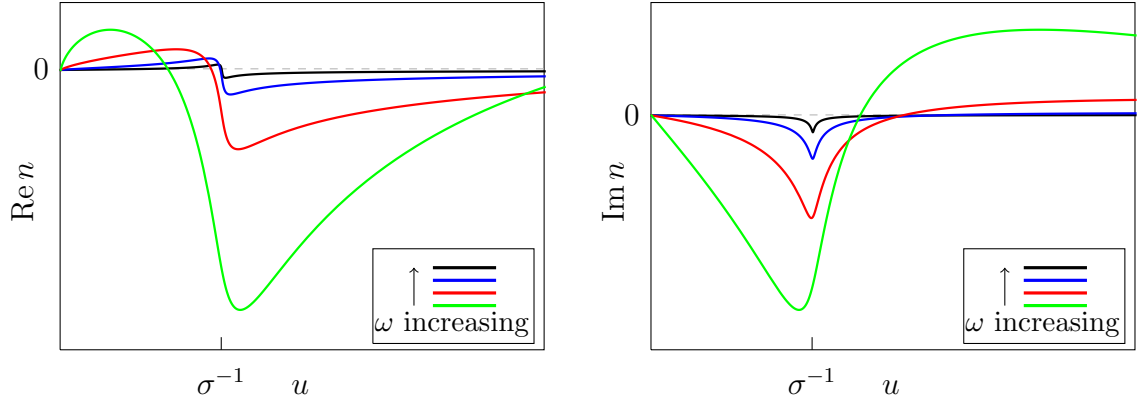

Calculating the vacuum polarization in the full QFT gives the following expression for the refractive index:
\EQ{
&\D n(u,\omega) = \frac{\a m^2}{16\pi\omega^2} \int_0^1 d\xi \\
&\times\left[
\int_{-\infty+i0^+}^u \frac{du'}{u-u'}~e^{-i z (u-u')}~\sqrt{\det \D(u,u')} ~-
\RE\int_{-\infty+i0^+}^u \frac{du'}{u-u'}~e^{-i z (u-u')} \right] \ ,
\label{cp}
}
with $z = m^2/2\xi(1-\xi)\omega$ as before, where the second term is a mass renormalisation
counter-term.\footnote{The $eA\phi^2$ theory in four dimensions requires a mass renormalisation.
The corresponding modification to the vacuum polaristion tensor $\Pi(x,x')$ produces the 
second term in (\ref{cp}), as explained in detail in ref.\cite{Hollowood:2011yh}, sections 5 and 7. 
Note that this means that keeping the $A$ field massless in this theory is a fine-tuning, 
unlike the case of QED where the real photon is kept massless by gauge invariance.
Compared to the formulae of \cite{Hollowood:2011yh}, we have always taken the initial value surface
to be $u_0 = -\infty$ here.}
Using the VVM determinant for the beam shockwave, we have
\EQ{
\Delta n(u,\omega)=\frac{\alpha m^2}{16\pi\omega^2}\int_0^1d\xi\,\int_{u}^{\infty-i0^+} dt\,
\Big[e^{-izt}\Big[(t+\sigma u(u-t))^{-1}-\RE e^{-iz t}t^{-1}\Big]\ ,
}
and performing the $t$ integral gives
\EQ{
\Delta n(u,\omega)=\frac{\alpha m^2}{16\pi\omega^2}\int_0^1d\xi\,
\Big\{\frac1{1-\sigma u}\exp\Big[\frac{i\sigma u^2 z}{1-\sigma u}\Big]\Gamma\Big(0,\frac{iuz}{1-\sigma u}\Big)-
\RE\Gamma(0,iuz)\Big\}\ .
\label{pp2}}
The $u$-dependence of the refractive index for different fixed values of the frequency is shown in fig.~\ref{h1}.

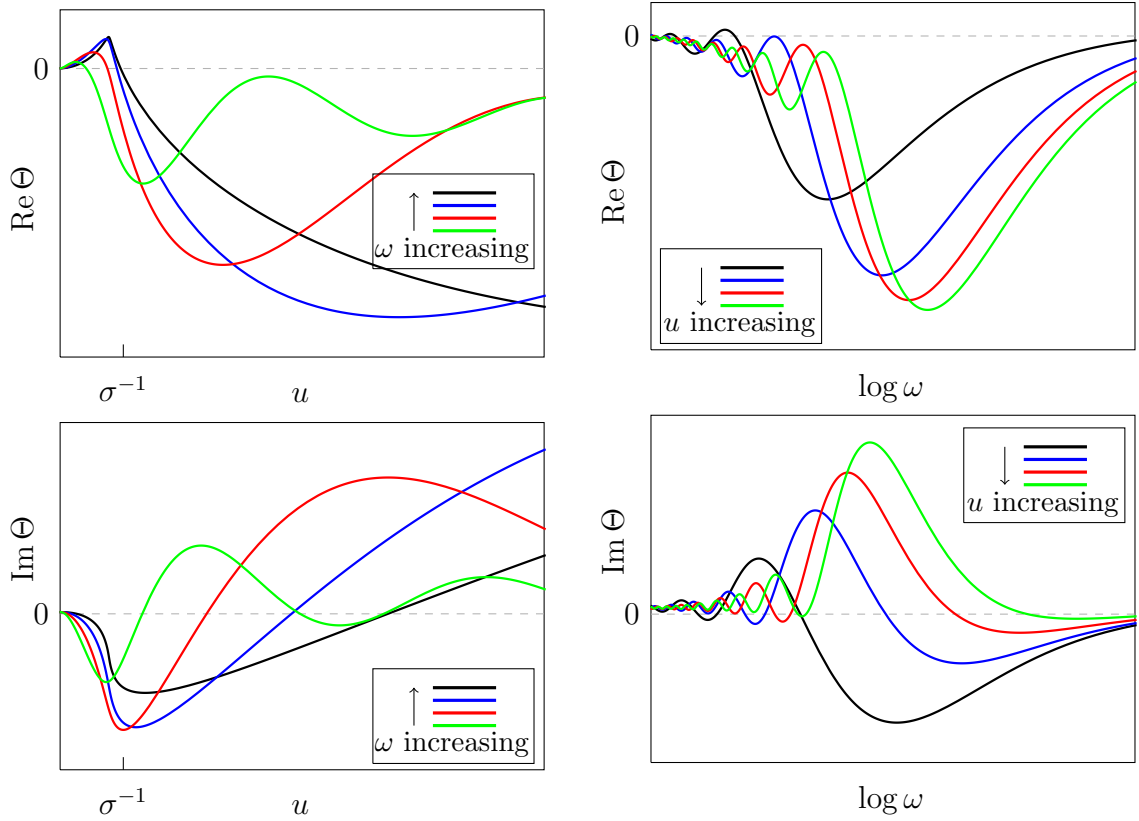
\begin{figure}[ht]
\begin{center}
\begin{tikzpicture}[scale=0.84]
\draw[-] (0,0.5) -- (7.66,0.5) -- (7.66,6) -- (0,6) -- (0,0.5);
\draw[-] (1,0.5) -- (1,0.7);
\node at (1,0) {$\sigma^{-1}$};
\node at (-0.3,5.07) {0};
\draw[dashed,black!30] (0,5.07) -- (7.66,5.07);
\pgftext[at=\pgfpoint{-0.15cm}{0.86cm},left,base]{\pgfuseimage{spic3}} 
\node at (3.8,-0.1) {$u$};
\node[rotate=90] at (-0.6,3) {$\RE\Theta$};
\begin{scope}[xshift=5.9cm,yshift=2.3cm,scale=1]
\draw (-0.95,-0.4) -- (1.6,-0.4) -- (1.6,1.1) -- (-0.95,1.1) -- (-0.95,-0.4);
\node at (0.3,-0.1) {\small $\omega$ increasing};
\draw[->] (-0.3,0.2) -- (-0.3,0.8);
\draw[very thick] (0,0.8) -- (1,0.8);
\draw[very thick,color=blue] (0,0.6) -- (1,0.6);
\draw[very thick,color=red] (0,0.4) -- (1,0.4);
\draw[very thick,color=green] (0,0.2) -- (1,0.2);
\end{scope}
\end{tikzpicture}
\hspace{0.2cm}
\begin{tikzpicture}[scale=0.84]
\draw[-] (0,0.5) -- (7.66,0.5) -- (7.66,6) -- (0,6) -- (0,0.5);
\node at (-0.3,5.47) {0};
\draw[dashed,black!30] (0,5.47) -- (7.66,5.47);
\pgftext[at=\pgfpoint{-0.15cm}{0.86cm},left,base]{\pgfuseimage{spic5}} 
\node at (3.8,-0.1) {$\log\omega$};
\node[rotate=90] at (-0.6,3) {$\RE\Theta$};
\begin{scope}[xshift=1.1cm,yshift=1cm,scale=1]
\draw (-0.95,-0.4) -- (1.6,-0.4) -- (1.6,1.1) -- (-0.95,1.1) -- (-0.95,-0.4);
\node at (0.3,-0.1) {\small $u$ increasing};
\draw[<-] (-0.3,0.2) -- (-0.3,0.8);
\draw[very thick] (0,0.8) -- (1,0.8);
\draw[very thick,color=blue] (0,0.6) -- (1,0.6);
\draw[very thick,color=red] (0,0.4) -- (1,0.4);
\draw[very thick,color=green] (0,0.2) -- (1,0.2);
\end{scope}
\end{tikzpicture}
\begin{tikzpicture}[scale=0.84]
\draw[-] (0,0.5) -- (7.66,0.5) -- (7.66,6) -- (0,6) -- (0,0.5);
\draw[-] (1,0.5) -- (1,0.7);
\node at (1,0) {$\sigma^{-1}$};
\node at (-0.3,2.97) {0};
\draw[dashed,black!30] (0,2.97) -- (7.66,2.97);
\pgftext[at=\pgfpoint{-0.15cm}{0.86cm},left,base]{\pgfuseimage{spic4}} 
\node at (3.8,-0.1) {$u$};
\node[rotate=90] at (-0.6,4) {$\IM\Theta$};
\begin{scope}[xshift=5.9cm,yshift=1cm,scale=1]
\draw (-0.95,-0.4) -- (1.6,-0.4) -- (1.6,1.1) -- (-0.95,1.1) -- (-0.95,-0.4);
\node at (0.3,-0.1) {\small $\omega$ increasing};
\draw[->] (-0.3,0.2) -- (-0.3,0.8);
\draw[very thick] (0,0.8) -- (1,0.8);
\draw[very thick,color=blue] (0,0.6) -- (1,0.6);
\draw[very thick,color=red] (0,0.4) -- (1,0.4);
\draw[very thick,color=green] (0,0.2) -- (1,0.2);
\end{scope}
\end{tikzpicture}
\hspace{0.2cm}
\begin{tikzpicture}[scale=0.84]
\draw[-] (0,0.5) -- (7.66,0.5) -- (7.66,6) -- (0,6) -- (0,0.5);
\node at (-0.3,2.85) {0};
\draw[dashed,black!30] (0,2.85) -- (7.66,2.85);
\pgftext[at=\pgfpoint{-0.15cm}{0.86cm},left,base]{\pgfuseimage{spic6}} 
\node at (3.8,-0.1) {$\log\omega$};
\node[rotate=90] at (-0.6,4) {$\IM\Theta$};
\begin{scope}[xshift=5.9cm,yshift=4.7cm,scale=1]
\draw (-0.95,-0.4) -- (1.6,-0.4) -- (1.6,1.1) -- (-0.95,1.1) -- (-0.95,-0.4);
\node at (0.3,-0.1) {\small $u$ increasing};
\draw[<-] (-0.3,0.2) -- (-0.3,0.8);
\draw[very thick] (0,0.8) -- (1,0.8);
\draw[very thick,color=blue] (0,0.6) -- (1,0.6);
\draw[very thick,color=red] (0,0.4) -- (1,0.4);
\draw[very thick,color=green] (0,0.2) -- (1,0.2);
\end{scope}
\end{tikzpicture}
\caption{The real and imaginary parts of the phase $\Theta(u,\omega)$ in $eA\phi^2$
theory as functions of $u$ and the frequency $\omega$.}
\label{h2}
\end{center}
\end{figure}

Once again, we can integrate \eqref{pp2} numerically to find the local phase $\Theta(u,\omega)$ and the corresponding 
coordinate shift. The results for $\Theta$ are shown in fig.~\ref{h2}, as functions of $u$ and $\omega$.
These plots show many features in common with those of QED but also significant differences, especially in the
$\omega$ dependence, related to the distinct UV behaviour of the $eA\phi^2$ theory.

Before commenting on these plots, we can again determine the high frequency behaviour analytically. We find
\EQ{
\Delta n(u,\omega\to\infty)=-\frac{\alpha m^2}{16\pi\omega^2}\cdot\frac{\sigma u}{\sigma u-1-i0^+}\log\omega
+{\cal O}(\omega^{-2})\ ,
}
while the high-frequency behaviour of the phase is
\EQ{
\Theta(u,\omega\to\infty)
=-\frac{\alpha m^2}{16\pi\sigma}\Big(\sigma u+\log\big|\sigma u-1\big|
+i\pi\vartheta(\sigma u-1)\Big)\frac{\log\omega}\omega+{\cal O}(\omega^{-1}) \ .
}

These expressions show several differences from the equivalent formulae for QED. First, notice that a relative factor of
$i$ difference in the integral expressions for $\Delta n(u,\omega)$, itself related to the different power counting,
effectively reverses the real and imaginary parts of the refractive index. This is evident in fig.~\ref{h1} where the
delta function-like behaviour at the focal point appears in $\IM \Delta n(u,\omega)$. However, power counting
also results in a different $\omega$-dependence. Here, $\Delta n(u,\omega) \sim \omega^{-2}\log\omega$ 
at high frequencies, so the near-singular behaviour at the focal point is softened and vanishes in the $\omega\to\infty$
limit.

This softening of the behaviour near the focal point is also evident in the plots in fig.~\ref{h2} showing the 
$u$-dependenceof the phase $\Theta(u,\omega)$. These also show the $u$-independent, linear $\omega$ dependence
of $\RE\Theta(u,\omega)$ at low frequencies implied by the effective Lagrangian (see \eqref{ck} above).
At high frequencies, however, we now find $\Theta(u,\omega) \sim \omega^{-1}\log\omega$, so the phase
itself also vanishes in this limit. The frequency dependence of the corresponding coordinate shift is qualitatively similar
to fig.~\ref{g2}.

Overall, therefore, the essential features of the refractive index and phase which ensure that causality is not violated
also appear in the $eA\phi^2$ theory.  However, its super-renormalizable nature implies a softer high-frequency
behaviour for the refractive index and scattering phase, while the coordinate shift $\Delta v$ vanishes at high frequency
as $\omega^{-2}\log\omega$. Once again, this demonstrates that causality is respected and, just as for QED, time machine
constructions do not work.

\section{Conclusions}\label{s8}

In this paper, we have demonstrated that (effective) actions for QFTs in gravitational backgrounds which, on their own, 
violate causality are not necessarily unphysical, but may be valid low-energy effective actions if they can be embedded 
in a causal, UV complete theory. Superluminality in a low-energy effective action {\it in curved spacetime} can
therefore not be used by itself to discard the theory as unphysical. The key question is whether there exists
a consistent UV completion. 

Notice the key r\^ole of gravity in this conclusion. For theories in flat space, the combination of the optical theorem
(which implies ${\rm Im}~n(\omega) > 0$) and the Kramers-Kr\"onig dispersion relation for the refractive index,
\begin{equation}
n(\infty) ~=~ n(0) - \frac{2}{\pi} \int_0^\infty \frac{d\omega}{\omega}~ {\rm Im}~n(\omega) \qquad \qquad 
{\rm (flat~space)}\ ,
\label{concl1}
\end{equation}
would imply that the high-frequency phase velocity exceeds its low-frequency limit, $v_\text{ph}(\infty) > v_\text{ph}(0)$.
The superluminal causality violations in the IR effective theory would therefore be inherited by its UV completion
and the theory would indeed be unphysical. In curved spacetime, however, the novel analytic properties of the 
relevant Green functions induced by the background geometry modify the dispersion relations and invalidate
this conclusion. In our example, this is evident in the branch cuts extending to the origin in the complex 
$\omega$-plane in the explicit expressions for the refractive indices in (\ref{db}) and (\ref{pp2}).
We will return to the issue of dispersion relations in curved spacetime theories in \cite{HS3}.

We explored these ideas in the challenging case of QFTs in a gravitational shockwave background, for which the classical 
null geodesics for a particle crossing the shock wavefront experience a discontinuous Shapiro time advance
$\Delta v_{\rm AS} < 0$ in Aichelburg-Sexl coordinates. The corresponding causality issues for the
classical, effective, and full UV-complete theories were interrogated in a two-shockwave time machine scenario. 
We showed that while a correct treatment of the shock wavefront, in accordance with the equivalence
principle, ensured that causality was respected at the classical level, the additional coordinate shift
$\Delta v_{\rm DH}$ implied by the effective action did permit a causality-violating, time machine trajectory.
However, the vanishing of the coordinate shift at high-frequency, $\Delta v(u,\omega\rightarrow\infty)\rightarrow 0$, 
ensures that causality is restored in the fundamental, UV complete theory.

To establish these results, we calculated the complete frequency dependence of the refractive index and phase
shift $\Theta(u,\omega)$ for a photon scattering from a gravitational shockwave in QED itself. 
First, in contrast to the prediction of the effective action, all the quantities $\Delta n(u,\omega)$, 
$\Theta(u,\omega)$ and $\Delta v(u,\omega)$ were shown to be {\it local} in the full theory,
depending on the lightcone distance beyond the shockwave. This smoothing out of the discontinuous jumps
associated with the effective action reflects the scale of the vacuum polarization cloud 
as it passes through the shockwave. Curiously though, in the high-frequency limit for QED in the beam shockwave, 
the shifts $\Delta v$ and $\Delta\Theta$ again become step functions, but this time taking place at the focal
point $u = 1/\sigma$ of the classical null geodesic congruence.
This behaviour was calculated analytically in (\ref{focal}), and is yet another reason contributing to the failure of the
shockwave time machine.
Similar discontinuities at the focal point were also found for the particle shockwave and $A\phi^2$ theory.

\begin{figure}[ht]
\begin{center}
\begin{tikzpicture}[scale=0.83]
\draw[-] (0,0.5) -- (7.66,0.5) -- (7.66,6) -- (0,6) -- (0,0.5);
\node at (-0.6,2.3) {\footnotesize $\D v_{\small\text{DH}}$};
\node at (-0.3,6) {0};
%\draw[dashed,black!30] (0,2.6) -- (7.66,2.6);
\pgftext[at=\pgfpoint{-0.15cm}{1.25cm},left,base]{\pgfuseimage{npic8}} 
\node at (3.8,-0.1) {$\log\omega$};
\node[rotate=90] at (-0.6,3.5) {$\RE\Delta v$};
\begin{scope}[xshift=5.9cm,yshift=1cm,scale=1]
\draw (-0.98,-0.4) -- (1.6,-0.4) -- (1.6,1.1) -- (-0.98,1.1) -- (-0.98,-0.4);
\node at (0.3,-0.1) {\small $u$ increasing};
\draw[<-] (-0.3,0.2) -- (-0.3,0.8);
\draw[very thick] (0,0.8) -- (1,0.8);
\draw[very thick,color=blue] (0,0.6) -- (1,0.6);
\draw[very thick,color=red] (0,0.4) -- (1,0.4);
\draw[very thick,color=green] (0,0.2) -- (1,0.2);
\end{scope}
\end{tikzpicture}
\hspace{0.1cm}
\begin{tikzpicture}[scale=0.84]
\draw[-] (0,0.5) -- (7.66,0.5) -- (7.66,6) -- (0,6) -- (0,0.5);
\node at (-0.3,6) {0};
\draw[dashed,black!30] (0,1.2) -- (7.66,1.2);
\node at (-0.6,1.2) {\footnotesize $-\frac\alpha{12}$};
\pgftext[at=\pgfpoint{-0.15cm}{1.25cm},left,base]{\pgfuseimage{npic7}} 
\node at (3.8,-0.1) {$\log\omega$};
\node[rotate=90] at (-0.6,3) {$\RE\Theta$};
\begin{scope}[xshift=1.1cm,yshift=2cm,scale=1]
\draw (-0.98,-0.4) -- (1.6,-0.4) -- (1.6,1.1) -- (-0.98,1.1) -- (-0.98,-0.4);
\node at (0.3,-0.1) {\small $u$ increasing};
\draw[<-] (-0.3,0.2) -- (-0.3,0.8);
\draw[very thick] (0,0.8) -- (1,0.8);
\draw[very thick,color=blue] (0,0.6) -- (1,0.6);
\draw[very thick,color=red] (0,0.4) -- (1,0.4);
\draw[very thick,color=green] (0,0.2) -- (1,0.2);
\end{scope}
\end{tikzpicture}
\caption{\footnotesize The effective coordinate shift $\RE\Delta v(u,\omega)$ and phase $\Theta(u,\omega)$
in QED as a function of $\log \omega$ at fixed lightcone distance $u$ from the (beam) shockwave.}
\label{conclfig}
\end{center}
\end{figure}
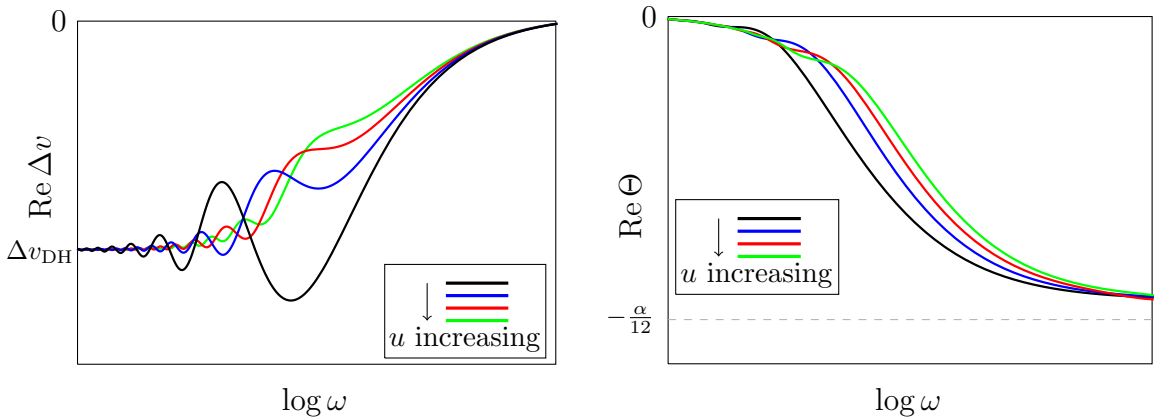
The frequency dependence of $\Delta v(u,\omega)$ and $\Theta(u,\omega)$ is most 
vividly illustrated in figs. \ref{g3} and \ref{g2}, which we reproduce here for convenience.
These plots make it clear how the coordinate shift $\Delta v(u,\omega)$ interpolates between its low-frequency
effective action value $\Delta v_{\rm DH}$ and zero at high frequency.

Notice that in QED, although we found that the refractive index becomes 1 in the high-frequency limit as required 
by causality, the phase shift itself asymptotically approaches a non-vanishing constant, as shown in fig.~\ref{conclfig}.
This initially surprising finding, which is nevertheless completely consistent with causality, led to a closer
inspection of the contrasting high-frequency behaviour in QED and the super-renormalizable scalar $A\phi^2$
theory in 4 dimensions. We found that the high-frequency behaviour of the phase shift and refractive index
(at fixed $u$ beyond the focal point) is
\SP{
\Delta n(u,\omega) &\sim - \frac{1}{\omega} \ , \qquad ~~~~
\Theta(u,\omega) \sim - \,\text{const} \qquad\qquad {\rm (QED})\\
\Delta n(u,\omega) &\sim  - \frac{\log\omega}{\omega^2}\ , \qquad 
\Theta(u,\omega) \sim  -  \frac{\log\omega}{\omega}  \qquad\qquad (A\phi^2) \ . 
\label{concl2}
}
Extrapolating this pattern suggests that non-renormalizable theories may exhibit non-causal high-frequency 
behaviour, and indeed, as will be demonstrated in \cite{HS2}, this turns out to be true.
This sheds further light on the question raised in the introduction \cite{Camanho:2014apa}, {\it viz.}~how overcoming the
causality problems inherent in the effective action could serve as a guide in constructing a consistent UV completion 
and confirms a close relationship between causality and renormalizability of the fundamental QFT. 

Finally, we comment briefly on the translation of our results to Planck energy scattering. This will be discussed in
more detail in the companion paper \cite{HS2}. 
First, in order to discuss scattering as such, we need to define the asymptotic past and future Minkowski spacetimes.
Recall that the full shockwave spacetime can be viewed as two Minkowski half-planes patched together along the surface 
$u=0$ with a displacement $\Delta v_{\rm AS}$. The classical Shapiro time advance depends on this patching, 
so giving a physical meaning to $\Delta v_{\rm AS}$ depends on making a physically motivated identification
of the past and future regions. 
For our discussion of scattering, we make the obvious choice implied by the Aichelburg-Sexl coordinates, 
thereby attributing physical significance to the classical phase shift $\omega \Delta v_{\rm AS} = - G s \log b^2/r_0^2$.

In this case, the scattering amplitude ${\mathcal A}(s,t)$ may be written in terms of the classical and quantum phase shifts as 
a Fourier transform,
\begin{equation} 
{\mathcal A}(s,t=-q^2) ~=~ - 2i s \int d^2b ~e^{i \vec{q}\cdot\vec{b}} ~\left[
\exp i \left(- \frac{s}{M_p^2} \log \frac{b^2}{r_0^2} +\Theta_\text{scat.}(\hat s) \right) - 1 \right] \ .
\label{concl3}
\end{equation}
Here, $s$ and $t$ are the usual Mandelstam variables and we have introduced the Planck mass $M_p$ through $G= 1/M_p^2$.
Crucially, $\Theta_\text{scat.}(\hat s)$, defined as the $u\rightarrow \infty$ limit of the phase $\Theta(u,\omega)$, is a
function of the single key variable
\begin{equation}
\hat s = \frac{s}{M_p^2} \left(\frac{\lambda_c}{b}\right)^2 \ ,
\label{concl4}
\end{equation}
which combines the scattering CM energy and the impact parameter. Notice especially the r\^ole played by the QFT scale $\lambda_c$
which characterises the vacuum polarization cloud. The equivalent r\^ole for the Reggeized UV completion in the case of
graviton scattering is played by the string scale $\lambda_s$ \cite{Camanho:2014apa,D'Appollonio:2015gpa}.

Here, we have shown that in the case of photon-shockwave scattering at near Planck energies, $\Theta_\text{scat.}(\hat s)$ is
an exactly calculable function in QFT for all values of $\hat s$, including the crucial high-$\hat s$ limit.
This demonstrates that the full amplitude ${\mathcal A}(s,t)$ is entirely compatible with causality for (super-) renormalizable
QFTs, despite the apparent causality problems associated with their IR effective actions.
Further discussion of the properties of the Planck energy scattering amplitude ${\mathcal A}(s,t)$ and its relation to UV completion
and renormalizability in QFT may be found in \cite{HS2}.

\section*{Acknowledgements}

\noindent This research was supported in part by STFC grant
ST/L000369/1.  We are grateful to Gabriele Veneziano for helpful
discussions and correspondence and the TH Division, CERN for hospitality
during key stages of this work.

\end{document}